\documentclass[]{pasj01}
\usepackage{url}
\usepackage{natbib}
\usepackage{color}
\usepackage{latexsym}
\usepackage{morefloats}
\usepackage{xspace}
\usepackage[usenames, dvipsnames]{xcolor}
\usepackage{colortbl}
\usepackage{multirow}
\usepackage{caption}
\usepackage{graphicx}
\usepackage{bm}
\DeclareGraphicsExtensions{.pdf,.png,.eps}
\def\avrg#1{\left\langle #1 \right\rangle}
\def\btheta{\boldsymbol{\theta}}
\def\etal{et~al.}
\begin{document}
\SetRunningHead{Murata et al.}{HSC CAMIRA cluster mass-richness relation}
\Received{}
\Accepted{}
%
\title{The mass-richness relation of optically-selected clusters from weak gravitational lensing and abundance with Subaru HSC first-year data}
\author{Ryoma~\textsc{Murata}\altaffilmark{1,2}}
\author{Masamune~\textsc{Oguri}\altaffilmark{3,2,1}}
\author{Takahiro~\textsc{Nishimichi}\altaffilmark{4,1}}
\author{Masahiro~\textsc{Takada}\altaffilmark{1}}
\author{Rachel~\textsc{Mandelbaum}\altaffilmark{5}}
\author{Surhud~\textsc{More}\altaffilmark{6, 1}}
\author{Masato~\textsc{Shirasaki}\altaffilmark{7}}
\author{Atsushi~J.~\textsc{Nishizawa}\altaffilmark{8}}
\author{Ken~\textsc{Osato}\altaffilmark{2}}
\altaffiltext{1}{Kavli Institute for the Physics and Mathematics of the Universe (WPI), The University of Tokyo Institutes for Advanced Study, The University of Tokyo, Kashiwa, Chiba 277-8583, Japan}
\altaffiltext{2}{Department of Physics, The University of Tokyo, Tokyo 113-0033, Japan}
\altaffiltext{3}{Research Center for the Early Universe, The University of Tokyo, Tokyo 113-0033, Japan}
\altaffiltext{4}{Center for Gravitational Physics, Yukawa Institute for Theoretical Physics, Kyoto University, Kyoto 606-8502, Japan}
\altaffiltext{5}{McWilliams Center for Cosmology, Department of Physics, Carnegie Mellon University, Pittsburgh, PA 15213, USA}
\altaffiltext{6}{The Inter-University Center for Astronomy and Astrophysics, Post bag 4, Ganeshkhind, Pune, 411007, India}
\altaffiltext{7}{National Astronomical Observatory of Japan (NAOJ), Mitaka, Tokyo 181-8588, Japan}
\altaffiltext{8}{Institute for Advanced Research, Nagoya University Furocho, Chikusa-ku, Nagoya, 464-8602, Japan}
\email{ryoma.murata@ipmu.jp}
%
%
\maketitle
\begin{abstract}
Constraining the relation between the richness $N$ and the halo mass $M$ 
over a wide redshift range for optically-selected clusters 
is a key ingredient for cluster-related science
in optical surveys, including the Subaru Hyper Suprime-Cam (HSC) survey.
We measure stacked weak lensing profiles 
around 1747 HSC CAMIRA clusters over a redshift range of $0.1\leq z_{\rm cl}\leq 1.0$ 
with $N\geq 15$ using
the HSC first-year shear catalog covering $\sim$$140$ ${\rm deg^2}$.
The exquisite depth and image quality of the HSC survey allow us to 
measure lensing signals around high-redshift clusters 
at $0.7\leq z_{\rm cl}\leq 1.0$ with a signal-to-noise ratio of $19$ 
within comoving radius range $0.5\lesssim R\lesssim 15 h^{-1}{\rm Mpc}$.
We constrain richness-mass relations $P(\ln N|M,z)$
of HSC CAMIRA clusters assuming a log-normal distribution without informative priors on model parameters,
by jointly fitting to the lensing profiles and abundance measurements
under both {\it Planck} and {\it WMAP} cosmological models.
We show that our model gives acceptable $p$-values
when we add redshift-dependent terms proportional to $\ln (1+z)$ and $[\ln (1+z)]^{2}$
in the mean and scatter relations of $P(\ln N|M,z)$.
Such terms presumably originate from the variation of photometric redshift errors as a function of redshift.
We show that constraints 
on the mean relation $\langle M|N \rangle$ are consistent between the {\it Planck} and {\it WMAP} models,
whereas the scatter values $\sigma_{\ln M|N}$ for the {\it Planck} model 
are systematically larger than those for the {\it WMAP} model. 
We also show that the scatter values for the {\it Planck} model increase toward lower richness values, 
whereas those for the {\it WMAP} model are consistent with constant values as a function of richness.
This result highlights the importance of the scatter in the mass-richness relation for cluster cosmology.
\end{abstract}
\section{Introduction} \label{sec:intro}
Clusters of galaxies are dominated by dark matter and therefore
are useful sites for cosmological studies 
since $N$-body simulations can predict cluster observables reasonably well.
The abundance and clustering of massive clusters and their time evolution 
are known to be sensitive to cosmological parameters 
such as the matter density ($\Omega_{\rm m}$), the normalization of the matter power spectrum ($\sigma_8$),
and dark energy 
\citep[see e.g.,][]{Whiteetal1993, Ekeetal1996, Kitayama&Suto1997, Haimanetal2001, Voit2005, Vikhlininetal2009, Mantzetal2010, Rozoetal2010, Allenetal2011, OguriandTakada2011, Weinbergetal2013, PlanckCollaboration2016b}.
Clusters of galaxies also play 
an important role in understanding galaxy formation physics
via their possible large environmental effect
\citep[e.g.,][]{Renzini2006, Kravtsov&Borgani2012}.

Clusters of galaxies can be identified in optical, X-ray, and radio/mm/submm wavelengths.
The recent development of wide-field optical imaging surveys
makes an optical selection of clusters particularly powerful.
This is because optical surveys take wide-field images with multiple photometric passbands
from which we can select clusters of galaxies efficiently
via the enhancement of galaxy number counts and derive photometric redshifts of clusters
\citep[e.g.,][]{Gladders&Yee2000}.
Ongoing and upcoming wide-field optical or infrared galaxy surveys
allow us to study the cosmology and galaxy formation physics
in great detail if systematic errors are under control.
Such surveys include the Kilo-Degree Survey \citep[KiDS;][]{deJongetal2013, Kuijkenetal2015},
the Dark Energy Survey \citep[DES; ][]{Flaugher2005, DESetal2016},
and the Subaru Hyper Suprime-Cam (HSC) survey \citep[][]{Aiharaetal2018a, Aiharaetal2018b} 
for ongoing surveys,
and the Large Synoptic Survey Telescope \citep[LSST;][]{Ivezicetal2008},
{\it Euclid} \citep{Laureijsetal2011},
and {\it Wide-Field Infrared Survey Telescope}
\citep[{\it WFIRST};][]{Spergeletal2015} for upcoming surveys.

Since theoretical predictions of cluster observables 
are primarily determined with respect to the halo mass for a given
cosmological model, we need a valid statistical model 
to connect the halo mass and observed mass proxy
in order
to make full use of cluster samples.
In optical surveys, a commonly-used mass proxy is an
optical richness, which roughly corresponds 
to the number of red-sequence member galaxies above some luminosity threshold
in each cluster \citep[e.g.,][]{Rozoetal2009}.
Well-calibrated and unbiased mass-richness relations allow us
to infer cluster masses from observed richness values.

Weak gravitational lensing provides a powerful means 
to constrain mass-observable relations of clusters.
It is the deflection of light due to the 
intervening matter density field along the line-of-sight direction
to produce a coherent distortion pattern 
in background galaxy shapes \citep[for reviews, see e.g.,][]{Bartelmann&Schneider2001, Kilbinger2015, Mandelbaum2018}.
Stacked weak lensing measurements statistically probe the projected average mass distribution of clusters
with equal sensitivity to the dark and baryonic matter.
Stacking shapes of background galaxies for a sample of clusters
enhances the signal-to-noise ratio of the measurements.
Previous studies have utilized the weak gravitational lensing effect
to constrain mass-observable relations, including mass-richness relations 
\citep[e.g.,][]{Johnstonetal2007, Leauthaudetal2010, Okabeetal2013, vonderLindenetal2014, 
Hoekstraetal2015, Battagliaetal2016, Simetetal2017, Melchioretal2017, Murataetal2018, Medezinskietal2018a, Miyatakeetal2018, McClintocketal2018}.

The Hyper Suprime-Cam Subaru Strategic Program (HSC-SSP) is a wide-field optical imaging survey with a 1.77 ${\rm deg}^2$ 
field-of-view camera on the 8.2-meter Subaru telescope 
\citep{Miyazakietal2012, Miyazakietal2015, Miyazakietal2018a, Komiyamaetal2018, Furusawaetal2018, Kawanomotoetal2018}.
The HSC survey is unique in its combination of depth 
and high-resolution image quality, 
which allows us to detect clusters of galaxies over a wide redshift range up to $z_{\rm cl} \sim 1$ and 
to measure stacked lensing profiles around such high-redshift clusters 
with a lower shape noise than 
that of other ongoing surveys such as KiDS and DES. 

In this paper, we present constraints
on the relation between the optical richness and halo mass 
of optically-selected HSC clusters in the redshift range 
$0.1 \leq z_{\rm cl} \leq 1.0$ \citep{Ogurietal2018a} 
detected by the CAMIRA cluster-finding algorithm \citep{Oguri2014}. 
For this purpose, we conduct
a joint analysis of the abundance and stacked lensing profiles
from the first-year data catalogs of the Subaru HSC-SSP survey.
For this work,
we adapt and apply a pipeline developed in \cite{Murataetal2018},
which was used for clusters at $0.1\leq z_{\rm cl}\leq 0.33$ 
in the Sloan Digital Sky Survey (SDSS)
selected by the redMaPPer cluster-finding algorithm \citep{Rykoffetal2014, Rozoetal2014, Rozoetal2015a, Rozoetal2015b, Rykoffetal2016}.
We model the probability distribution of the richness 
for a given halo mass and redshift $P(\ln N|M, z)$ 
without informative prior distributions for richness-mass parameters. 
We then use Bayes theorem to calculate the mass-richness relation $P(\ln M|N)$ in each redshift bin.
In order to accurately model the abundance and stacked lensing profiles,
we use \textsc{Dark Emulator} \citep{Nishimichietal2018}, which is 
constructed from a suite of high-resolution $N$-body simulations.
We employ an analytic model for the covariance matrix describing statistical errors for the abundance and stacked lensing profiles.
We validate the analytic covariance matrix against realistic mock shear and halo catalogs \citep{Shirasakietal2019}.

The structure of this paper is as follows.
In Section~\ref{sec:data}, we briefly describe the Subaru HSC data and catalogs used in our richness-mass relation analysis.
In Section~\ref{sec:measure}, 
we describe measurements of the cluster abundance and stacked cluster lensing profiles.
In Section~\ref{sec:modeling}, 
we summarize model ingredients for our richness-mass relation analysis.
In Section~\ref{sec:results}, we show the resulting constraints on the mass-richness relation.
We discuss the robustness of our results in Section~\ref{sec:discussion}.
We conclude in Section~\ref{sec:conclusion}.

Throughout this paper we use natural units where the speed of light is set equal to unity, $c=1$.
We use $M\equiv M_{\rm 200m}= 4\pi (R_{200{\rm m}})^3\bar{\rho}_{\rm m0}\times 200/3$ for the halo mass definition, 
where $R_{\rm 200m}$ is the spherical halo boundary comoving radius within which 
the mean mass density is $200$ times the present-day mean mass density. 
We note that in this paper we use a radius and density in comoving coordinates rather than in physical coordinates.
We adopt the standard flat $\Lambda$-dominated cold dark matter model 
as the fiducial cosmological model
with the parameters from the \textit{Planck15} (hereafter the {\it Planck}) result \citep{PlanckCollaboration2016}:
$\Omega_{\rm b0}h^2=0.02225$ and $\Omega_{\rm c0}h^2=0.1198$ for the density parameters of baryon and cold dark matter, respectively,
$\Omega_{\Lambda}=0.6844$ for the cosmological constant,
$\sigma_8=0.831$ for the normalization of the matter fluctuation,
and $n_s=0.9645$ for the spectral index.
We also use cosmological parameters consistent with those from {\it WMAP9} (hereafter the {\it WMAP}) 
results \citep{Hinshawetal2013}
to compare with the results when assuming the {\it Planck} cosmological parameters:
$\Omega_{\rm b0}h^2=0.02254$, 
$\Omega_{\rm c0}h^2=0.1142$, 
$\Omega_{\Lambda}=0.721$,
$\sigma_8=0.82$,
and $n_s=0.97$.
%
\section{HSC first-year dataset} \label{sec:data}
%
\subsection{HSC-SSP survey}
HSC
is a wide-field prime focus camera with a $1.5$ ${\rm deg}$ diameter field-of-view
mounted on the 8.2-meter Subaru telescope 
\citep{Miyazakietal2012, Miyazakietal2015, Miyazakietal2018a, Komiyamaetal2018, Furusawaetal2018, Kawanomotoetal2018}.
With its unique combination of a wide field-of-view,
a large aperture of the primary mirror, and 
excellent image quality, 
HSC enables us to measure lensing signals out to a relatively high redshift.
Under the Hyper Suprime-Cam Subaru Strategic Program \citep[HSC-SSP;][]{Aiharaetal2018a},
the HSC is conducting a multi-band wide-field imaging survey over six years with 300 nights of Subaru time.
The HSC-SSP survey consists of three layers: Wide, Deep, and UltraDeep.
The Wide layer is designed for weak lensing science and aims at covering 1400 deg$^2$ of 
the sky with five broad bands, ${\it grizy}$, 
with a $5 \sigma$ point-source depth of $r\sim 26$.
The $i$-band images are taken when the seeing is better
since $i$-band images are used for galaxy shape measurements for weak lensing analysis,
resulting in a median PSF FWHM of $\sim$$0_{.}^{''}58$ for $i$-band images 
for the HSC first-year shear catalog described in Section~\ref{sec:data:lensing}.
The software pipelines that reduce the data are described in \cite{Boschetal2018}.

While the HSC-SSP Data Release 1 \citep[][]{Aiharaetal2018b} is
based on data taken on 61.5 nights between March 2014 and November 2015,
in this paper we use HSC cluster, lensing shear, and photometric redshift catalogs 
based on the S16A internal release data of the HSC-SSP survey
that was taken during March 2014 through April 2016, about 90 nights in total.

\subsection{HSC CAMIRA cluster catalog} \label{sec:data:cluster}
We use the CAMIRA (Cluster finding Algorithm based on Multi-band Identification of Red-sequence gAlaxies)
cluster catalog from the S16A internal release data of the HSC-SSP Wide dataset presented in \cite{Ogurietal2018a}, 
which was constructed using 
the CAMIRA algorithm \citep{Oguri2014}.
The CAMIRA algorithm is a red-sequence cluster finder based on a stellar population synthesis model \citep{Bruzual&Charlot2003}
to predict colors of red-sequence galaxies at a given redshift 
and to compute likelihoods of being red-sequence galaxies as a function of redshift.
In addition, \cite{Ogurietal2018a} calibrated the stellar population synthesis model with spectroscopic galaxies to improve its accuracy.
In the CAMIRA algorithm, the richness corresponds to the number of red member galaxies with stellar mass $M_{*}\gtrsim 10^{10.2} M_{\odot}$ (roughly corresponding to a luminosity range of $L \gtrsim 0.2 L_{*}$) within a circular aperture with radius $R \lesssim 1\ h^{-1} {\rm Mpc}$ in physical coordinates.
The CAMIRA algorithm does not include a richness-dependent scale radius to define the richness, unlike the redMaPPer algorithm \citep{Rykoffetal2012}.
The HSC images are deep enough to detect cluster member galaxies down to $M_{*}\sim 10^{10.2}
M_{\odot}$ even at a cluster redshift $z_{\rm cl}\sim 1$,
which allows a reliable cluster detection at such high redshifts without a richness incompleteness correction.
The algorithm employs a spatially-compensated filter such that the background level is automatically subtracted in deriving the richness.
The CAMIRA algorithm identifies a brightest cluster galaxy (BCG) for each cluster candidate 
that is defined by a peak in the three-dimensional richness map in RA, dec, and redshift space \citep{Oguri2014, Ogurietal2018a}. 
The cluster centers are defined as the locations of the 
identified BCGs. 
The mask-corrected richness $N$ and
cluster photometric redshift $z_{\rm cl}$ are 
refined iteratively during the BCG identification process.
The offset of BCG positions from matched X-ray cluster centers is investigated in \cite{Ogurietal2018a}.
The bias and scatter in photometric cluster redshifts of $\Delta z_{\rm cl}/(1+z_{\rm cl})$
are shown to be better than $0.005$ and $0.01$ respectively with 4$\sigma$ clipping over most of the redshift range
by using available spectroscopic redshifts of BCGs.
We use the cluster catalog without applying a bright-star mask \citep{Ogurietal2018a}.

The catalog contains $1921$ clusters of $0.1 \leq z_{\rm cl} \leq 1.1$ and richness $N \geq 15$
with almost uniform completeness and purity over the sky region.
We use 1747 CAMIRA clusters with $15 \leq N \leq 200$ and $0.1 \leq z_{\rm cl} \leq 1.0$.
The total area for the CAMIRA clusters is estimated to be $\Omega_{\rm tot}=232.8$ ${\rm deg}^2$ \citep{Ogurietal2018a}, 
which is larger than the area for the lensing shear catalog presented in Section~\ref{sec:data:lensing}.
The CAMIRA algorithm calculates the mask area fraction $f_{\rm mask}$ to correct for the richness 
and adopts the minimum mask fraction value to reject detections \citep{Oguri2014}.
Thus areas for clusters with lower redshifts are slightly smaller since the richness is defined within a circular aperture 
with a radius $R \lesssim 1\ h^{-1} {\rm Mpc}$ in physical coordinates 
and clusters at lower redshift have a higher rejection rate from the mask cut.
We use a random catalog to estimate this effect by injecting clusters at the catalog level 
into the footprint to calculate the rejection rate from the masking cut as a function of cluster redshift.
We then define a weighting function to quantify the masking effect as

\begin{equation}
w_{\mathrm{rand}}(z_{\rm cl})=\frac{n_{\rm sample}(z_{\rm cl})}{n_{\rm keep}(z_{\rm cl})},
\label{eq:w_r}
\end{equation}
where $n_{\rm sample}(z_{\rm cl})$ and $n_{\rm keep}(z_{\rm cl})$ are the number of injected clusters 
and the number of clusters not rejected by the masking cut, respectively \citep[similarly defined in][]{Murataetal2018}.
In addition, we define the effective area of the CAMIRA cluster catalog at a given redshift as
\begin{equation}
\Omega_{\rm eff}(z_{\rm cl})=\frac{\Omega_{\rm tot}}{w_{\rm rand}(z_{\rm cl})},
\label{eq:Omega_eff}
\end{equation}
to account for the detection efficiency as a function of redshift in measurements below,
although this effect is not very large for the CAMIRA clusters with $w_{\rm rand}(z_{\rm cl}=0.1)=1.02$ at most. 
Here we note that $\Omega_{\rm eff}(z_{\rm cl}) \leq \Omega_{\rm tot}$.
We use the same random catalog to measure lensing signals for the subtraction of systematics in Section~\ref{sec:measurements:lens}.
\subsection{HSC weak lensing shear catalog} \label{sec:data:lensing}
We employ the HSC first-year shear catalog \citep{Mandelbaumetal2018a, Mandelbaumetal2018b} 
based on the S16A internal release data
for weak lensing measurements around HSC CAMIRA clusters described in Section~\ref{sec:data:cluster}.
The galaxy shapes are measured on coadded $i$-band images 
with the re-Gaussianization moment-based method \citep{HirataSeljak2003}
and fully described in \cite{Mandelbaumetal2018b}\footnote{In this method, galaxy shapes are defined in terms of distortion: 
$(e_1, e_2)=(e \cos 2\phi, e \sin 2\phi)$ with $e=(a^2 - b^2)/(a^2 + b^2)$
where $a$ and $b$ are the major and minor axes of galaxy shape respectively, 
and $\phi$
indicates the position angle with respect to the RA/dec coordinate system
\citep{Bernstein&Jarvis2002}. }.
This method has been applied extensively to SDSS data,
and thus the systematics of the method are well understood 
\citep{Mandelbaumetal2005, Reyesetal2012, Mandelbaumetal2013}.
Both shape uncertainties and biases are estimated per galaxy
with simulations created using an open source software package \textsc{GalSim} \citep{Roweetal2015}
with galaxy samples from the {\it Hubble Space Telescope} COSMOS survey \citep[see][for more details]{Mandelbaumetal2018b}.
More specifically, 
\cite{Mandelbaumetal2018b} estimate multiplicative bias $m$, additive bias $(c_1, c_2)$,
intrinsic root-mean-square ellipticity $e_{\rm rms}$,
and shape measurement error $\sigma_e$ for the galaxy ensemble in the simulation, and define
interpolation functions to produce an estimate of that quantity for each galaxy in the real data.
The systematic uncertainty in the overall shear calibration is estimated to be $0.01$ in \cite{Mandelbaumetal2018b}.

\cite{Mandelbaumetal2018a} 
applied a number of cuts to satisfy the requirements for carrying out first-year weak lensing
cosmology analyses. 
For example, the catalog is constructed using regions of sky with approximately full depth 
in all five bands to ensure the homogeneity of the sample. 
\cite{Mandelbaumetal2018a} also limited the 
\texttt{cmodel} magnitude \citep[see][for the definition of \texttt{cmodel} magnitude]{Boschetal2018} 
with ${\it i}_{\rm cmodel}<24.5$,
which is conservative compared to the {\it i}-band magnitude limit of $\sim$$26.4$
\citep[$5 \sigma$ for point sources;][]{Aiharaetal2018a}.
As a result, 
the first-year shear catalog covers $\Omega_{\rm lens}= 136.9~ {\rm deg}^2$ in total
with six distinct fields (XMM, GAMA09H, GAMA15H, HECTOMAP, VVDS, and WIDE12H).
We note that the area coverage of the shear catalog is smaller than that of the CAMIRA clusters
($\Omega_{\rm tot}=232.8$ ${\rm deg}^2$) due to the conservative cuts for the shape measurements.
\cite{Mandelbaumetal2018a}, \cite{Ogurietal2018b}, and \cite{Hikageetal2019}
performed extensive null tests of the shear catalog 
to show that the shear catalog satisfies the requirements 
for HSC first-year weak lensing analyses using cosmic shear and galaxy-galaxy lensing.
Even after relatively conservative cuts, 
the HSC first-year shear catalog includes galaxy shapes
with a high source number density, $24.6$ (raw) 
and $21.8$ (effective) ${\rm arcmin}^{-2}$ \citep{Mandelbaumetal2018a}, 
which enables us to measure lensing signals around high redshift HSC CAMIRA clusters ($z_{\rm cl} \leq 1.0$).
\subsection{HSC photometric redshift catalog} \label{sec:data:photoz}
We use a photometric redshift (photo-$z$) catalog \citep{Tanakaetal2018}
for the source galaxies in Section~\ref{sec:data:lensing} 
estimated from the S16A internal release data of
the HSC five broadband photometry.
In the HSC survey, 
several different codes are employed to estimate the photometric redshifts:
a machine-learning code based on a self-organizing map (\texttt{MLZ}),
a classical template-fitting code (\texttt{Mizuki}),
an empirical polynomial fitting code \citep[\texttt{DEmP};][]{Hsieh&Yee2014},
an extended (re)weighing method to find the nearest neighbors in color/magnitude space
from a reference spectroscopic redshift sample (\texttt{NNPZ}),
a neural network code using PSF-matched aperture (afterburner) photometry (\texttt{Ephor AB}),
and a hybrid code combining machine learning with template fitting \citep[\texttt{FRANKEN-Z};][Speagle et al. {\it in prep.}]{Speagleetal2019}
The codes are trained, validated, and tested with spectroscopic and grism redshifts 
as well as COSMOS 30-band data with high accuracy photo-$z$ \citep{Ilbertetal2009, Laigleetal2016} 
in \cite{Tanakaetal2018}.
The photo-$z$ estimation is most accurate in the range $0.2 \lesssim z \lesssim 1.5$, where  
the HSC filter set straddles the 4000${\rm \AA}$ break \citep{Tanakaetal2018}.

Among these catalogs from different codes, 
we choose \texttt{MLZ} as the fiducial photometric redshift catalog for the lensing measurement 
in Section~\ref{sec:measurements:lens}, 
while we also use the other photometric redshift catalogs to check the robustness 
of our results to our choice of \texttt{MLZ} (see Section~\ref{subsubsec:diffphotoz}).
In this paper, we use the redshift probability distribution functions (PDFs), $P(z)$, 
and randomly (i.e., Monte Carlo) sampled point estimates drawn from the full PDFs, $z_{\rm mc}$ from \cite{Tanakaetal2018} 
for the lensing measurements and the source galaxy selection.
We also use the best point estimates $z_{\rm best}$ \citep[see section 4.2 of][]{Tanakaetal2018} 
only for lensing covariance estimation as described in Appendix~\ref{appendix:analyticcov}.
We correct for the effect of photometric redshift bias on the lensing measurements 
(More et al. {\it in prep})
using COSMOS 30-band photo-$z$ data (see Section~\ref{sec:measurements:lens}).
\section{Measurement} \label{sec:measure}
We describe the measurement method for the cluster abundance in Section~\ref{sec:measurement:abundance}
and the stacked cluster lensing profile in Section~\ref{sec:measurements:lens}.
\subsection{Cluster abundance} \label{sec:measurement:abundance}
We use the abundance of CAMIRA clusters in given richness and redshift bins 
as cluster observables to constrain the richness-mass relation of the clusters.
We divide the CAMIRA clusters into 12 bins for the abundance measurements
with four richness and three redshift bins as shown in Table~\ref{tab:binnig}.
We use the point estimate of richness and redshift of the clusters
to calculate the abundance
(i.e., we ignore any errors in the richness estimation and the cluster photometric redshift).

We measure the abundance of the clusters in each bin corrected for
the detection efficiency \citep{Murataetal2018} as
\begin{equation}
\widehat{N}_{\alpha, \beta}= \sum_{l; {N_l\in N_{\alpha}, z_l\in z_{\beta} }}\frac  {\Omega_{\rm tot}}{\Omega_{\rm eff}(z_l)},
\label{eq:N_alpha}
\end{equation}
where $N_{\alpha}$ and $z_{\beta}$ denote the $\alpha$-th richness bin and $\beta$-th redshift bin
in Table~\ref{tab:binnig}, respectively. The summation runs over
all clusters in each richness and redshift bin.
The factor $\Omega_{\rm tot}/\Omega_{\rm eff}$ corrects for the detection efficiency as discussed in Section~\ref{sec:data:cluster}.
Equation~(\ref{eq:N_alpha}) gives an estimate of the abundance of the clusters
we could observe for the survey area of $\Omega_{\rm tot}$ without the mask effect.
We then do not need to include the mask effect in the model prediction given in Section~\ref{sec:modeling:abundance}.

The numbers of clusters in the $\alpha$-th richness and $\beta$-th redshift bin before and after the 
detection efficiency correction are given in Table~\ref{tab:binnig} 
by $N_{\alpha, \beta}^{\rm uncorr}$ and $\widehat{N}_{\alpha, \beta}$, respectively.
The correction is less than $\sim$$1 \%$ for all bins.
%
\begin{table*}[t]
   \caption{Binning scheme for the CAMIRA clusters and characteristics of each bin for the abundance and lensing measurements.$^{*}$
            }
  \begin{center}
    \begin{tabular*}{0.80 \textwidth}{ cccccccccccc } \hline\hline
      \centering
 Abundance bin &  $\alpha$ & $\beta$ & $N_{\mathrm{min} }$ & $N_{\mathrm{max} }$ & $\left< N \right>$ & $z_{\mathrm{cl, min} }$ & $z_{ \mathrm{cl, max} }$
    & $\left<z_{\rm cl} \right>$
    & $N^{\rm uncorr}_{\alpha, \beta }$
    & $\widehat{N}_{\alpha, \beta}$\\ \hline
 1 &  1 & 1 & 15 & 20 & 17.1 & 0.1 & 0.4 & 0.27 & 255 & 258.3 \\
 2 &  2 & 1 & 20 & 30 & 24.1 & 0.1 & 0.4 & 0.27 & 208 & 210.7 \\
 3 &  3 & 1 & 30 & 60 & 38.2 & 0.1 & 0.4 & 0.26 & 92 & 93.3 \\
 4 &  4 & 1 & 60 & 200 & 77.4 & 0.1 & 0.4 & 0.26 & 9 & 9.1 \\
 5 &  1 & 2 & 15 & 20 & 17.2 & 0.4 & 0.7 & 0.56 & 301 & 301.7 \\
 6 &  2 & 2 & 20 & 30 & 24.0 & 0.4 & 0.7 & 0.53 & 210 & 210.6 \\
 7 &  3 & 2 & 30 & 60 & 38.9 & 0.4 & 0.7 & 0.54 & 79 & 79.2 \\
 8 &  4 & 2 & 60 & 200 & 73.1 & 0.4 & 0.7 & 0.50 & 7 & 7.0 \\
 9 &  1 & 3 & 15 & 20 & 17.0 & 0.7 & 1.0 & 0.84 & 339 & 339.2 \\
 10 &  2 & 3 & 20 & 30 & 23.8 & 0.7 & 1.0 & 0.83 & 181 & 181.1 \\
 11 &  3 & 3 & 30 & 60 & 36.3 & 0.7 & 1.0 & 0.84 & 65 & 65.0 \\
 12 &  4 & 3 & 60 & 200 & 64.7 & 0.7 & 1.0 & 1.0 & 1 & 1.0 \\  \hline
\\
 \hline \hline
      Lensing bin &  $\alpha$ & $\beta$ & $N_{\mathrm{min} }$ & $N_{\mathrm{max} }$ & $\left< N \right>$ & $z_{\mathrm{cl, min} }$ & $z_{ \mathrm{cl, max} }$
    & $\left<z_{\rm cl} \right>$
    & $N^{\rm uncorr}_{\alpha, \beta }$
    & $\widehat{N}_{\alpha, \beta}$\\ \hline
 1 &  1 & 1 & 15 & 20 & 17.1 & 0.1 & 0.4 & 0.27 & 255 & 258.3 \\
 2 &  2 & 1 & 20 & 30 & 24.1 & 0.1 & 0.4 & 0.27 & 208 & 210.7 \\
 3 &  3 & 1 & 30 & 200 & 41.7 & 0.1 & 0.4 & 0.26 & 101 & 102.4 \\
 4 &  1 & 2 & 15 & 20 & 17.2 & 0.4 & 0.7 & 0.56 & 301 & 301.7 \\
 5 &  2 & 2 & 20 & 30 & 24.0 & 0.4 & 0.7 & 0.53 & 210 & 210.6 \\
 6 &  3 & 2 & 30 & 200 & 41.6 & 0.4 & 0.7 & 0.53 & 86 & 86.2 \\
 7 &  1 & 3 & 15 & 20 & 17.0 & 0.7 & 1.0 & 0.84 & 339 & 339.2 \\
 8 &  2 & 3 & 20 & 30 & 23.8 & 0.7 & 1.0 & 0.83 & 181 & 181.1 \\
 9 &  3 & 3 & 30 & 200 & 36.7 & 0.7 & 1.0 & 0.84 & 66 & 66.0 \\ \hline
\end{tabular*}
\end{center}
\tabnote{
  $^{*}$
  Here $\alpha$ and $\beta$ denote the bin number for richness and redshift, respectively.
  Each bin is defined by $N_{\rm min}$, $N_{\rm max}$, $z_{\rm cl, min}$, and $z_{\rm cl, max}$, and
  $\left< N \right>$ and $\left< z_{\rm cl} \right>$ give the mean values of richness and redshift. 
  $N^{\rm uncorr}_{\alpha, \beta }$ and $\widehat{N}_{\alpha, \beta}$ are
  the abundance measurements without and with the correction discussed in Section~\ref{sec:measurement:abundance}, respectively.
}
\label{tab:binnig}
\end{table*}
%
\subsection{Stacked cluster lensing profile}\label{sec:measurements:lens}
We cross-correlate the positions of CAMIRA clusters with the shapes of background galaxies
to measure the average excess mass density profile around the clusters (hereafter the stacked lensing profile).
We follow the procedure in \cite{Mandelbaumetal2018a, Mandelbaumetal2018b} 
to estimate the stacked lensing profile for a sample of CAMIRA clusters
for $\alpha$-th richness and $\beta$-th redshift bin in Table~\ref{tab:binnig} 
as
\begin{eqnarray}
&&\widehat{ \Delta\Sigma}_{l, \alpha, \beta}(R)=
\frac{1}{1+\widehat{m}_{l, \alpha, \beta}(R)} 
\left[\frac{1}{2  \widehat{\mathcal{R}}_{l, \alpha, \beta} (R)} \right.\nonumber\\
&&\times \left.\frac{1}{N_{ls}^{\alpha, \beta}(R)} \left.
\sum_{l,s; N_l\in N_\alpha, z_l\in z_\beta} w_{ls}
\left\langle \Sigma_{\rm cr}^{-1}\right\rangle_{ls}^{-1} e_{+}({\btheta}_s)\right|_{R=\chi_l|\btheta_l-\btheta_s|}\right.\nonumber\\
&&\left.\left.
-\frac{1}{N_{ls}^{\alpha, \beta}(R)}
\sum_{l,s; N_l\in N_\alpha, z_l\in z_\beta} w_{ls}
\left\langle \Sigma_{\rm cr}^{-1}\right\rangle_{ls}^{-1}
c_{+}({\btheta}_s)\right|_{R=\chi_l|\btheta_l-\btheta_s|}
\right], \nonumber\\
\label{eq:lensestimator}
\end{eqnarray}
where the subscripts $l$ and $s$ stand for 
{\it lens} (cluster) and {\it source}, respectively, 
and $e_{+}$ and $c_{+}$ are the tangential component of the source galaxy
ellipticity and the additive bias with respect to the cluster center, respectively.
The summation runs over all pairs of clusters and source galaxies 
after a source selection cut described below 
in a given comoving transverse separation bin of
$R=\chi_l|\btheta_l-\btheta_s|$, 
where $\chi_l$ is the comoving distance to each cluster,
and $\btheta_l$ and $\btheta_s$ are angular positions of the lenses and the sources, respectively.
We use 11 radial bins that are equally spaced logarithmically from  $0.42$ $h^{-1} {\rm Mpc}$
to $14.0$ $h^{-1} {\rm Mpc}$ in comoving coordinates. 
We use  area-weighted mean values of comoving radii for the representative radial values.
The value for the first inner bin is $0.51~h^{-1}{\rm Mpc}$. 
We do not use lensing profiles at $<0.42h^{-1}$Mpc
to avoid a possible dilution effect \citep{Medezinskietal2018b}
and an increased blending effect 
(Murata et al. {\it in prep.}) at such small radii. 

The critical surface mass density is defined for a system of lens and source for a flat universe as
\begin{equation}
\Sigma_{\rm cr}^{-1}(z_l,z_s)
=4\pi G (1+z_l)\chi_l \left( 1-\frac{\chi_l }{\chi_s}\right)\  (z_l \leq z_s)
\label{eq:sigma_cr}
\end{equation}
and $0$ when $z_l>z_s$, where
$G$ is the gravitational constant.
We average this over the source photometric redshift PDF for each lens-source pair 
as
\begin{equation}
\left\langle \Sigma_{\rm cr}^{-1}\right\rangle_{ls}=\displaystyle \int_{z_l}^{\infty} {\rm d}z_s \Sigma_{\rm cr}^{-1}(z_l, z_s) P(z_s).
\label{eq:weightedsigmacr}
\end{equation}
In addition, the lens-source pair weight is given as
\begin{equation}
w_{ls}=\frac{\Omega_{\rm tot}}{\Omega_{\rm  eff}(z_l)}
\left\langle \Sigma_{\rm cr}^{-1}\right\rangle_{ls}^2
w_s,
\label{eq:lsweight}
\end{equation}
and $w_s$ is the source weight defined as
\begin{equation}
w_s=\frac{1}{\sigma_e^2+e_{\rm rms}^2},
\label{eq:shape_weight}
\end{equation}
where $\sigma_e$ is the measurement error of galaxy ellipticity
and $e_{\rm rms}$ is the intrinsic per-component root-mean-square ellipticity estimated in \cite{Mandelbaumetal2018b}
for each source galaxy.
The lensing weight factor $\Omega_{\rm tot}/\Omega_{\rm eff}(z_l)$
corrects for the effective area difference as a function of cluster redshift
as in the abundance estimator in equation~(\ref{eq:N_alpha}).
We note that this correction to the lensing estimator is small (less than $1\%$)
for all bins.
The denominator in equation~(\ref{eq:lensestimator}) is the weighted number of
lens-source pairs in each separation bin computed as
\begin{equation}
N_{ls}^{\alpha, \beta}(R)
=\left.\sum_{l,s;
N_l \in N_\alpha, z_l \in z_\beta}w_{ls}\right|_{R=\chi_l|\btheta_l-\btheta_s|}.
\end{equation}
The multiplicative bias correction is estimated as
\begin{equation}
  \widehat{m}_{l, \alpha, \beta}(R)=\frac{ \displaystyle \left. \sum_{l,s; N_l\in N_\alpha, z_l\in z_\beta} m_s w_{ls}\right|_{R=\chi_l|\btheta_l-\btheta_s|} }
          { \displaystyle \left. \sum_{l,s; N_l\in N_\alpha, z_l\in z_\beta} w_{ls} \right|_{R=\chi_l|\btheta_l-\btheta_s|} },
\label{eq:m}
\end{equation}
where $m_s$ is the estimated multiplicative bias for each source galaxy \citep{Mandelbaumetal2018b}.
The shear responsivity factor 
represents the statistically-averaged response of galaxy distortions
to small shears \citep{Kaiseretal1995, Bernstein&Jarvis2002},
and
is measured as
\begin{equation}
\widehat{\mathcal{R}}_{l, \alpha, \beta}(R)=
  1-\frac{ \displaystyle \left. \sum_{l,s; N_l\in N_\alpha, z_l\in z_\beta} e_{\rm rms}^2 w_{ls} \right|_{R=\chi_l|\btheta_l-\btheta_s|} }
  { \displaystyle \left. \sum_{l,s; N_l\in N_\alpha, z_l\in z_\beta} w_{ls} \right|_{R=\chi_l|\btheta_l-\btheta_s|} },
\label{eq:responsivity}
\end{equation}
which is found to be around $0.83$ for all the bins.

Here we describe the source selection cut.
As shown in e.g., \cite{Medezinskietal2018b},
it is important to use a secure background galaxy sample
for cluster weak lensing in order to minimize the
dilution effect by cluster member galaxies especially
at inner radii.
We use the {\it Pcut} method \citep{Oguri2014}
as our fiducial background source selection cut.
In the {\it Pcut} method, we select source galaxies whose photometric redshift PDFs
lie mostly beyond the cluster redshift plus some threshold $\Delta z$:
$\int_{z_l+ \Delta z}^{ \infty } {\rm d}z P(z)>0.98$.
We apply an additional cut on the randomly sampled redshift from PDFs, $z_{\rm mc}<2.5$,
for each lens-source pair in equation~(\ref{eq:lensestimator}).
We use $\Delta z=0.1$ for the fiducial cut
since \cite{Medezinskietal2018b}
shows the possible dilution effect is negligible for the {\it Pcut} method with this threshold value
at $R\gtrsim 0.5 h^{-1} {\rm Mpc}$.
In Section~\ref{subsubsec:diffphotoz},
we also check the robustness of our choice of this source selection cut
by using the {\it Pcut} method with $\Delta z=0.2$ and
the {\it color-color cut} in \cite{Medezinskietal2018b}.
The {\it color-color cut} uses the color-color space of $g-i$ vs $r-z$ for HSC,
where cluster red-sequence member galaxies can be well isolated from background galaxies.

In addition, 
we correct for selection bias from the lower cut on the resolution factor 
at $R_2=0.3$ in the shape catalog \citep{Mandelbaumetal2018b} as
\begin{equation}
  \widehat{m}_{ {\rm sel,} l, {\alpha, \beta}}(R)=A_{\rm sel} p_{l, \alpha, \beta}(R_2=0.3; R),
\label{eq:selectionbias}
\end{equation}
where $A_{\rm sel}=0.0087$ 
and $p_{l, \alpha, \beta}(R_2=0.3; R)$ is calculated by summation of lens-source weights, $w_{ls}$, in each of the radial, richness, and redshift bins.
This correction is found to be $m_{\rm sel} \sim 0.01$ for all bins. 
We then correct for the redshift variation of the intrinsic shape noise \citep{Mandelbaumetal2018b} 
by employing $m=0.03$ for $1.0 \leq z_{\rm best} \leq 1.5$
and $m=-0.01$ for the other redshift ranges, and by averaging this over lens-source pairs in each bin. 
This correction is found to be a positive multiplicative bias, but less than $0.01$ for all bins.

We also correct for systematic bias effects of the photometric redshift on the lensing measurements
by comparing the critical surface mass density from the photometric redshift estimates 
against that from the accurate photometric redshifts in the COSMOS 30-band catalog \citep{Ilbertetal2009, Laigleetal2016}
with the weak lensing weight $w_{ls}$ multiplied by a self-organizing map weight $w_{\rm SOM}$ 
which adjusts the COSMOS 30-band photo-$z$ sample to mimic the HSC source galaxy sample (More et al. {\it in prep.}).
We assume that the COSMOS 30-band photometric redshift estimates are sufficiently accurate
due to the larger numbers of bands.
This method is based on \cite{Nakajimaetal2012} 
in which the method was applied to SDSS data,
and was also applied in \cite{Miyatakeetal2018} (see their equation 11 for more details).
The debias factor is found to be $m \sim 0.01$ for CAMIRA clusters with $0.1 \leq z_{\rm cl} \leq 0.7$ and 
$m\sim 0.02$ for CAMIRA clusters with $0.7 \leq z_{\rm cl} \leq 1.0$
for the fiducial photo-$z$ code \texttt{MLZ} and the fiducial {\it Pcut} with $\Delta z=0.1$.
The debias factor is similar
for the other photo-$z$ catalogs and photo-$z$ cuts.
Our result indicates that the photo-$z$ bias correction is not very large 
when we average the critical surface mass density
over the photo-$z$ PDF in equation~(\ref{eq:weightedsigmacr}) with
the {\it Pcut} method or the {\it color-color cut}
for the shape catalog.
We also estimate the uncertainties of these correction factors for each richness and redshift bin
using  
the jackknife resampling technique \citep{Efron1982} with ten subsamples of the COSMOS 30-band catalog, 
where we recalculate the SOM weight for each jackknife resampled subsample.
We estimate the photo-$z$ bias correction uncertainties as
$\sigma_{\alpha, \beta, {\rm photoz}} \sim 0.001, 0.002, 0.004$ for $\beta=1,2,3$, respectively,
for the fiducial photo-$z$ catalog and source selection cut. 
We note that we ignore the impact of photo-$z$ biases, outliers, and the limited field variance of galaxies due to the small area
in the COSMOS 30-band catalog, 
thus the uncertainties might be underestimated.
These values are used for marginalization together with 
the systematic uncertainty in the overall calibration of the shear $\sigma_{\rm shear}=0.01$ \citep{Mandelbaumetal2018b}
in Section~\ref{sec:modeling:lensing}
\footnote{
We note that we apply this photo-$z$ correction for all photo-$z$ catalogs
except for \texttt{FRANKEN-Z} 
since we do not have the photo-$z$ estimates 
for the galaxies with HSC photometry in the COSMOS 30-band catalog 
from \texttt{FRANKEN-Z}.
}.

After the above corrections for $\widehat{ \Delta\Sigma}_{l, \alpha, \beta}(R)$, 
we also subtract the lensing measurement around random points as
\begin{equation}
\widehat{ \Delta\Sigma}_{\alpha, \beta}(R)=\widehat{ \Delta\Sigma}_{l, \alpha, \beta}(R)-\widehat{ \Delta\Sigma}_{r, \alpha, \beta}(R),
\label{eq:lensest}
\end{equation}
where $\widehat{ \Delta\Sigma}_{r, \alpha, \beta}(R)$ replaces the clusters with random points 
in the estimator for $\widehat{ \Delta\Sigma}_{l, \alpha, \beta}(R)$.
This subtraction allows us to measure the excess mass density profile with respect
to the background density as stressed in \cite{Sheldonetal2004} and \cite{Mandelbaumetal2005} 
\citep[also see][for a recent detailed study]{Singhetal2017}.
The random subtraction can also correct for an additive bias 
due to shear systematics including point-spread function ellipticity errors \citep{Mandelbaumetal2005}.
We use a random catalog of the CAMIRA clusters presented in Section~\ref{sec:data:cluster}
with the same richness and redshift distributions as the data,
and 
there are 100 times as many random points as real clusters in each redshift and richness bin.
%
\section{Forward modeling of cluster observables} \label{sec:modeling} 
We adopt a {\it forward} modeling approach to model the abundance and stacked lensing profiles
\citep[e.g.,][]{Zuetal2014, Murataetal2018, Costanzietal2018b} for a fixed cosmological model.
In this approach, we model the probability distribution of the richness for a given halo mass and redshift, $P(\ln N|M, z)$.
An alternative approach models the probability distribution of halo mass for a given richness and redshift, $P(\ln M|N, z)$,
as in some previous works \citep[e.g.,][]{Baxteretal2016, Simetetal2017, Melchioretal2017, McClintocketal2018}.
\subsection{Richness-mass relation} \label{sec:modeling:massrichness}
Following \cite{LimaandHu2005}, we assume that the probability distribution of the 
{\it observed} richness for halos with a fixed mass and redshift is given 
by a log-normal distribution as
\begin{equation}
P(\ln N|M, z) = \frac{1}{\sqrt{2\pi}\sigma_{\ln N|M, z}}
\exp\left( -\frac{x^2(N,M,z)}{2\sigma_{\ln N|M, z}^2}\right),
\label{eq:p_lambda}
\end{equation}
where $x(N, M, z)$ models the mean relation of $\ln N$ parametrized by four model parameters, 
$A$, $B$, $B_z$, and $C_z$ as
\begin{eqnarray}
x(N, M, z)& &\equiv \ln N-\left[ A+B \ln \left(\frac{M}{M_{\mathrm{pivot} }}\right) \right. \nonumber \\
& & \left. + B_z \ln \left(\frac{1+z}{1+z_{\mathrm{pivot} }}\right) + C_z \left[ \ln \left(\frac{1+z}{1+z_{\mathrm{pivot} }}\right) \right]^2 \right]. \nonumber \\
\label{eq:lambda_M}
\end{eqnarray}
Hence $x(N, M, z)=0$ gives the mean relation of $\ln N$ (also the median relation for $N$) as
\begin{eqnarray}
\langle \ln N\rangle (M, z)& &\equiv \int_{-\infty}^{+\infty} {\rm d}\ln N~ P(\ln N|M, z)\ln N \nonumber\\
&&=  A+B \ln \left(\frac{M}{M_{\mathrm{pivot} }}\right)\nonumber \\  
&& + B_z \ln \left(\frac{1+z}{1+z_{\mathrm{pivot} }}\right) + C_z \left[ \ln \left(\frac{1+z}{1+z_{\mathrm{pivot} }}\right) \right]^2. \nonumber \\
\label{eq:mean_relation}
\end{eqnarray}
We adopt $M_{\rm pivot}=3 \times 10^{14}h^{-1} M_{\odot}$ for the pivot mass scale
and $z_{\rm pivot}=0.6$ for the pivot cluster redshift.
We discuss the validity of including redshift evolution parameters $B_z$ and $C_z$ in Section~\ref{sec:discussion:redshiftevolution}.
In this work, we ignore errors on  $z_{\rm cl}$ for simplicity 
given the small bias and scatter of the cluster redshifts \citep{Ogurietal2018a} as described in Section~\ref{sec:data:cluster}.

In addition, we assume that the scatter of the richness around the mean relation
at a fixed halo mass and redshift can be parametrized by four parameters,
$\sigma_0$, $q$, $q_z$, and $p_z$ as 
\begin{eqnarray}
\sigma_{\ln N|M, z}& &= \sigma_0 + q \ln\left(\frac{M}{M_{\mathrm{pivot} } }\right) \nonumber\\ 
&& + q_z \ln \left(\frac{1+z}{1+z_{\mathrm{pivot} }}\right) + p_z \left[ \ln \left(\frac{1+z}{1+z_{\mathrm{pivot} }}\right) \right]^2.
\label{eq:scatter_M}
\end{eqnarray}
We also discuss the validity of including redshift evolution parameters, $q_z$ and $p_z$, in Section~\ref{sec:discussion:redshiftevolution}.
We only consider the parameter regions of $\{\sigma_0, q, q_z, p_z\}$ that result in $\sigma_{\ln N|M, z}>0$ for all halo mass and redshift 
for the parameter estimation.
In this treatment, $\sigma_{\ln N|M, z}$ should be effectively regarded as a {\it total} scatter,
including contributions from the intrinsic scatters, the richness measurement errors, 
the halo orientation effect \citep[e.g.,][]{Dietrichetal2014}, the projection effect \citep[e.g.,][]{Costanzietal2018},
and any other source of observational scatter that may be present.

In summary, we model the richness-mass relation in equation~(\ref{eq:p_lambda}) 
with eight model parameters. We constrain these parameters and check whether this model can reproduce measurements 
of the cluster abundance and stacked lensing profiles simultaneously with an acceptable value of $\chi^{2}_{\rm min}/{\rm dof}$.

\subsection{Abundance in richness and redshift bins} \label{sec:modeling:abundance}
Once we fix the richness-mass relation parameters of $P(\ln N|M, z)$,
we can predict the abundance of CAMIRA clusters for a given cosmology.
For the $\alpha$-th richness bin ($N_{\alpha, {\rm min}} \leq N \leq N_{\alpha, {\rm max}} $)
and the $\beta$-th redshift bin ($z_{\beta, {\rm min}} \leq z \leq z_{\beta, {\rm max}} $),
the abundance of the clusters for the total survey area is given as
\begin{eqnarray}
N_{\alpha, \beta}
&=&  \Omega_{\rm tot}
\int_{z_{\rm \beta, min}}^{z_{\rm \beta, max}}\!\mathrm{d}z~ \frac{ \mathrm{d^2} V}{ \mathrm{d}z \mathrm{d}\Omega} \nonumber \\
&\times&
\int_{M_{\rm min}}^{M_{\rm max}}\!\!\mathrm{d}M~
\frac{\mathrm{d}n}{\mathrm{d} M} 
\int_{\ln N_{\alpha, {\rm min} }}^{\ln N_{\alpha,{\rm max} } } \!\!\ {\rm d} \ln N ~
P(\ln N|M, z) \nonumber \\
&=& \Omega_{\rm tot}  \int_{z_{\rm \beta, min}}^{z_{\rm \beta, max}}\!\!\
\mathrm{d}z~ \frac{\mathrm{\chi}^{2}(z)}{H(z)}\nonumber \\
&\times& \int_{M_{\rm min }}^{M_{\rm max }}\!\!\mathrm{d}M~
\frac{\mathrm{d}n}{\mathrm{d} M}\,
S(M, z|N_{\alpha,{\rm min}}, N_{\alpha,{\rm max}}),
\label{eq:richnessfunc_model}
\end{eqnarray}
where $\chi^2(z)/H(z)$ denotes the comoving volume per unit redshift interval and unit steradian,
and $\mathrm{d}n/\mathrm{d}M$ is the halo mass function at redshift $z$ for a fixed cosmological model.
The selection function of halo mass at a fixed redshift 
in the richness bin
is calculated by integrating the log-normal distribution of $P(\ln N|M, z)$ over the richness range as
\begin{eqnarray}
&&S(M, z|N_{\alpha,{\rm min}},N_{\alpha,{\rm max}}) \nonumber \\
&&\equiv \int_{\ln N_{\alpha,{\rm min}}}^{\ln N_{\alpha,\rm{max}}}
\!\! \mathrm{d}\ln N~ P(\ln N|M, z) \nonumber \\
&&=
\frac{1}{2}
\left[
{\rm erf}\left(\frac{x(N_{\alpha,{\rm max}}, M, z)}{\sqrt{2}\sigma_{\ln N|M, z}}\right)
-{\rm erf}\left(\frac{x(N_{\alpha,{\rm min}}, M, z)}{\sqrt{2}\sigma_{\ln N|M, z}}\right)\right],\nonumber \\
\label{eq:selection_func}
\end{eqnarray}
where {\rm erf}$(x)$ is the error function.

\subsection{Stacked cluster lensing profile in richness and redshift bins} \label{sec:modeling:lensing}
The stacked lensing profile of halos with mass $M$ at redshift $z_l$ probes
the average radial profile of the matter distribution around halos, $\rho_{\rm hm}(r; M, z_l)$.
Assuming statistical isotropy in the cluster detections, 
the average matter distribution around the halos is one-dimensional 
as a function of separation from the halo center $r$,
where $r$ is in comoving coordinates.
We express the average matter density profile with the cross-correlation function 
between the halo distribution and the matter density fluctuation field, $\xi_{\rm hm}(r; M, z_l)$, as
\begin{equation}
\rho_{\rm hm}(r;M,z_l)=\bar{\rho}_{\rm m0}
\left[1+\xi_{\rm hm}(r;M,z_l)\right].
\label{eq:rho_hm}
\end{equation}
We note that we use the present-day mean matter density, $\bar{\rho}_{\rm m0}$, 
since we use the comoving density.
The cross-correlation function is connected to the cross-power spectrum, $P_{\rm hm}(k; M, z_l)$,
via the Fourier transform as 
\begin{equation}
\xi_{\rm hm}(r;M,z_l)=\int_{0}^{\infty}\!\!\frac{k^2\mathrm{d}k}{2\pi^2}~ P_{\rm hm}(k;M,z_l)j_0(kr),
\end{equation}
where $j_0(x)$ is the zeroth-order spherical Bessel function.
The surface mass density profile is obtained from 
a projection of the three-dimensional matter profile along the line-of-sight direction as
\begin{eqnarray}
\Sigma(R;M,z_l)&=&\bar{\rho}_{\rm m0}\int_{-\infty}^{\infty}\!\mathrm{d}
\chi~\xi_{\rm hm}\left(r=\sqrt{R^2+\chi^2}; M, z_l\right)\nonumber\\
&=&\bar{\rho}_{\rm m0}\int_{0}^{\infty}\!\frac{k\mathrm{d}k}{2\pi}~P_{\rm hm}(k;M,z_l)J_0(kR),
\label{eq:Sigma_M}
\end{eqnarray}
where $J_0(x)$ is the zeroth-order Bessel function and $R$ is the projected separation from the halo center in comoving coordinates.
The excess surface mass density profile around halos,
which is the direct observable from the stacked cluster lensing measurement,
is given as
\begin{eqnarray}
\Delta\Sigma(R;M,z_l)&=&
\avrg{\Sigma(R;M,z_l)}_{<R}-\Sigma(R;M,z_l)  \nonumber\\
&=&
\bar{\rho}_{\rm m0} \int_{0}^{\infty}\!\frac{k\mathrm{d}k}{2\pi}~
P_{\rm hm}(k;M,z_l)J_2(kR),
\label{eq:dSigma_M}
\end{eqnarray}
where $\avrg{\Sigma(R;M,z_l)}_{<R}$ is the average of $\Sigma(R; M, z_l)$ 
within a circular aperture of radius $R$, and $J_2(x)$ is a second-order Bessel function.

We can compute the model prediction for the stacked lensing profile
accounting for the distribution of halo masses and redshifts for CAMIRA clusters
in the $\alpha$-th richness and $\beta$-th redshift bin
as
\begin{eqnarray}
\Delta\Sigma_{\alpha, \beta}(R)
&=&
\frac{1}{N_{\Delta\Sigma}^{\alpha, \beta}}
\int_{z_{\beta, \rm min}}^{z_{\beta, \rm max}}\!\!\mathrm{d}z\!\!
~ \frac{\mathrm{\chi}^{2}(z)}{H(z)} w_{l}^{\alpha, \beta}(z)
\int_{M_{\rm min}}^{M_{\rm max} }
\!\!\mathrm{d}M~
\nonumber \\
&& \times
\frac{\mathrm{d}n}{\mathrm{d}M}S(M, z|N_{\alpha,{\rm min}},N_{\alpha,{\rm max}})
\nonumber \\
&& \times
\Delta\Sigma(R; M, z)\nonumber\\
&& \times
\left[1 +
\left\langle\frac{1}{\Sigma_{\mathrm{cr}}}\right\rangle_{\alpha, \beta}\hspace{-0.5em}(R)~
\Sigma(R; M, z) \right].
\label{eq:lensing_model}
\end{eqnarray}
The term in square brackets accounts for the non-linear contribution of 
reduced shear, which might not be negligible at very small radii 
\citep[e.g.,][]{Johnstonetal2007}, where $\langle1/\Sigma_{\mathrm{cr}}\rangle_{\alpha, \beta}(R)$
is measured from pairs of CAMIRA clusters and source galaxies
in each radial bin for the $\alpha$-th richness and $\beta$-th redshift cluster bin.
The lens redshift weight of $w_{l}^{\alpha, \beta}(z)$ is introduced to account for the weight distribution 
of lens redshift in the lensing measurement, and is calculated as
\begin{equation}
w_{l}^{\alpha, \beta}(z)=
\sum_{l,s; N_l \in N_\alpha, z_l \in z_\beta, z_l\in z}
w_{ls}.
\label{eq:wlensformodel}
\end{equation}
More specifically, we compute $w_{l}^{\alpha, \beta}(z)$ as follows.
First we divide the lens-source pairs into 12 lens redshift bins for each $\alpha$-th richness and $\beta$-th redshift bin,
which are linearly spaced in $z_l \in [z_{\beta, {\rm min} }, z_{\beta, {\rm max} }]$.
We then estimate the weight in equation~(\ref{eq:wlensformodel}) over all sources in each lens redshift bin.
We interpolate the weight values linearly as a function of lens redshift.
The normalization factor in the denominator of equation~(\ref{eq:lensing_model}) is 
similar to the abundance prediction in equation~(\ref{eq:richnessfunc_model}), but
is defined accounting for the lens redshift weight as
\begin{eqnarray}
N_{\Delta\Sigma}^{\alpha, \beta}
&=& \int_{z_{\beta, \rm min}}^{z_{\beta, \rm max}}\!\!\mathrm{d}z
~ \frac{\mathrm{\chi}^{2}(z)}{H(z)} w_{l}^{\alpha, \beta}(z)
\int_{M_{\rm min}}^{M_{\rm max}}\!\!\mathrm{d}M~
\nonumber\\
&&\times
\frac{\mathrm{d}n}{\mathrm{d}M}
S(M, z|N_{\alpha,{\rm min}},N_{\alpha,{\rm max}}).
\label{eq:N_dsigma}
\end{eqnarray}
This lens redshift weight changes the model prediction by less than $1$--$2 \%$ in the lensing profile amplitude
compared to that without accounting for the lens redshift weight.

The identified BCGs as cluster centers can be 
off-centered from the true halo centers \citep[e.g.,][]{Linetal2004, Rozoetal2014, Oguri2014, Ogurietal2018a}. 
We marginalize over the effect of off-centered clusters on the lensing profiles in equation~(\ref{eq:lensing_model}) 
by modifying the halo-matter cross-power spectrum in equations~(\ref{eq:Sigma_M}) and~(\ref{eq:dSigma_M}) 
for CAMIRA clusters in the $\alpha$-th richness and 
$\beta$-th redshift bin as
\begin{eqnarray}
P_{\rm hm}(k; M, z_l)
&\rightarrow&
\left[
f_{{\rm cen}}^{\alpha, \beta}+(1-f_{{\rm cen}}^{\alpha, \beta})\tilde{p}_{\rm off}(k;R_{\beta, \rm off})
\right] \nonumber \\
&\times& P_{\rm hm}(k; M, z_l),
\label{eq:pk_off}
\end{eqnarray}
following \cite{OguriandTakada2011} and \cite{Hikageetal2012, Hikageetal2013}.
Here $f_{{\rm cen}}^{\alpha, \beta}$ is
a parameter to model the fraction of centered clusters
in the $\alpha$-th richness and $\beta$-th redshift
bin, and $(1-f_{{\rm cen}}^{\alpha, \beta})$ is
the fraction of off-centered clusters.
While \cite{Murataetal2018} assigned a model parameter describing the centering fraction
independently for each lensing bin, 
we employ an empirical parametrization $f_{\rm cen}^{\alpha,\beta}$ 
that depends on the average richness and redshift in each bin 
(see Table~\ref{tab:binnig}) to reduce the number of model parameters as
\begin{equation}
f_{\rm cen}^{\alpha, \beta}=f_{0} 
+f_{N} \ln{ \left( \frac{ \langle N \rangle_{\alpha,\beta}  }{N_{\rm pivot} } \right) }
+f_{z} \ln{ \left( \frac{1+\langle z_{\rm cl}\rangle_{\alpha,\beta} }{1+z_{\rm pivot} } \right) },
\label{eq:fcenparam}
\end{equation}
with three model parameters ($f_{0}$, $f_{N}$, and $f_{z}$),
where we use $N_{\rm pivot}=25$ and $z_{\rm pivot}=0.6$
and we restrict the parameter regions of $f_{0}$, $f_{N}$, and $f_{z}$ such that $0<f_{{\rm cen}}^{\alpha, \beta}<1$ 
for all richness and redshift bins.
The function $p_{\rm off}(r; R_{\beta, \rm off})$
is the normalized one-dimensional radial profile of detected centers by the CAMIRA algorithm
with respect to the true halo center, 
for which we assume a Gaussian distribution \citep[e.g.,][]{Johnstonetal2007, OguriandTakada2011} 
given
as $p_{\rm off}(r; R_{\beta, \rm off})\propto
\exp(-r^2/2 R^{2}_{\beta, \rm off})$,
where $R_{\beta, \rm off}$ is a parameter to model
the typical off-centering radius in the $\beta$-th lens redshift bins.
The Fourier transform of this function is denoted as
$\tilde{p}_{\rm off}(k;R_{\beta, \rm off})=\exp(-k^2 R_{\beta, \rm off}^2/2 )$.
Since the aperture radius of CAMIRA clusters
is independent of the richness, with
aperture size of $\sim$$1\ h^{-1}{\rm Mpc}$ in physical coordinates \citep{Ogurietal2018a},
we use one model parameter ($R_{\beta, \rm off}$) for 
each $\beta$-th redshift bin, which is common for 
all richness bins,
with a flat prior from $10^{-3}$ to $0.5 \times (1+\langle z_{\rm cl}\rangle_{\beta} ) h^{-1} {\rm Mpc}$.
We note that \cite{Ogurietal2018a} investigated
the offset distribution of centers detected using the CAMIRA algorithm from centers of matched X-ray clusters,
and showed that
in most cases the offset
is less than $0.5 h^{-1} {\rm Mpc}$ in physical transverse distances.

\begin{figure*}
  \begin{center}
    \includegraphics[width=17.0cm]{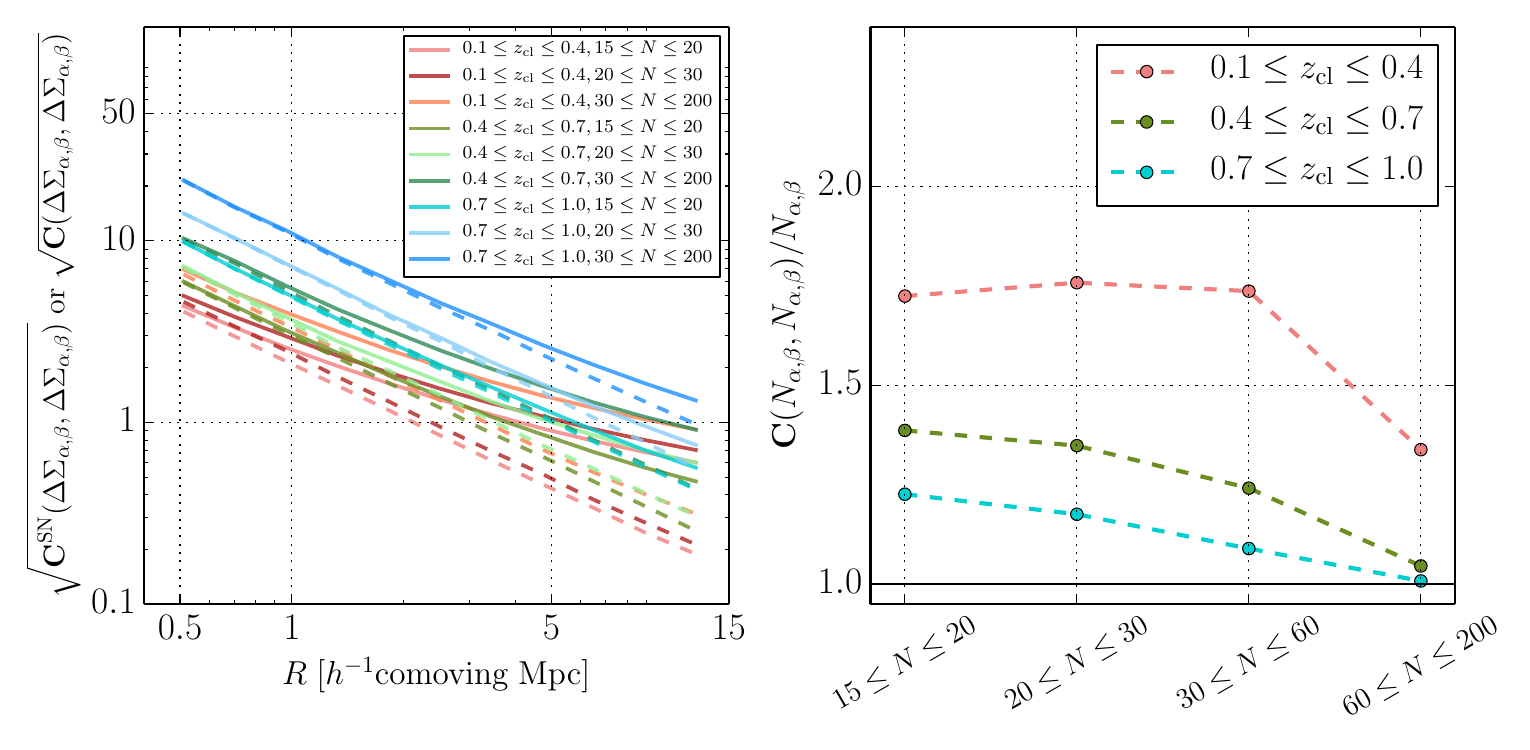}
  \end{center}
  \caption{
    Diagonal components of the covariance matrix for the stacked lensing profiles and abundance measurements for CAMIRA clusters
    in the fiducial analysis with the photometric redshift catalog \texttt{MLZ}, the fiducial photometric redshift source selection {\it Pcut} with $\Delta z=0.1$,
    and {\it Planck} cosmological parameters.
    We estimate the shape noise covariance ${\bf C}^{\rm SN}$ for the lensing measurements by
    randomly rotating the shapes in the real shape catalog,
    and we use analytic covariances to estimate other components of the sample covariance and Poisson noise
    where we use the richness-mass parameters of $\{A, B, B_z, C_z, \sigma_0, q, q_z, p_z\}=\{3.16, 0.92, -0.13, 4.17, 0.29, -0.12, -0.02, 0.52\}$,
    which are best-fit parameters with the simpler covariance (see Appendix~\ref{appendix:analyticcov} for more details).
    Left: solid curves show the square root of the diagonal elements of 
    the full covariance ${\bf C}$ for the lensing measurements in each redshift and richness bin
    and dashed curves show the shape noise contribution ${\bf C}^{\rm SN}$ for comparison.
    Right: the ratio of the full covariance of the abundances in each redshift and richness bin,
    relative to the Poisson contribution. The denominators come from the model prediction of the abundances $N_{\alpha, \beta}$
    at the same richness-mass relation parameters as above:
    $N_{\alpha,1}=\{255.0, 191.2, 105.8, 17.0\}$,
    $N_{\alpha,2}=\{319.6, 198.4, 75.2,  5.0\}$,
    $N_{\alpha,3}=\{340.6, 185.2, 54.5,  2.0\}$.
    }
\label{fig:Cij_fid_Planck}
\end{figure*}
%
\begin{figure}
  \begin{center}
    \includegraphics[width=8.5cm]{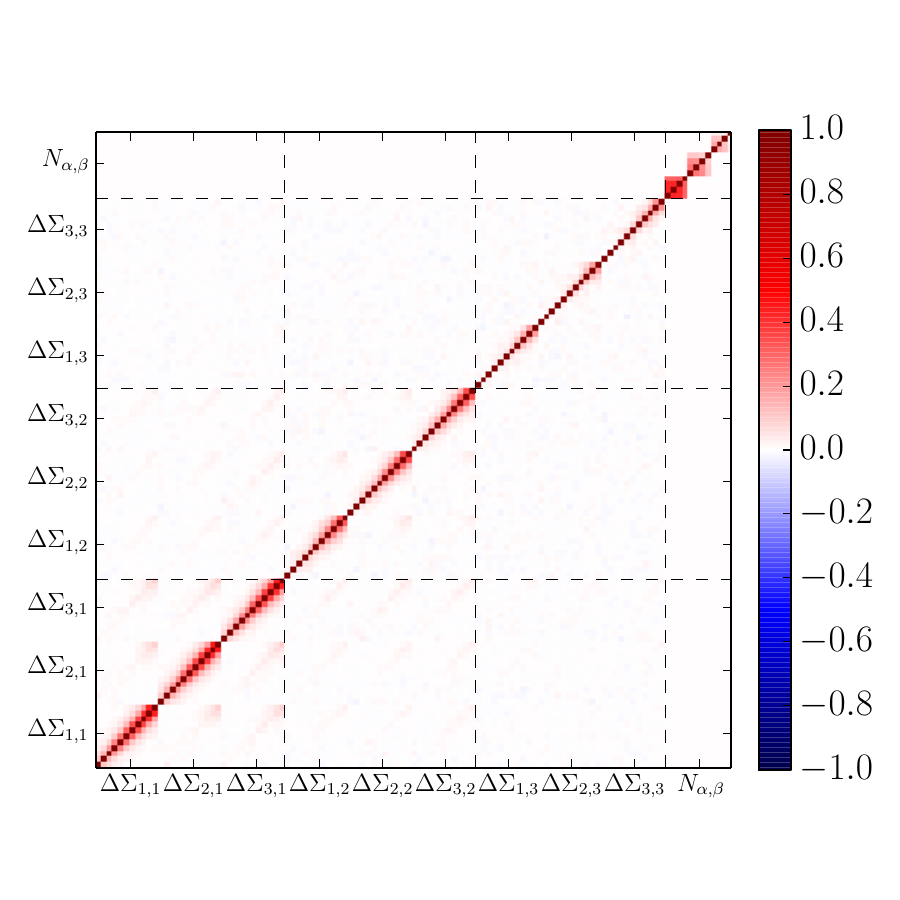}
  \end{center}
  \caption{ Correlation coefficient $r_{ij}$ (equation~\ref{eq:rrr}) of the covariance matrix in Figure~\ref{fig:Cij_fid_Planck}
            for the fiducial setup.
            The order of data vectors for $N_{\alpha, \beta}$ is the same as in Table~\ref{tab:binnig}.
            In the fiducial analysis, we do not include the cross-covariance between the lensing and abundance measurements
            since its effect on parameter estimation is small. See Appendix~\ref{appendix:analyticcov} for more details.
           }
\label{fig:rij_fid_Planck}
\end{figure}

We also marginalize over the shape calibration and photometric redshift bias uncertainties
by introducing a nuisance parameter $m_{\rm lens}$ following a Gaussian prior distribution
with a mean of zero and a standard deviation of $0.01$
to change the lensing model prediction as
\begin{equation}
\Delta\Sigma_{\alpha, \beta}(R) \to (1+m_{\alpha,\beta}) \Delta\Sigma_{\alpha, \beta}(R),
\label{eq:lensingbias}
\end{equation}
where $m_{\alpha,\beta}=m_{\rm lens}\times \sqrt{\sigma_{\rm shear}^2 + \sigma_{\alpha,\beta, \rm photoz}^2 }/0.01$.
\cite{Mandelbaumetal2018b} show that 
the systematic uncertainty in the overall calibration of the shear
is $\sigma_{\rm shear}=0.01$,
and we use the photometric redshift bias uncertainty of $\sigma_{\alpha,\beta, \rm photoz}$
derived in Section~\ref{sec:measurements:lens}.
We expect a large correlation in the residual shape calibration and photometric redshift bias
among different richness and redshift bins, and thus we conservatively use just one parameter $m_{\rm lens}$ for all bins.

Once the halo mass function ${\rm d}n(M, z)/{\rm d}M$
and the three-dimensional halo-matter cross-correlation $\xi_{\rm hm}(r; M, z)$
or $P_{\rm hm}(k; M, z)$ are provided for a given cosmological model (see Section~\ref{sec:simulation:emulator}),  
we can calculate the model prediction of the stacked lensing profile
in each richness and redshift bin.
The model is specified by 15 parameters in total for the entire cluster
sample of $15 \leq N \leq 200$ and $0.1 \leq z_{\rm cl} \leq 1.0$:
eight parameters $\{ A, B, B_z, C_z, \sigma_0, q, q_z, p_z\}$
for the richness-mass relation $P(\ln N|M, z)$, six parameters 
$\{f_{0}, f_N, f_z, R_{1, \rm off}, R_{2, \rm off}, R_{3, \rm off}\}$ for the
off-centering effect, and one nuisance parameter ($m_{\rm lens}$) for the uncertainty in lensing amplitudes.
We use the FFTLog algorithm \citep{Hamilton2000} for Fourier transforms,
which allows a fast, but sufficiently accurate and precise computation of the model prediction.
\subsection{ $N$-body simulation based halo emulator for the mass function and the lensing profile} \label{sec:simulation:emulator}
The model predictions must be accurate in order to estimate the model parameters in an unbiased way.
For this purpose, cosmological $N$-body simulations are one of the methods commonly used in the literature.
Here, we use the database generated by the \textsc{Dark~Quest} campaign \citep{Nishimichietal2018}
to predict the halo mass function and the halo-matter cross-correlation function.

\cite{Nishimichietal2018} develop a scheme called \textsc{Dark~Emulator}
for predicting statistical quantities of halos,
including the mass function, the halo-matter cross-correlation function, 
the halo-halo auto-correlation function as
a function of halo mass, redshift, separation length, and cosmological parameters,
based on a series of high-resolution, cosmological $N$-body simulations.
The simulation suite is composed of
cosmological {\it N}-body simulations for 101 cosmological models within a flat $w$CDM framework,
which are sampled around the {\it Planck} cosmological parameters.
The simulations trace the nonlinear evolution of $2048^3$ particles in a box size of
$1~$or$~2 h^{-1}{\rm Gpc}$ on a side
with mass resolution of $\sim$$10^{10} h^{-1}M_{\odot}$ 
or $\sim 8\times10^{10}h^{-1}M_\odot$, respectively. 
There are 21 redshift bins
for each simulation realization across the range $0 \leq z \leq 1.47$.
To identify dark matter halos,
\textsc{Rockstar} \citep{Behroozietal2013} is employed.
The halo mass defined for the simulations includes all particles
within a distance $R_{\rm 200m}$ from the halo center.
The minimum halo mass of the emulator is $10^{12} h^{-1} M_{\odot}$.
Throughout this paper, we set $M_{\rm min}=10^{12} h^{-1}M_{\odot}$
and $M_{\rm max}=2 \times 10^{15} h^{-1} M_{\odot}$ for the minimum and maximum halo masses,
respectively, to evaluate the halo mass integration in the model predictions of
abundances and stacked lensing profiles.

For the {\it Planck} cosmological model, 
instead of the prediction of the \textsc{Dark~Emulator} code,
we use a simpler interpolation scheme for a fixed cosmological model.
Specifically, we employ exactly the same cosmological parameters as their fiducial cosmological model, 
for which 24 independent realizations of high-resolution runs are available.
The relevant statistics are interpolated as a function of mass and redshift \citep[see][]{Murataetal2018}.
On the other hand, 
we use \textsc{Dark~Emulator} presented in \cite{Nishimichietal2018}
to interpolate over the cosmological parameter space 
and compute the predictions for the \textit{WMAP} cosmological model.
This relies on data compression based on Principal Component (PC) Analysis followed by 
Gaussian Process Regression for each PC coefficient.
This is done for a subset of 80 cosmological models.
Note that the realizations for the {\it Planck} cosmology 
are not used in the regression, 
but rather are used as part of the validation set to assess the performance of the emulator.

We estimate impacts of the uncertainties in the emulator on constraints of the mass-richness parameters as follows.
For the {\it Planck} cosmology, we shift the halo mass function and the halo-matter
cross-correlation function in each bin of halo mass, redshift, and separation length
by one standard deviation uncertainty from the 24 realizations to opposite sides,
and we calculate $\chi^{2}$ at the best-fit mass-richness relation parameters for the fiducial analysis below.
We find that $\chi^{2}=106.9$ while we have $\chi^{2}=107.0$ for the fiducial emulator as shown in Table~\ref{tab:fidparams},
suggesting that the emulator precision for the {\it Planck} cosmology is high enough for our analysis.
For the {\it WMAP} cosmology, we estimate the impacts 
using the outputs of \textsc{Dark~Emulator} for the {\it Planck} cosmology 
by comparing it with the fiducial emulator for the {\it Planck} cosmology. 
Since the realizations for the {\it Planck} cosmology are not used in the regression, 
differences between the 24 realizations and the \textsc{Dark~Emulator} 
reveal the typical impact of the uncertainties in the emulator from the \textsc{Dark~Emulator}
for the {\it WMAP} cosmology.
The errors for the model parameters are consistent between the two emulators.
On the other hand, a shift for the median value of $A$ is $\sim 0.35$ compared to the error width, while
shifts for the other parameters are smaller than $\sim 0.15$ compared to the error widths.
This suggests that systematic uncertainties from the emulator for the {\it WMAP} cosmology
increase the error for $A$ by $\sim 6\%$ and the errors for other parameters by $\sim 1\%$.
Since these impacts of the emulator precision do not change 
the constraints on the mass-richness relation parameters very significantly,
we ignore these uncertainties in this paper for simplicity.

\begin{table*}[t]
  \captionsetup{justification=centering}
  \caption{ The model parameters, priors, and 
            parameter estimations from our joint analysis of lensing and abundance measurements.$^*$
             }
  \begin{center}
    \begin{tabular*}{\textwidth}{ccccc} \hline\hline
      \centering
      Parameter                     &  Description & Prior & Median and Error                & Median and Error \\
                                    &              &       & {\it Planck}    &  {\it WMAP}    \\  \hline
      $A$ & $\langle \ln N\rangle$ at pivot mass scale and pivot redshift & ${\rm flat}[2, 5]$ & $3.15^{+0.07}_{-0.08}$ & $3.36^{+0.05}_{-0.06}$ \\ 
      $B$ & Coefficient of halo mass dependence in $\langle \ln N\rangle$ & ${\rm flat}[0, 2]$ & $0.86^{+0.05}_{-0.05}$ & $0.83^{+0.03}_{-0.03}$ \\ 
      $B_z$ & Coefficient of linear redshift dependence in $\langle \ln N\rangle$ & ${\rm flat}[-50, 50]$ & $-0.21^{+0.35}_{-0.42}$ & $-0.20^{+0.26}_{-0.34}$ \\ 
      $C_z$ & Coefficient of square redshift dependence in $\langle \ln N\rangle$ & ${\rm flat}[-50, 50]$ & $3.61^{+1.96}_{-2.23}$ & $3.51^{+1.32}_{-1.59}$ \\ 
      $\sigma_0$ & $\sigma_{\ln N|M, z}$ at pivot mass scale and pivot redshift & ${\rm flat}[0, 1.5]$ & $0.32^{+0.06}_{-0.06}$ & $0.19^{+0.07}_{-0.07}$ \\ 
      $q$ & Coefficient of halo mass dependence in $\sigma_{\ln N|M, z}$ & ${\rm flat}[-1.5, 1.5]$ & $-0.08^{+0.05}_{-0.04}$ & $-0.02^{+0.03}_{-0.03}$ \\ 
      $q_z$ & Coefficient of linear redshift dependence in $\sigma_{\ln N|M, z}$ & ${\rm flat}[-50, 50]$ & $0.03^{+0.31}_{-0.30}$ & $0.23^{+0.37}_{-0.35}$ \\ 
      $p_z$ & Coefficient of square redshift dependence in $\sigma_{\ln N|M, z}$ & ${\rm flat}[-50, 50]$ & $0.70^{+1.71}_{-1.60}$ & $1.26^{+1.77}_{-1.49}$ \\ 
      $f_0$ & Centering fraction at pivot richness and redshift & $0<f_{\rm cen}^{\alpha,\beta}<1$ & $0.68^{+0.05}_{-0.06}$ & $0.68^{+0.05}_{-0.06}$ \\ 
      $f_N$ & Coefficient of richness dependence in $f_{\rm cen}^{\alpha,\beta}$ & $0<f_{\rm cen}^{\alpha,\beta}<1$ & $0.33^{+0.10}_{-0.09}$ & $0.33^{+0.10}_{-0.09}$ \\ 
      $f_z$ & Coefficient of redshift dependence in $f_{\rm cen}^{\alpha,\beta}$ & $0<f_{\rm cen}^{\alpha,\beta}<1$ & $-0.14^{+0.35}_{-0.34}$ & $-0.19^{+0.34}_{-0.34}$ \\ 
      $R_{1,\rm off}$ & Off-centering radius for the first redshift bin & ${\rm flat}[10^{-3}, 0.64]$ & $0.39^{+0.10}_{-0.09}$ & $0.38^{+0.10}_{-0.09}$ \\ 
      $R_{2, \rm off}$ & Off-centering radius for the second redshift bin & ${\rm flat}[10^{-3}, 0.78]$ & $0.55^{+0.12}_{-0.11}$ & $0.52^{+0.12}_{-0.10}$ \\ 
      $R_{3, \rm off}$ & Off-centering radius for the third redshift bin & ${\rm flat}[10^{-3}, 0.92]$ & $0.59^{+0.17}_{-0.17}$ & $0.58^{+0.17}_{-0.15}$ \\ 
      $m_{\rm lens}$ & Marginalization parameter of the lensing amplitudes & ${\rm Gauss}(0, 0.01)$ & $0.00^{+0.01}_{-0.01}$ & $0.00^{+0.01}_{-0.01}$ \\ \hline
      $\chi^2_{\rm min}/{\rm dof}$ &   &   & $107.0/97$ & $106.6/97$ \\ \hline
    \end{tabular*}
  \end{center}
  \tabnote{$^*$ In the fiducial analysis, we vary 15 parameters while fixing the cosmological parameters 
           to either {\it Planck} or {\it WMAP}.
           We use flat priors for all richness-mass parameters and off-centering parameters, denoted as ${\rm flat}[x, y]$,
           with the region between $x$ and $y$. We use a Gaussian prior for marginalizing over parameters of the lensing amplitudes, 
           with mean value 0 and standard deviation 0.01 (see around equation~\ref{eq:lensingbias} for more details on the implementation).
           We additionally restrict the scatter parameter space to $\sigma_{\ln N|M,z}>0$ 
           for the range of halo masses and redshifts we consider, 
           $10^{12}\leq M/[h^{-1}M_{\odot}]\leq2\times 10^{15}$ and $0.1 \leq z\leq 1.0$. 
           We also restrict the off-centering fraction parameters space to $0<f_{\rm cen}^{\alpha,\beta}<1$
           for all richness and redshift bins.
           The maximum ranges of priors for the off-centering radii 
           correspond to $R<0.5 h^{-1}{\rm Mpc}$ in physical coordinates
           using the mean redshift value in each redshift bin from Table~\ref{tab:binnig}.
           The column labeled as ``Median and Error'' denotes the median and the 16th and 84th percentiles of the posterior distribution.
           We also show the minimum chi-square ($\chi^{2}_{\rm min}$) with the number of degrees of freedom (dof) at the bottom row.
           Since the lensing marginalization parameter is determined strongly by the prior above, 
           we do not include it as a free parameter when calculating dof 
           (i.e., dof$=97=111-14$, where $111$ is the total number of data points 
           and $14$ is the total number of richness-mass relation parameters and off-centering parameters).
           The correlations among the parameters are shown in Appendix~\ref{appendix:mcmccontour}.
           We adapt $M_{\rm pivot}=3 \times 10^{14}h^{-1} M_{\odot}$, $z_{\rm pivot}=0.6$, and
           $N_{\rm pivot}=25$ for the pivot values.
            }
  \label{tab:fidparams}
\end{table*}
\subsection{Covariance} \label{sec:simulation:mockcat}
We must estimate the covariance describing the 
statistical uncertainties of the stacked lensing profiles and the abundance.
The covariance consists of shape noise covariance from the finite number of lens-source pairs, 
Poisson noise for the abundance from the finite number of the clusters,
and sample covariance from an imperfect sampling of 
the fluctuations in large-scale structure within a finite survey volume 
\citep[e.g.,][]{HuandKravtsov2003, Takada&Bridle2007, OguriandTakada2011, Takada&Hu2013, Hikage&Oguri2016, Takahashietal2018, Shirasaki&Takada2018}.
We estimate the shape noise covariance for the stacked lensing profiles 
directly from the data \citep[e.g.,][]{Murataetal2018}
by repeating lensing measurements (equation \ref{eq:lensestimator}) for
source galaxies with their orientations randomized $15,000$ times.
The shape noise covariance estimated in this manner
accounts for 
the survey geometry and the inhomogeneous distribution of source galaxies.
As discussed in Appendix~\ref{appendix:analyticcov},
we adopt an analytic halo model \citep{Cooray&Sheth2002}
to compute 
the sample covariance and the covariance 
for the abundance
assuming that the distribution of clusters and lensing fields obey a Gaussian distribution.
We note that we compute the shape noise covariance and the sample covariance
for each setup of cosmological parameters, photo-$z$ catalog, and source selection cut
as described in Appendix~\ref{appendix:analyticcov}.

In Appendix~\ref{sec:appendix:compmock},
we validate the analytic covariance model for the sample covariance 
by comparing it with the
covariance estimation from
2268 mock catalogs of the source galaxies and the CAMIRA clusters for the HSC footprint,
which are generated from full-sky ray-tracing simulations with halo catalogs
\citep{Shirasakietal2019} based on methods described in \cite{Shirasakietal2017} \citep[see also][]{Shirasaki&Yoshida2014}.
The mock catalogs
are based on full-sky, light-cone cosmological simulations
constructed from sets of $N$-body simulations \citep{Takahashietal2017}
with {\it WMAP} cosmological parameters \citep{Hinshawetal2013}.
The lensing effects at a given angular position
are computed by a ray-tracing simulation through the foreground matter distribution
from the multiple lens-plane algorithm \citep{HamanaMellier2001, Shirasakietal2015}.
Each source plane is given in \textsc{HEALPix} format \citep{Gorskietal2005}
with an angular resolution of about $0.43$ arcmin.
The mock galaxy shape catalog
accounts for 
various effects as in the real data, including
survey geometry, 
the inhomogeneous angular distribution of source galaxies,
statistical uncertainties in the photo-$z$ estimation of each galaxy, and
variations in the lensing weight from observational conditions and galaxy properties,
since the mock catalog is constructed based on the {\it real} shape catalog \citep[see][for more details]{Shirasakietal2019}.
We also construct mock catalogs of the CAMIRA clusters by using the catalog of halos
in each light-cone simulation realization in \cite{Takahashietal2017}, which
are identified with \textsc{Rockstar} \citep{Behroozietal2013}.
We refer the readers to Appendix~\ref{sec:appendix:compmock} for more details.

Figure~\ref{fig:Cij_fid_Planck} shows the diagonal components of the covariance matrix 
for the fiducial analysis with {\it Planck} cosmological parameters,
the fiducial photo-$z$ catalog, and the fiducial source selection cut.
The left panel compares the full covariance with the shape noise covariance 
for the stacked lensing profiles.
The sample covariance starts to become the dominant contribution 
even from $R=2 h^{-1}{\rm Mpc}$ for the lowest redshift bin ($0.1 \leq z_{\rm cl} \leq 0.4$)
due to the high source density of the HSC shear catalog \citep[see also][]{Miyatakeetal2018}.
The right panel shows diagonal components of the covariance for the abundance compared to the Poisson term.
The sample variance contributions in the abundance for the lower redshift bins are higher 
than 
those for the higher redshift bins.

Figure~\ref{fig:rij_fid_Planck} shows the correlation coefficient matrix defined as
\begin{equation}
r_{i j}\equiv \frac{ {\bf C}( D_i,  D_j) }{ \sqrt{ {\bf C}( D_i,  D_i){\bf C}( D_j,  D_j) } },
\label{eq:rrr}
\end{equation}
where $D_i$ is the $i$-th component of the data vector {\bf D}.
There are large correlations among neighboring bins especially for large radii and lower redshift,
since 
the same large-scale structure causes spatially-correlated fluctuations 
in the stacked lensing profile 
and the abundance measurements
especially among neighboring radii or different richness bins for each redshift bin
\citep{Takada&Bridle2007, OguriandTakada2011}.

We then calculate the signal-to-noise ratio with the measurements and the covariance for the fiducial analysis 
with the {\it Planck} cosmological parameters.
The total signal-to-noise ratio from the stacked lensing profiles and abundance measurements
is $61.0$, and the signal-to-noise ratios for the stacked lensing profiles and the abundance 
are $53.0$ and $30.3$, respectively.
The lensing signal-to-noise ratios in each redshift bin are
$36.6$, $34.2$, and $19.3$ from the lowest to highest redshift bins.
%
\section{Results} \label{sec:results}
In this section, we show the posterior distribution of the parameters,
the joint probability $P_{\beta}(\ln M, \ln N)$ in each redshift bin,
the mass-richness relation $P_{\beta}(\ln M|N)$ in each redshift bin, 
and the richness-mass relation $P(\ln N|M, z)$, 
from the joint analysis of the model parameters based on the abundance and lensing profiles under the fiducial setups.
In this work, we fix the cosmological parameters
to the {\it Planck} cosmology \citep{PlanckCollaboration2016} 
or the {\it WMAP} cosmology \citep{Hinshawetal2013} as described in Section~\ref{sec:intro}
to investigate how and to what degree the difference of the cosmological parameters 
affects the constraints on the mass-richness relation without informative priors on the parameters,
and to gain insight into cluster cosmology by comparing the results.
%
\subsection{Posterior distribution of parameters} \label{sec:results:parameterest}
\begin{figure*}[t]
  \begin{center}
    \includegraphics[width=0.99 \textwidth]{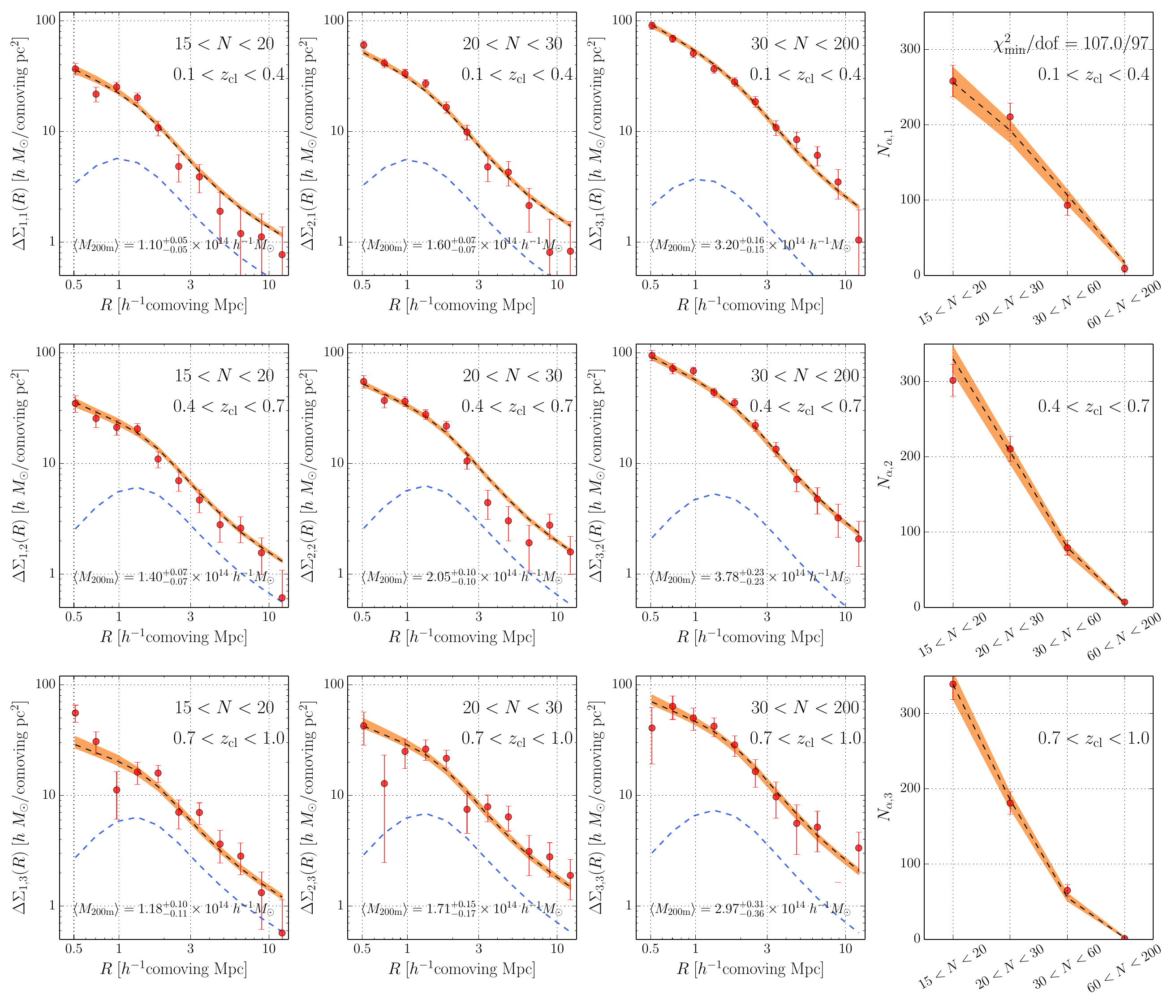}
  \end{center}
  \caption{ The measurements and fitting results of the stacked lensing and abundance with the {\it Planck} cosmological parameters.
            Points with error bars show the measurements,
            and the shaded regions show the 16th and 84th percentiles of the model predictions 
            from the MCMC chains.
            The error bars denote the diagonal components of the covariance matrix (see also Figure~\ref{fig:Cij_fid_Planck}).
            The black dashed curves show the model predictions at the best-fit parameters.
            The light-blue dashed curves show 
            contributions from off-centered clusters to lensing profiles at the best-fit model parameters.
            We show the minimum value of the reduced chi-square in the upper-right corner. 
            We also give 16th, 50th (median), and 84th percentiles of the mean mass $\langle M_{\rm 200m} \rangle$
            in each lensing panel from the MCMC chains.
            }
\label{fig:fitting_fid_Planck}
\end{figure*}
%
\begin{figure*}[t]
  \begin{center}
    \includegraphics[width=0.99 \textwidth]{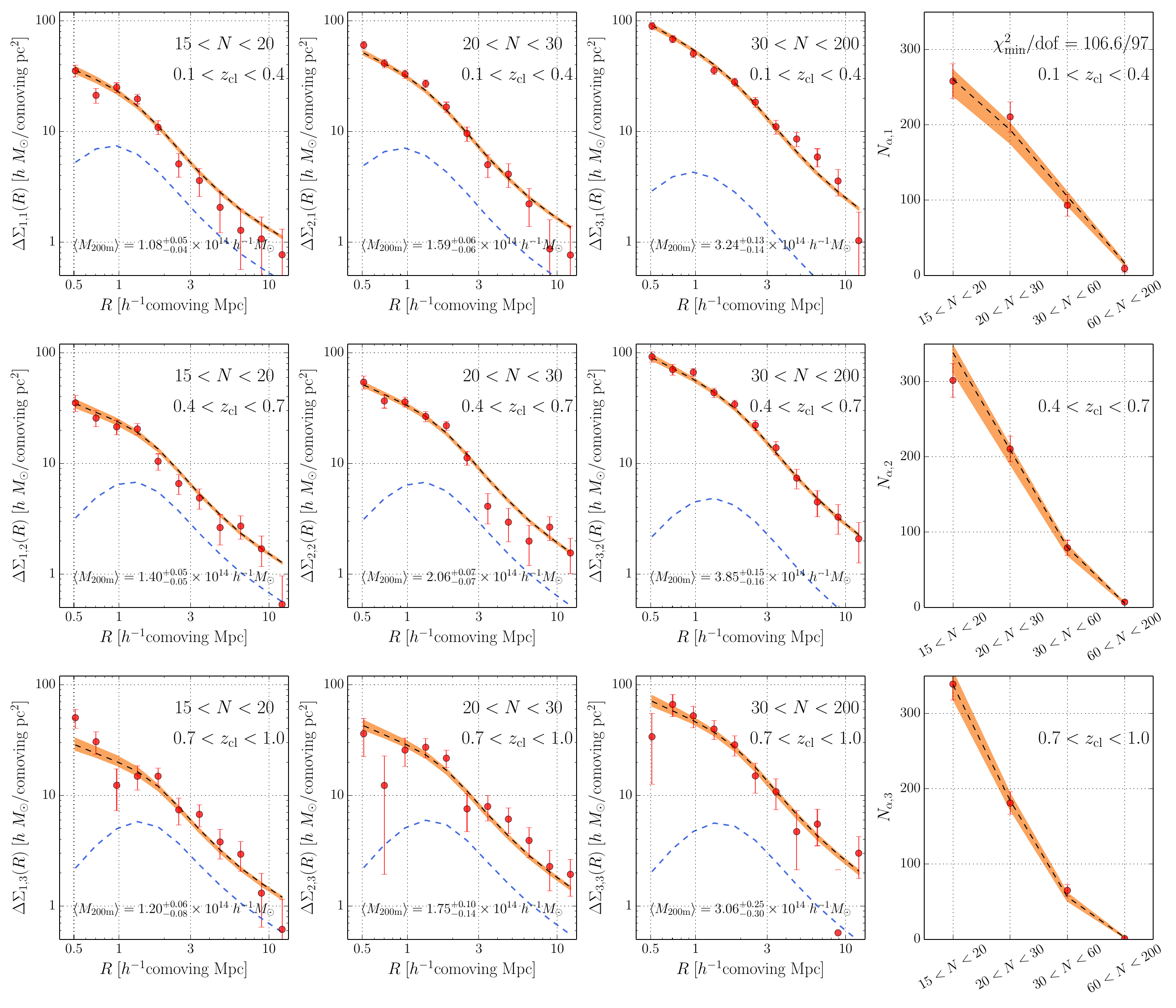}
  \end{center}
  \caption{ Same as Figure~\ref{fig:fitting_fid_Planck}, but for the {\it WMAP} cosmological parameters. }
\label{fig:fitting_fid_WMAP}
\end{figure*}
We constrain the model parameters
by comparing the model predictions with the measurements of the abundance and the lensing profiles.
We perform Bayesian parameter estimation assuming a Gaussian form for the likelihood,
$\mathcal{L}\propto |{\bf C}|^{-1/2} \exp(-\chi^2/2)$, with
\begin{equation}
 \chi^2 =
 \sum_{i, j}~ \Bigl[ {\bf D} - {\bf D}^{\rm model}\Bigr]_i\left({\bf C}^{-1}\right)_{ij}
    \Bigl[{\bf D} - {\bf D}^{\rm model} \Bigr]_j,
\end{equation}
where ${\bf D}$ is a data vector that consists of the lensing profiles 
and the abundance in different radial, richness, and redshift bins, 
${\bf D}^{\rm model}$ is the model prediction, 
and ${\bf C}^{-1}$ is the inverse of the covariance matrix (see Section~\ref{sec:simulation:mockcat}).
We use 11 radial bins ${\bf R}$
in each richness and redshift bin for the stacked lensing profile.
The indices $i$ and $j$ run from $1$ 
to the total number of data points ($111$ for the fiducial analysis).
We perform the parameter estimation 
with the affine-invariant Markov Chain Monte Carlo (hereafter MCMC) sampler of \cite{Goodman&Weare2010}
as implemented in the python package \textsc{emcee} \citep{Foreman-Mackeyetal2013}.

In Table~\ref{tab:fidparams}, we summarize the results of the parameter estimation,
including a description of each parameter, priors, the median and 68$\%$ credible level
interval after removing the burn-in chains and marginalizing over the other parameters, 
and $\chi^{2}_{\rm min}/{\rm dof}$ to show goodness-of-fit 
under ${\it Planck}$ or ${\it WMAP}$ cosmology.
In fitting we employ uninformative flat priors
for all of the richness-mass relation parameters 
and the off-centering parameters. 
We also show the 68$\%$ and 95$\%$ credible level 
contours in each two-parameter subspace, and one-dimensional posterior distributions
in Appendix~\ref{appendix:mcmccontour}. 
Each richness-mass relation parameter is well constrained by the joint analysis compared to its prior.
From Table~\ref{tab:fidparams} and Appendix~\ref{appendix:mcmccontour},
we find that constraints on $A$, $\sigma_0$ and $q$ are systematically different 
between {\it Planck} and {\it WMAP} cosmologies 
mainly due to the differences in their halo mass functions.
On the other hand, constraints on the other parameters are similar to each other.
The result for the {\it WMAP} cosmology prefers 
a higher mean normalization $A$ and lower scatter normalization $\sigma_0$
than for the {\it Planck} cosmology.
In addition, the result for {\it Planck} cosmology prefers negative $q$ values 
(i.e., larger scatter at the lower halo mass)
more significantly than the {\it WMAP} cosmology, although it is still a $<2 \sigma$ preference.
For both cosmological models, 
off-centering parameters are constrained well compared to their priors. 
The centering fraction at the pivot richness and redshift is constrained as $f_0=0.68^{+0.05}_{-0.06}$,
which is consistent with $f_{\rm cen}=0.68 \pm 0.09$ in the analysis of \cite{Ogurietal2018a}, who  
estimated the centering fraction without redshift or richness dependences 
by comparing CAMIRA cluster centers with X-ray centroids.
The richness dependence of the off-centering fraction $f_N$ prefers positive values with a high significance,
indicating that the higher richness clusters are centered better than the lower richness clusters.
This might be partly because CAMIRA obtained lower richness values for off-centered
clusters given that the richness is computed around
the identified BCG using a circular aperture.
On the other hand, the redshift dependence parameter $f_z$ is consistent with zero.
The off-centering radius parameter for each redshift bin $R_{\beta, {\rm off}}$ 
is constrained compared to its prior, but only marginally.

We check the validity of our model by monitoring the 
$\chi^2$ of the best-fitting models.
Since the posterior distribution of the nuisance parameter $m_{\rm lens}$ 
for lensing amplitudes is strongly determined by its prior, 
we do not include it as a parameter when calculating the number of degrees-of-freedom,
thus ${\rm dof}=111-14=97$ where $111$ is the total number of data points for the fitting
and $14$ is the total number of richness-mass relation parameters 
and off-centering parameters.
We find that $\chi^{2}_{\rm min}/{\rm dof}=107.0/97$ ($p$-${\rm value}=0.23$) 
for the {\it Planck} cosmology, and 
$\chi^{2}_{\rm min}/{\rm dof}=106.6/97$ ($p$-${\rm value}=0.24$) 
for the {\it WMAP} cosmology, both of which are acceptable $p$-values.
This indicates that we cannot distinguish between
{\it Planck} and {\it WMAP} cosmologies from the 
abundance and lensing measurements, partly because
of our adoption of a flexible richness-mass relation.

We also show the comparison of the model predictions from the MCMC chains 
with the measurements of the lensing profiles and abundance 
in Figures~\ref{fig:fitting_fid_Planck} and \ref{fig:fitting_fid_WMAP} 
for {\it Planck} and {\it WMAP} cosmologies, respectively.
The figures show that 
the model predictions reproduce the lensing profiles and abundance 
simultaneously for both sets of cosmological parameters 
with the fiducial richness-mass relation model.
\subsection{Joint probability $P_{\beta}(\ln M, \ln N)$ and mass-richness relation $P_{\beta}(\ln M| N)$} \label{sec:result:joint}
%
\begin{figure*}
  \begin{center}
    { \includegraphics[width=5.62cm]{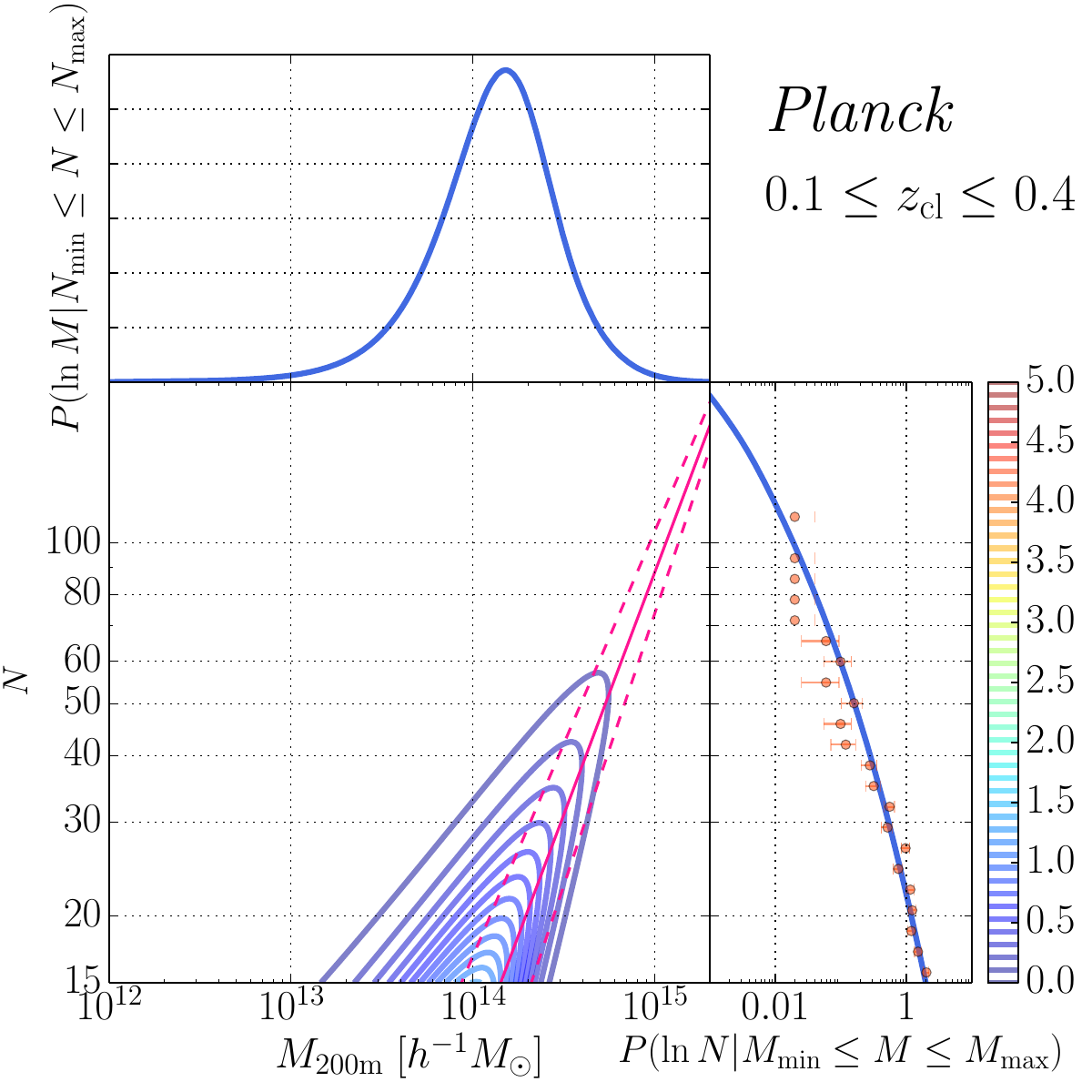}
      \includegraphics[width=5.62cm]{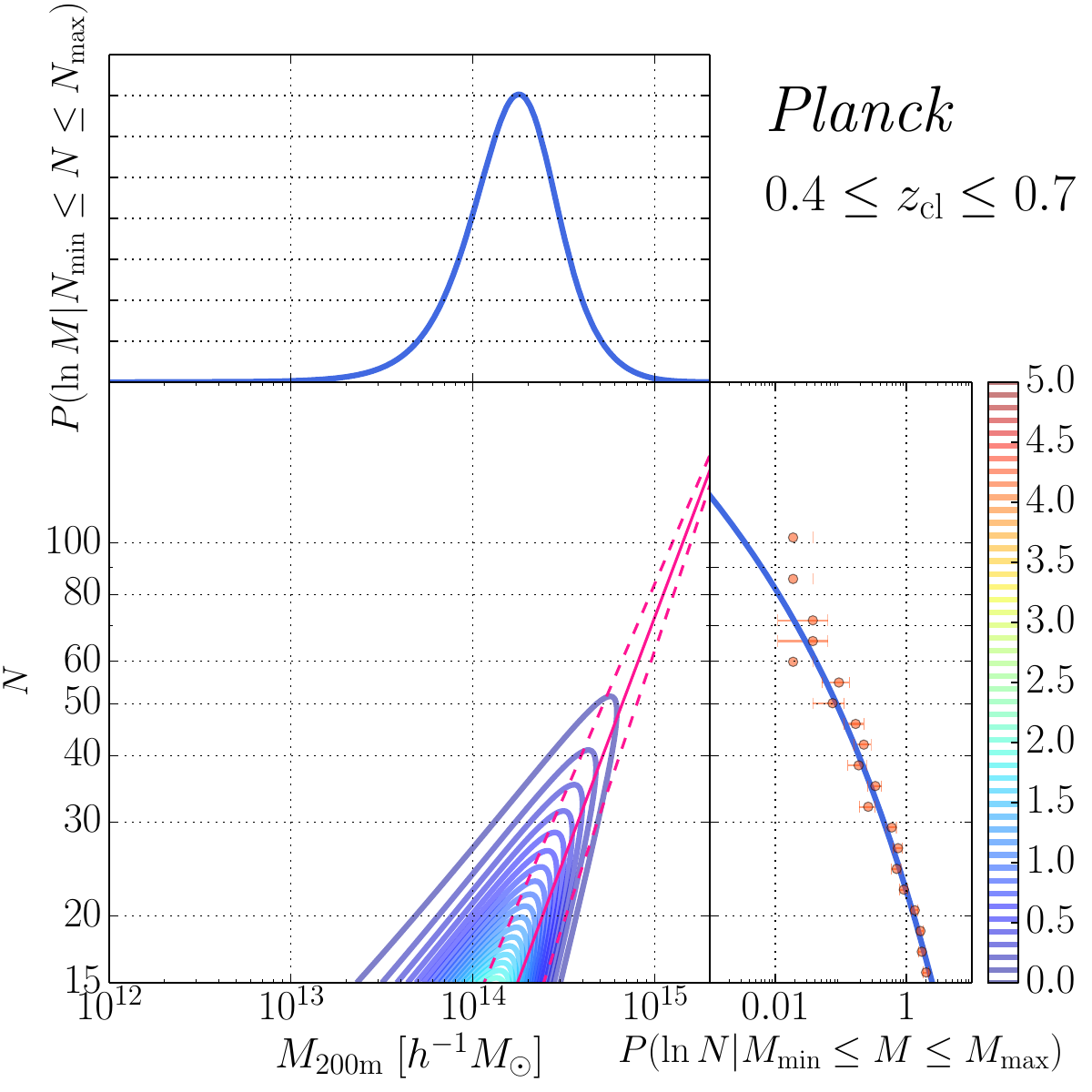}
      \includegraphics[width=5.62cm]{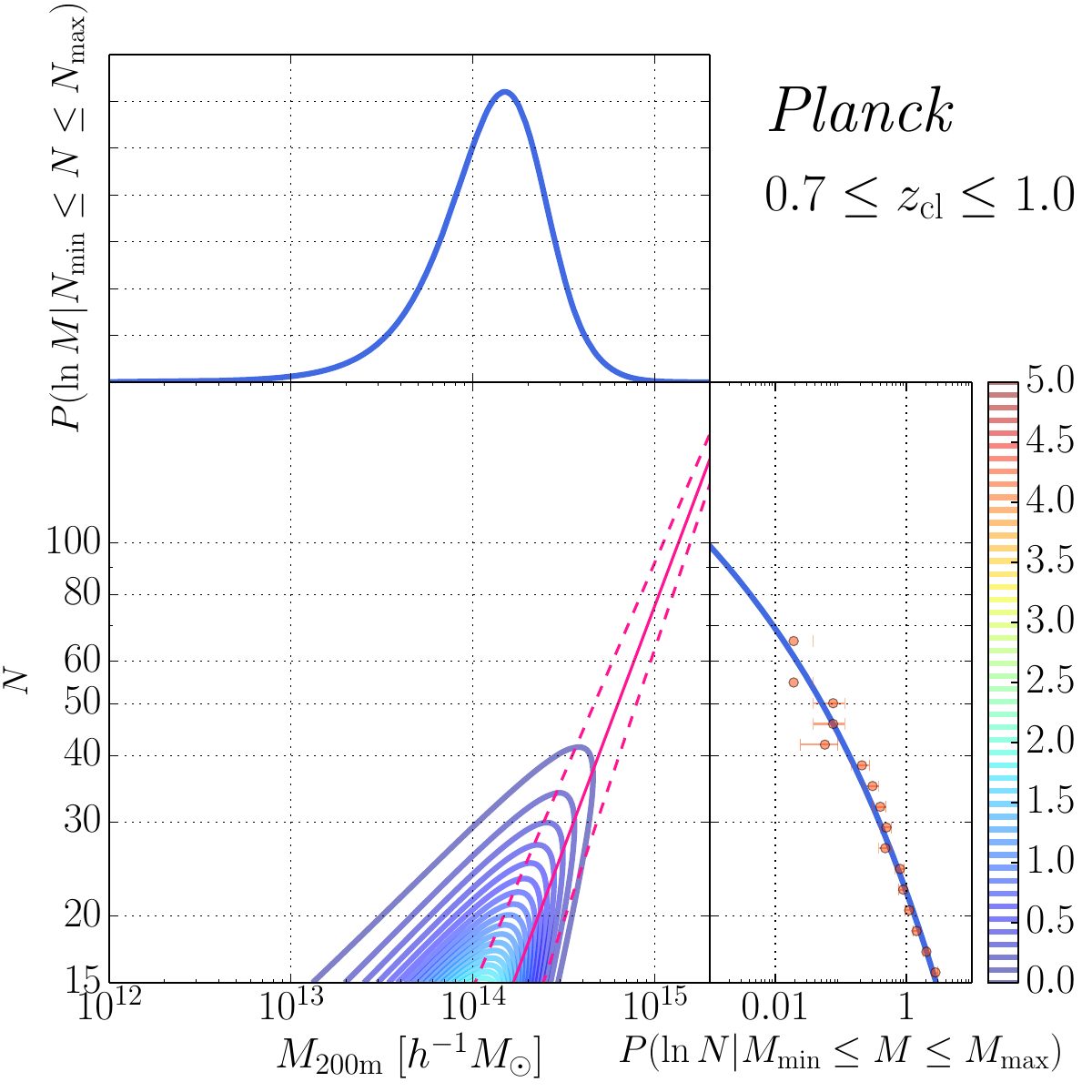}
      \\
      \includegraphics[width=5.62cm]{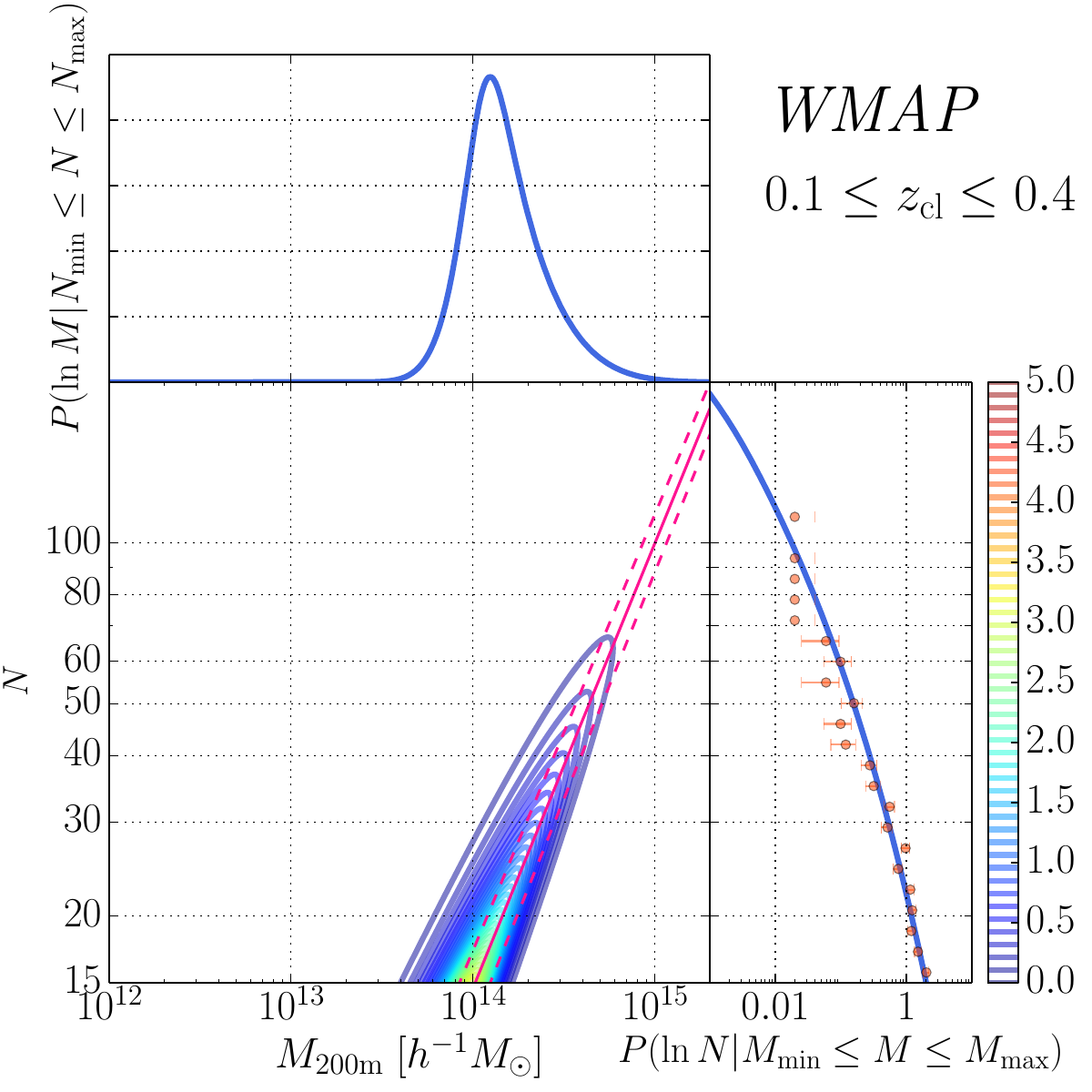}
      \includegraphics[width=5.62cm]{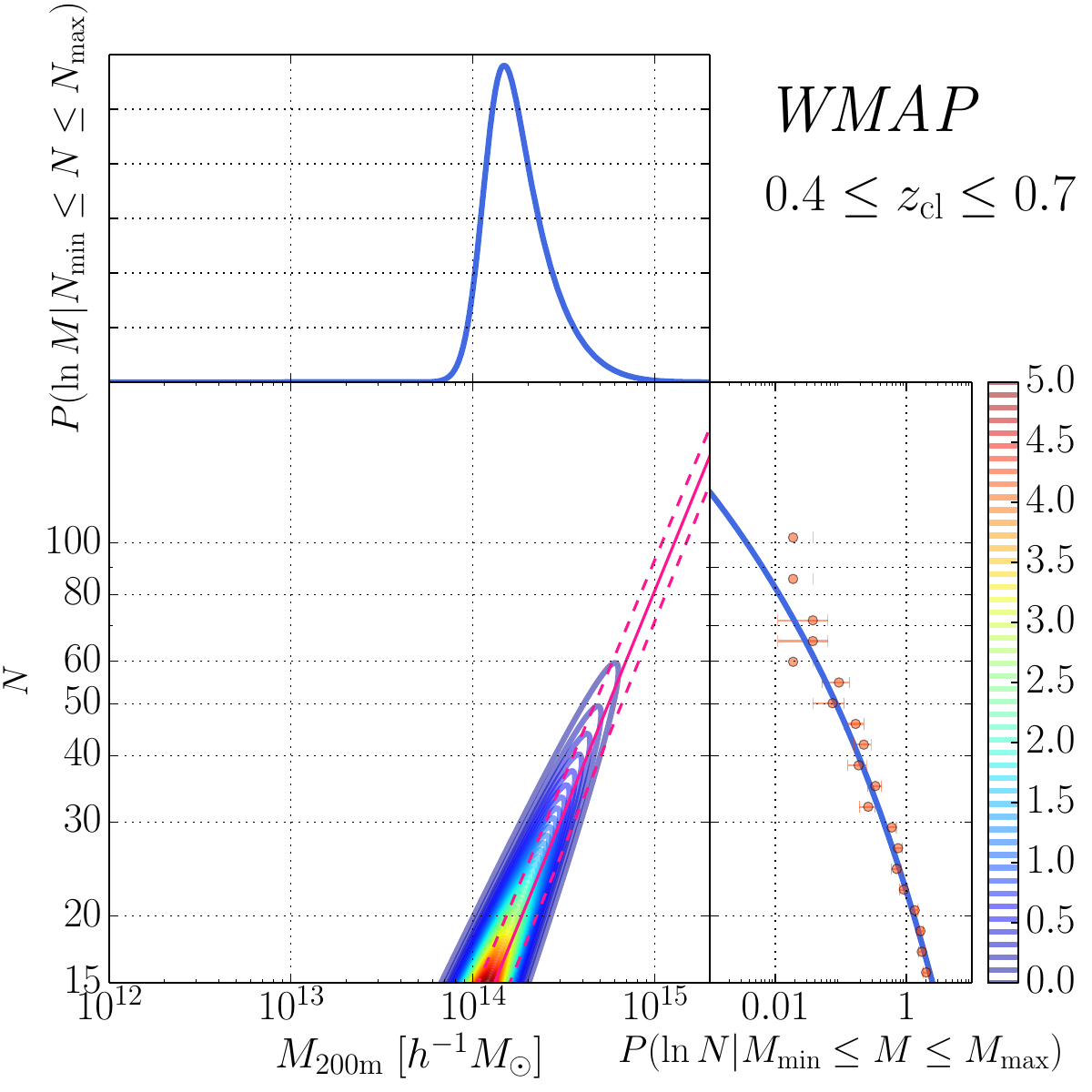}
      \includegraphics[width=5.62cm]{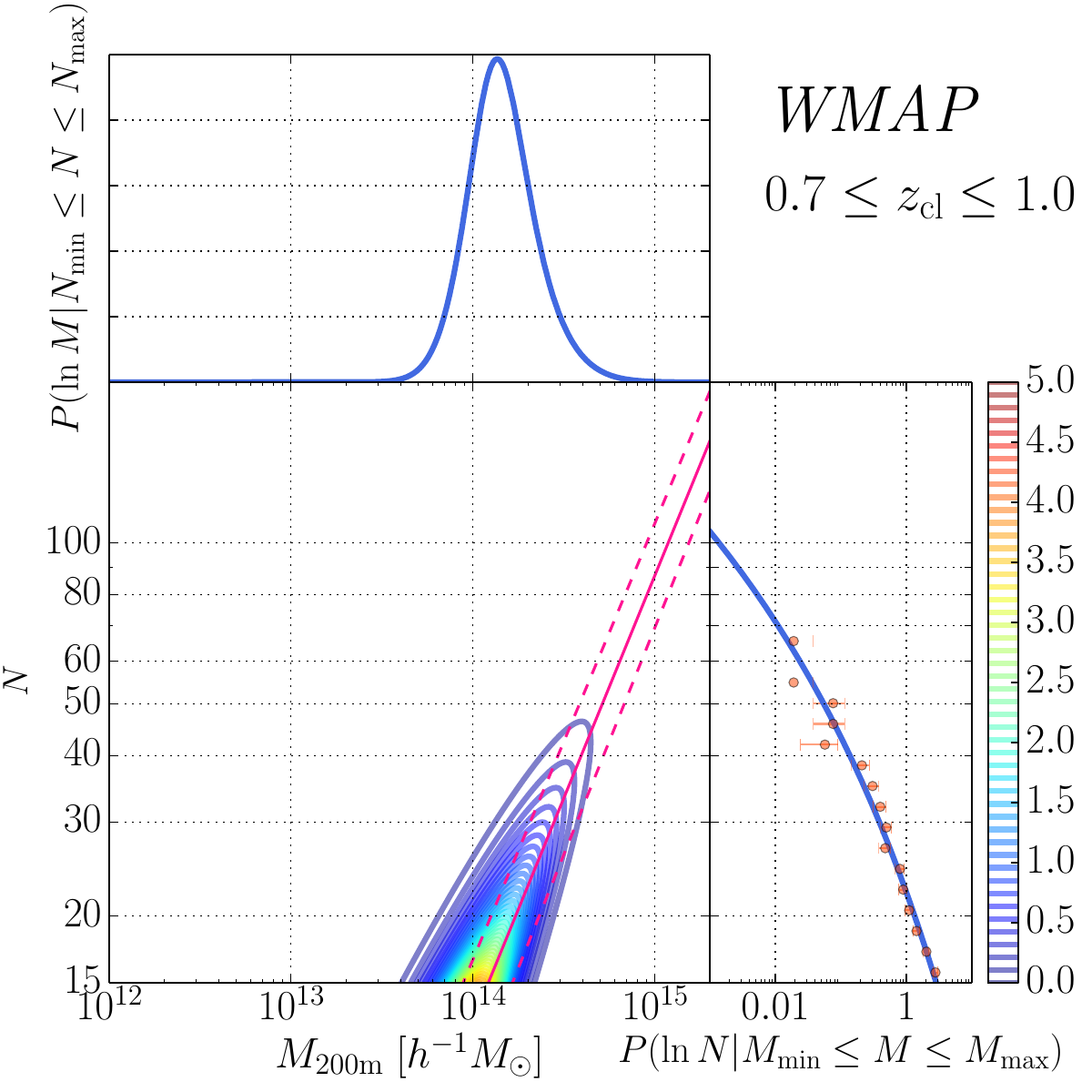}
      }
  \end{center}
  \caption{
            Joint probability distribution $P_{\beta}(\ln M, \ln N)$ defined in equation~(\ref{eq:joint})
            from the best-fit parameters in the fiducial analysis
            in Figures~\ref{fig:fitting_fid_Planck} and \ref{fig:fitting_fid_WMAP}.
            The upper panels show the result for the {\it Planck} cosmological parameters for each redshift bin,
            whereas the lower panels show the result for the {\it WMAP} cosmological parameters.
            The solid line shows the best-fit model of $\langle \ln N \rangle(M, z)$ and the dashed lines show the 16th and 84th percentiles of the richness
            distribution at a fixed mass (i.e., the width of $\sigma_{\ln N|M,z}$).
            Here, for simplicity, we use $z=0.3, 0.6, 0.9$ as representative values for the redshift bins.
            The solid line in the right panel of each plot
            shows the probability distribution of the richness defined in equation~(\ref{eq:Pr_integrate}), 
            and 
            points with error-bars denote measurements with 
            Poisson errors in finer richness binning.
            The top panel of each plot shows
            the probability distribution of the halo mass defined in equation~(\ref{eq:PMtot}).
            }
\label{fig:joint_fid_PlanckWMAP}
\end{figure*}
%
\begin{figure*}
  \begin{center}
    {\includegraphics[width=8.5cm]{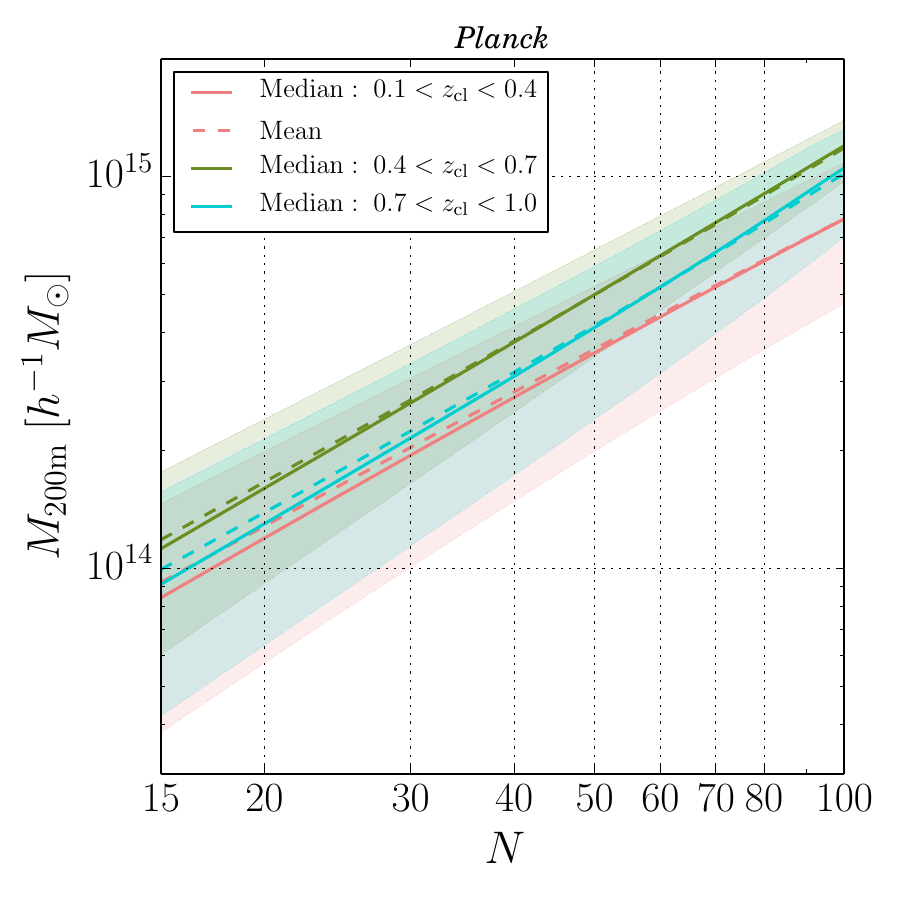}
     \includegraphics[width=8.5cm]{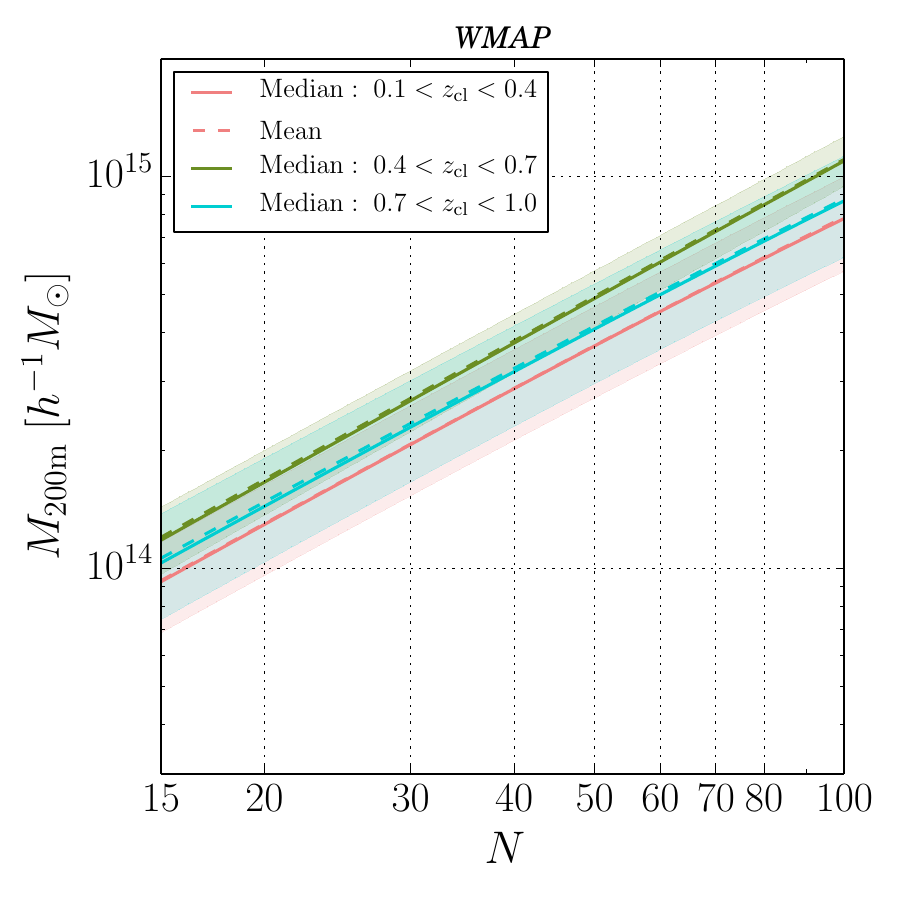}
      }
  \end{center} 
  \caption{ Conditional probability distribution $P_{\beta}(\ln M|N)$ defined in equation~(\ref{eq:bayes_zbins})
from the best-fit richness-mass relation parameters from the fiducial analysis, for all three redshift bins.
Left and right panels show results for {\it Planck} and
{\it WMAP} cosmological parameters, respectively.
The solid line denotes the median of the mass distribution at a fixed richness for each redshift bin, 
and the dashed lines denote the mean mass at a fixed richness $\langle M|N \rangle$ (see equation~\ref{eq:MgivenN}).  
The shaded regions show the range of the 16th and 84th percentiles of the mass distribution at a fixed richness. }
\label{fig:PMgivenN_best_fid_PlanckWMAP}
\end{figure*}
%
\begin{figure*}
  \begin{center}
    {\includegraphics[width=8.5cm]{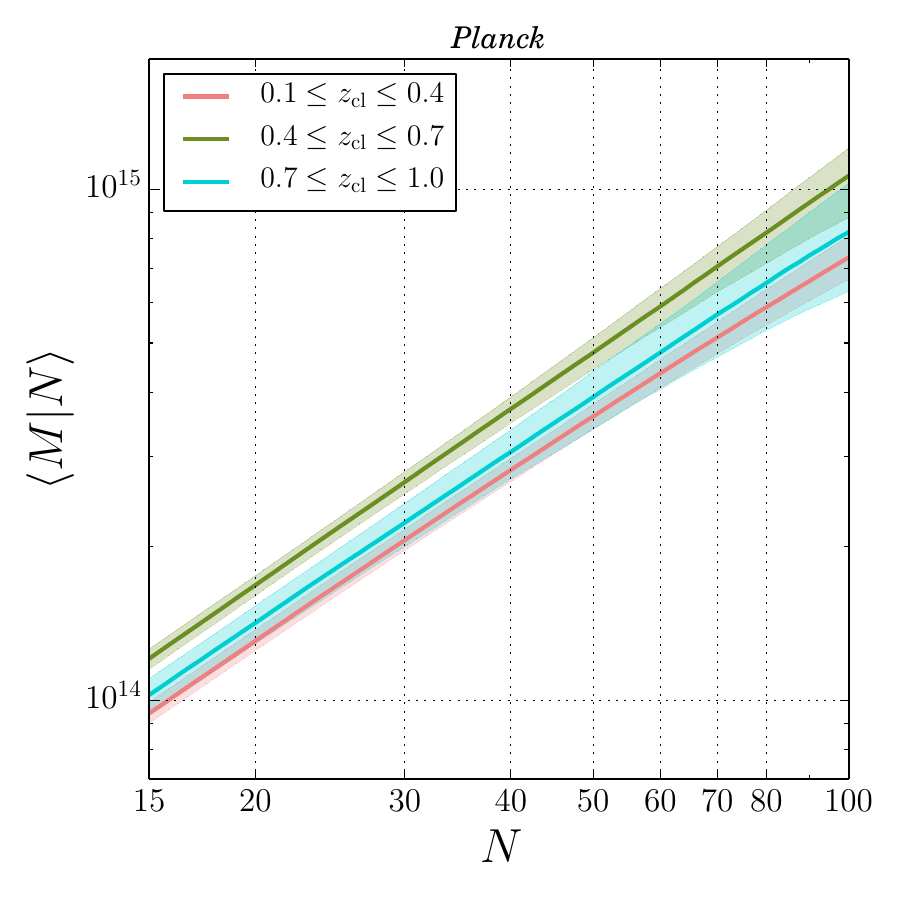}
     \includegraphics[width=8.5cm]{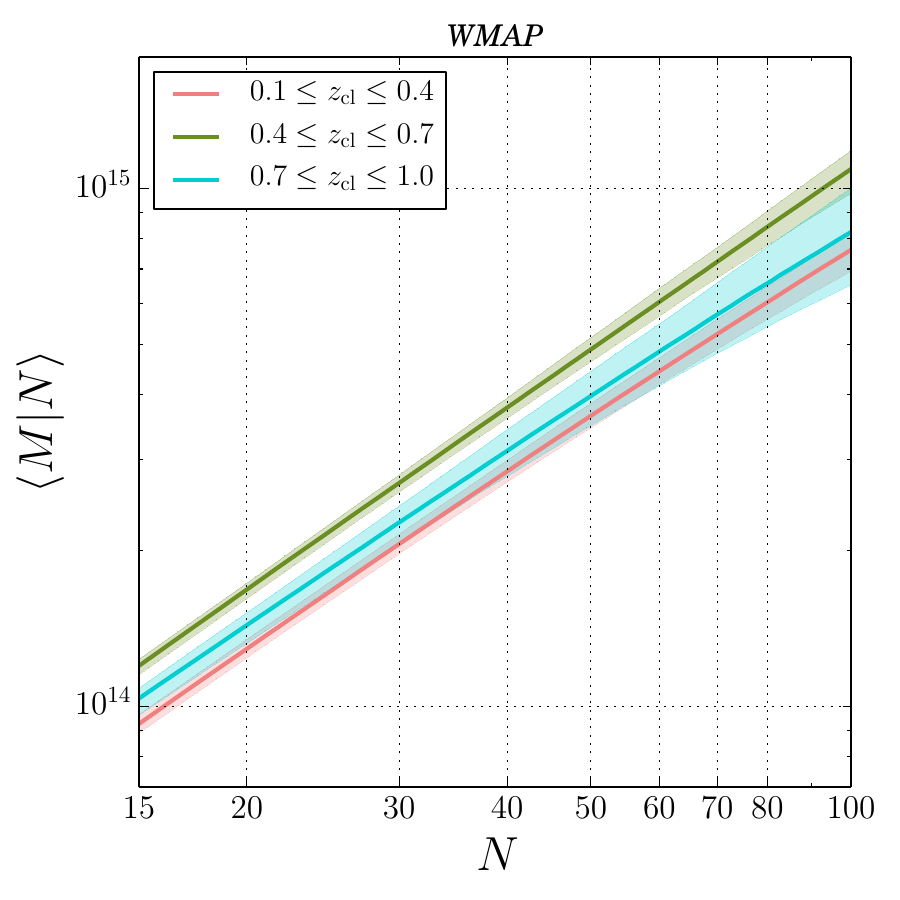}
     \\
     \includegraphics[width=8.5cm]{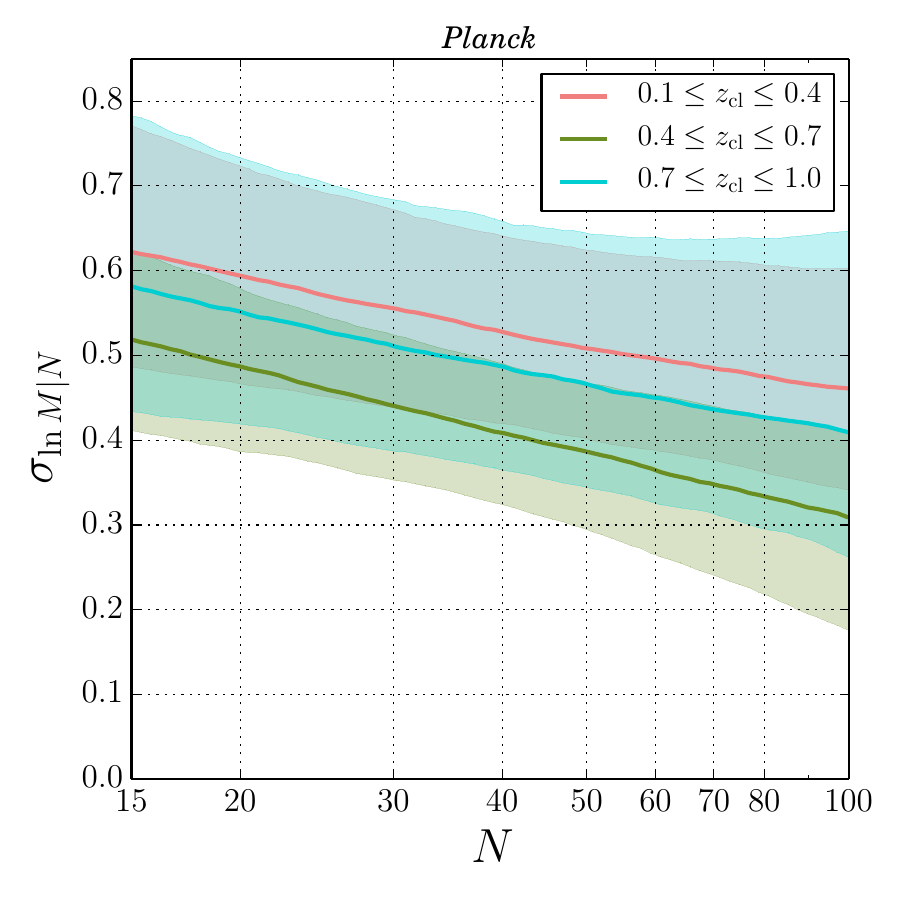}
     \includegraphics[width=8.5cm]{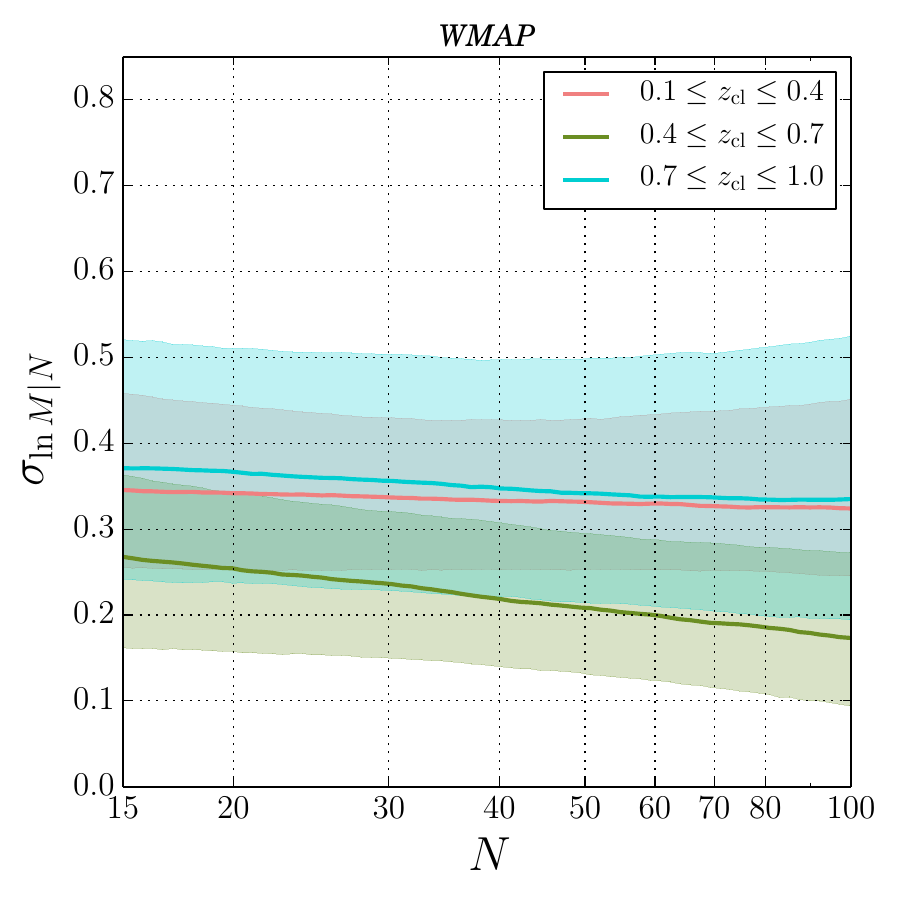}
      }
  \end{center}
  \caption{ The median and the 16th and 84th percentiles 
            of the mean and scatter values of $\langle M|N \rangle$ and $\sigma_{\ln M|N}$
            from the MCMC chains for the fiducial analysis,
            for all three redshift bins.
            Note that this figure differs from 
            Figure~\ref{fig:PMgivenN_best_fid_PlanckWMAP}, which shows 
            $\langle M|N \rangle$ and $\sigma_{\ln M|N}$ at the best-fit parameters only. 
            The left panels show the results for the {\it Planck} cosmological parameters
            and the right panels for the {\it WMAP} cosmological parameters.
            We note that $\sigma_{\ln M|N}$ is defined by the half width of the $68\%$ percentile region of 
            the mass distribution
            at a fixed richness (see text for more details).            
  }
\label{fig:meanscatterchains_fid_PlanckWMAP}
\end{figure*}
We then calculate the joint probability distribution of halo mass and richness in each redshift bin 
after averaging over the redshift range with volume weight ${\rm d}^{2}V/{\rm d}z{\rm d}\Omega=\chi^2(z)/H(z)$ as
\begin{eqnarray}
  P_{\beta}( \ln M,\ln N)&& 
  \propto
  \int_{z_{\beta, \rm min} }^{ z_{\beta, \rm max} }
  \mathrm{d}z~ \frac{ \chi^{2}(z) }{ H(z) } P(\ln N| M, z) P( \ln M| z)\nonumber \\
  && \hspace{-1.0em} \propto
  \int_{z_{\beta, \rm min} }^{ z_{\beta, \rm max} }
  \mathrm{d}z~ \frac{ \chi^{2}(z) }{ H(z) } P(\ln N| M, z) \frac{ {\rm d}n(M, z) }{ {\rm d}\ln M }
  \label{eq:joint}
\end{eqnarray}
where $P(\ln M| z)$ is the probability distribution of the halo mass for a given redshift,
and thus is proportional to the halo mass function ${\rm d}n(M, z)/{\rm d}\ln M$.
The normalization factor is determined in the range of 
$10^{12} \leq M/[h^{-1} M_{\odot}] \leq 2\times 10^{15}$ 
and $15 \leq N \leq 200$, and we restrict the domain of the joint probability to this range. 
We use the best-fit model parameters of the fiducial analysis in Table~\ref{tab:fidparams}
for $P(\ln N|M, z)$.
The contours in Figure~\ref{fig:joint_fid_PlanckWMAP} 
show the joint probability distribution (equation~\ref{eq:joint}) in each redshift bin 
for the {\it Planck} or {\it WMAP} cosmological parameters.
We obtain the distribution of $\ln N$ and $\ln M$ in the range of richness ($N_{\rm min}=15$, $N_{\rm max}=200$) 
for each redshift bin by integrating the joint probability distribution 
along the halo mass and the richness directions respectively 
as
\begin{eqnarray}
& &P_{\beta}(\ln N|M_{\rm min} \leq M \leq M_{\rm max}) \nonumber \\
&=&\int_{\ln M_{\rm min}}^{\ln M_{\rm max}} \mathrm{d}\ln M~ P_{\beta}(\ln M, \ln N)
\label{eq:Pr_integrate}
\end{eqnarray}
and
\begin{eqnarray}
& &P_{\beta}(\ln M|N_{\rm min} \leq N \leq N_{\rm max}) \nonumber \\
&=&\int_{\ln N_{\rm min}}^{\ln N_{\rm max}} \mathrm{d}\ln N~ P_{\beta}(\ln M, \ln N).
\label{eq:PMtot}
\end{eqnarray}
The richness distributions $P_{\beta}(\ln N|M_{\rm min} \leq M \leq M_{\rm max})$
shown in Figure~\ref{fig:joint_fid_PlanckWMAP} 
indicate that the model reproduces the observed richness function at much finer bins 
than those used for the analysis.
The joint probability contours in Figure~\ref{fig:joint_fid_PlanckWMAP} 
show that scatter widths of the mass at a given fixed richness 
for the {\it WMAP} cosmology are smaller than those for the {\it Planck} cosmology for all redshift bins,
and that the widths for the middle redshift bin ($0.4 \leq z_{\rm cl} \leq 0.7$) are the smallest among the redshift bins
for both cosmologies.
We discuss the origin of these results in Section~\ref{sec:discussion:redshiftevolution}.

Since we constrain the richness-mass relation $P(\ln N|M,z)$ from a joint analysis, 
we can compute the mass-richness relation $P(\ln M|N, z)$ using Bayes theorem as
\begin{eqnarray}
P(\ln M|N, z)&=&
\frac{P(\ln N|M, z)P(\ln M| z)}{ \displaystyle \int_{\ln M_{\rm min} }^{\ln M_{\rm max}} {\rm d}\ln M~ P(\ln N|M, z)P(\ln M| z) }\nonumber \\
&=&
\frac{ \displaystyle P(\ln N|M, z)\frac{ {\rm d}n(M, z) }{ {\rm d}\ln M}}{ \displaystyle \int_{\ln M_{\rm min} }^{\ln M_{\rm max}} {\rm d}\ln M~ P(\ln N|M, z)\frac{ {\rm d}n(M, z) }{ {\rm d}\ln M}}
\label{eq:bayes}
\end{eqnarray}
where we use the halo mass function for $P(\ln M|z)$.
We then average the mass-richness relation over the redshift range with volume weight for each redshift bin as
\begin{equation}
P_{\beta}(\ln M|N)=
\frac{ \displaystyle \int_{z_{\beta, {\rm min} }}^{z_{\beta, {\rm max} }}{\rm d}z~ \frac{ \chi^{2}(z) }{ H(z) } P(\ln M|N, z)}
{ \displaystyle \int_{z_{\beta, {\rm min} }}^{z_{\beta, {\rm max} }}{\rm d}z~ \frac{ \chi^{2}(z) }{ H(z)}}.
\label{eq:bayes_zbins}
\end{equation}
We obtain the mean mass for a given richness in each redshift bin as
\begin{equation}
  \langle M|N \rangle_{\beta}={ \displaystyle \int_{\ln M_{\rm min}}^{\ln M_{\rm max}} }{\rm d}\ln M~ P_{\beta}(\ln M|N) M.
\label{eq:MgivenN}
\end{equation}
\begin{figure*}
  \begin{center}
    { \includegraphics[width=8.5cm]{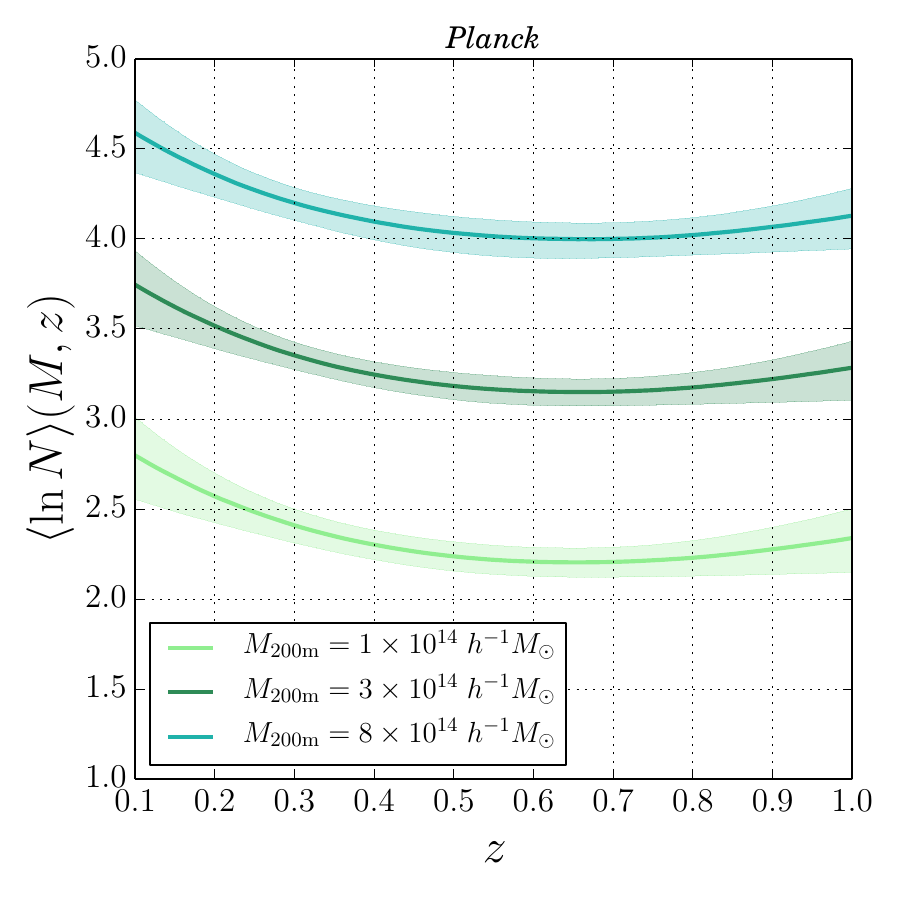}
      \includegraphics[width=8.5cm]{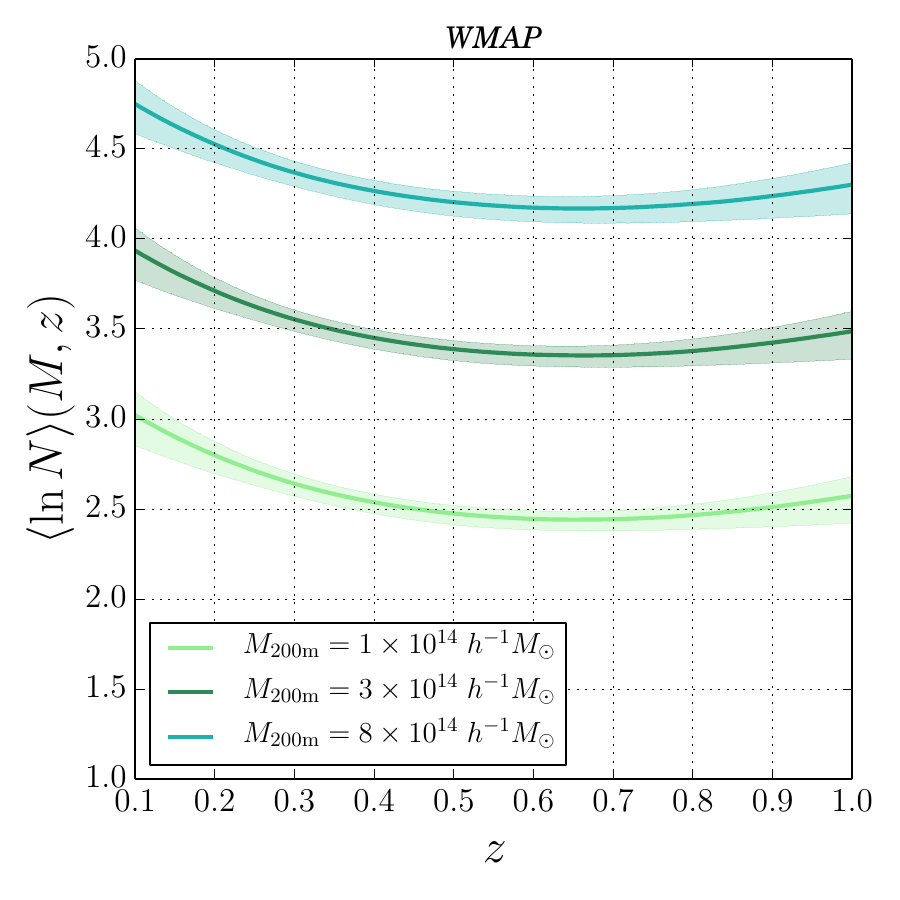}
      \\
      \includegraphics[width=8.5cm]{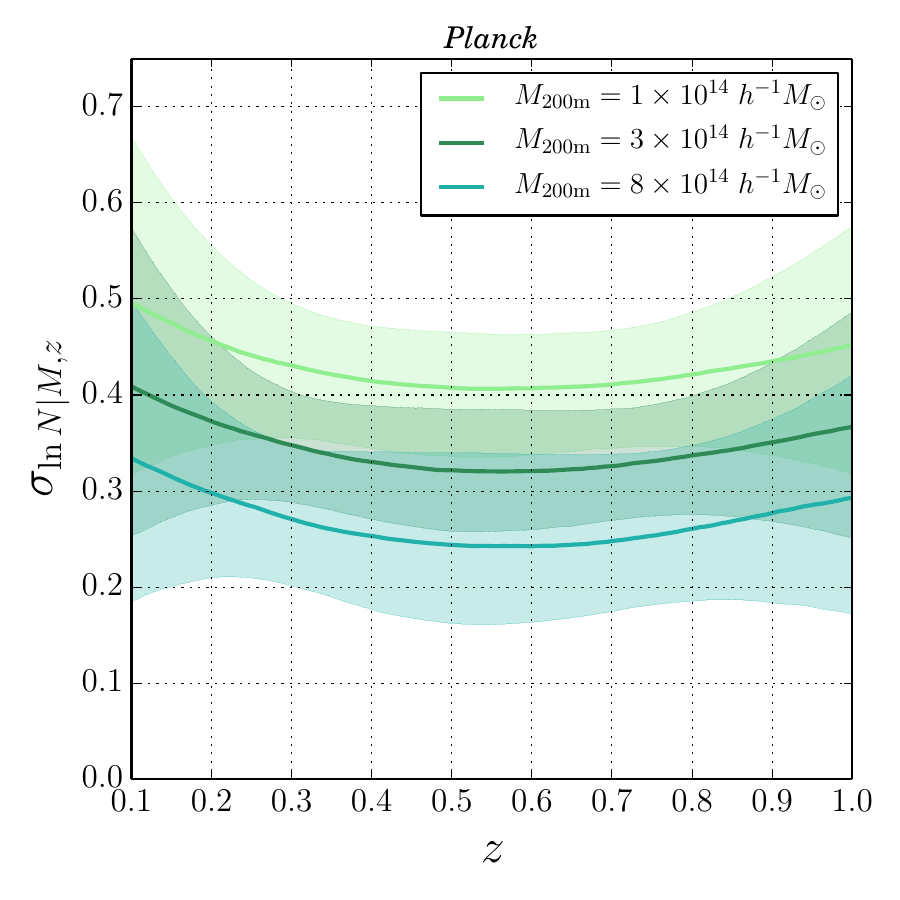}
      \includegraphics[width=8.5cm]{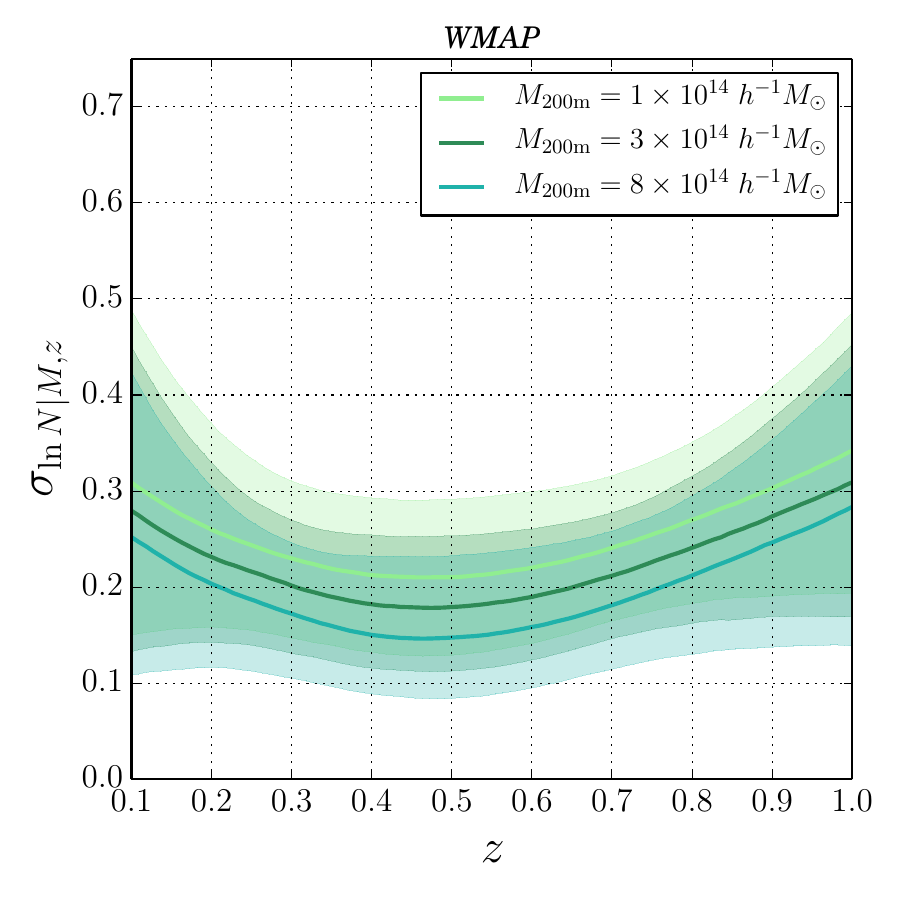}
      }
  \end{center}
  \caption{ The mean and scatter relations of the richness distribution 
    given halo masses as a function of redshift.
    The left panels show the results for the {\it Planck} cosmological parameters
    and the right panels for the {\it WMAP} cosmological parameters.
    We show the results for $M/(10^{14} h^{-1} M_{\odot})=1,3,8$
    as representative halo masses.
    The solid lines show the median values at fixed redshift and the shaded regions
    show the 16th and 84th percentiles from the MCMC chains.}
\label{fig:lnNsigmalnN_fid_PlanckWMAP}
\end{figure*}
Figure~\ref{fig:PMgivenN_best_fid_PlanckWMAP} shows 
the median, mean, and 16$\%$ and 84$\%$ percentile region of the mass distribution at
fixed richness values
in each redshift bin for {\it Planck} and {\it WMAP} cosmologies, using
the best-fit richness-mass relation parameters of the fiducial analysis.

We also show the constraint on the mean relation computed in equation~(\ref{eq:MgivenN}) 
and the scatter relation $\sigma_{\ln M|N}$ in Figure~\ref{fig:meanscatterchains_fid_PlanckWMAP}
from the MCMC chains after marginalizing over the model parameters 
(i.e., not only at the best-fit parameters as shown in Figure~\ref{fig:PMgivenN_best_fid_PlanckWMAP}).
Since the model generally predicts a skewed distribution of halo mass for a fixed richness value in $\ln M$ space,
we define $\sigma_{\ln M|N}$ as the half width of the 68$\%$ percentile region 
of $P_{\beta}(\ln M|N)$ for each redshift bin 
as $\sigma_{\ln M|N}=(\ln M_{84} - \ln M_{16} )/2$, 
where $M_{84}$ and $M_{16}$ 
are masses corresponding to the 84th and 16th percentiles 
of $P_{\beta}(\ln M|N)$ at a fixed richness,
respectively.
The mean relations for {\it Planck} and {\it WMAP} cosmologies
are consistent with each other given the error bars for all the redshift bins, 
and the mean relation of $0.4 \leq z_{\rm cl} \leq 0.7$ 
has a larger amplitude than in the other redshift bins
for both cosmologies with relatively high significance given the error bars.
We constrain the mean relations at $N=25$ with $\sim 4 \%$ precision for 
$0.1 \leq z_{\rm cl} \leq 0.4$ and $0.4 \leq z_{\rm cl} \leq 0.7$, and 
$\sim 8\%$ precision for $0.7 \leq z_{\rm cl} \leq 1.0$.
The scatter relation for the {\it Planck} cosmology increases toward lower richness values
for all redshift bins, 
whereas the scatter for the {\it WMAP} cosmology is consistent 
with a constant value as a function of richness for all redshift bins.
The scatter values for the {\it Planck} cosmology are systematically
larger than those for the {\it WMAP} cosmology.
This result is qualitatively consistent with the one obtained from cosmological analysis of SDSS redMaPPer clusters in \cite{Costanzietal2018b}, 
which shows larger scatter values for larger 
$S_8=\sigma_8(\Omega_{\rm m}/0.3)^{0.5}$ values 
($S_8=0.85$ for the {\it Planck} and $S_8=0.79$ for the {\it WMAP} cosmology in this work),
although the scatter modeling method is different from ours.
The scatter values in the middle redshift bin ($0.4 \leq z_{\rm cl} \leq 0.7$)
are also lower than those in the other redshift bins.
We discuss the origin of this result in Section~\ref{sec:discussion:redshiftevolution}.
%
\begin{figure*}
  \begin{center}
    { \includegraphics[width=5.6cm]{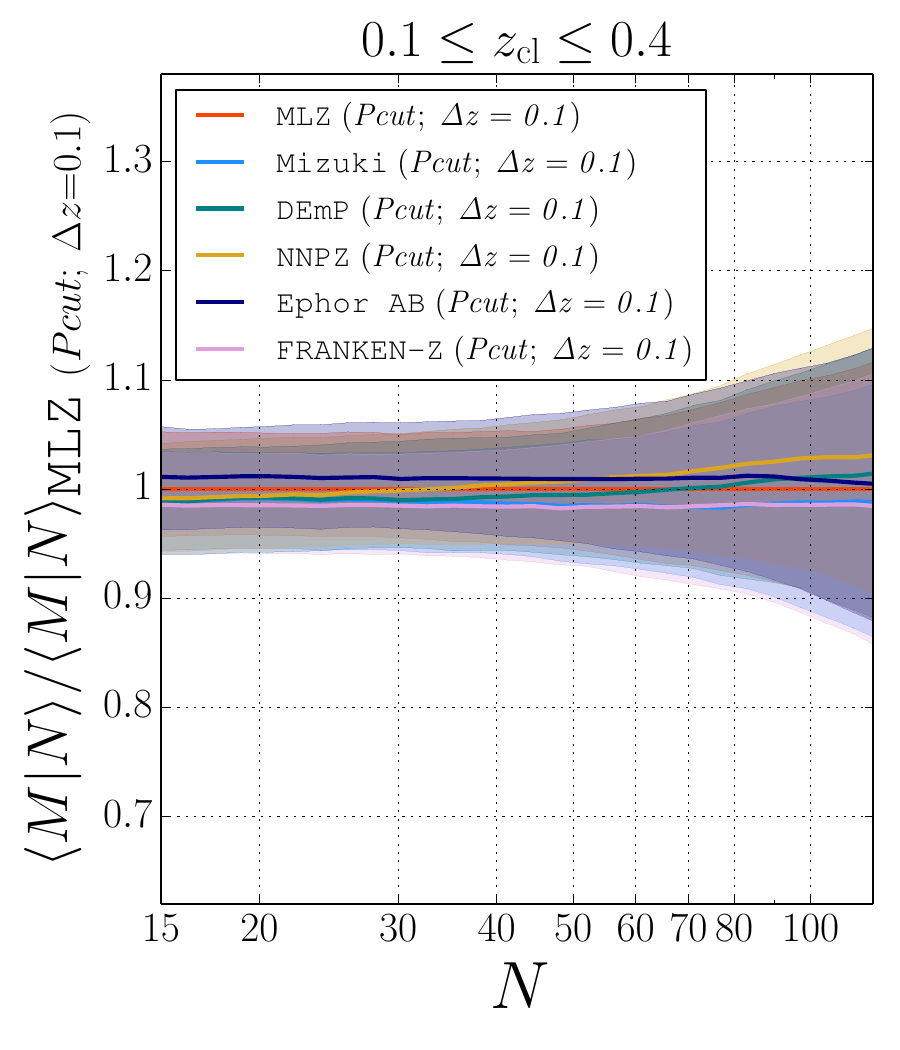}
      \includegraphics[width=5.6cm]{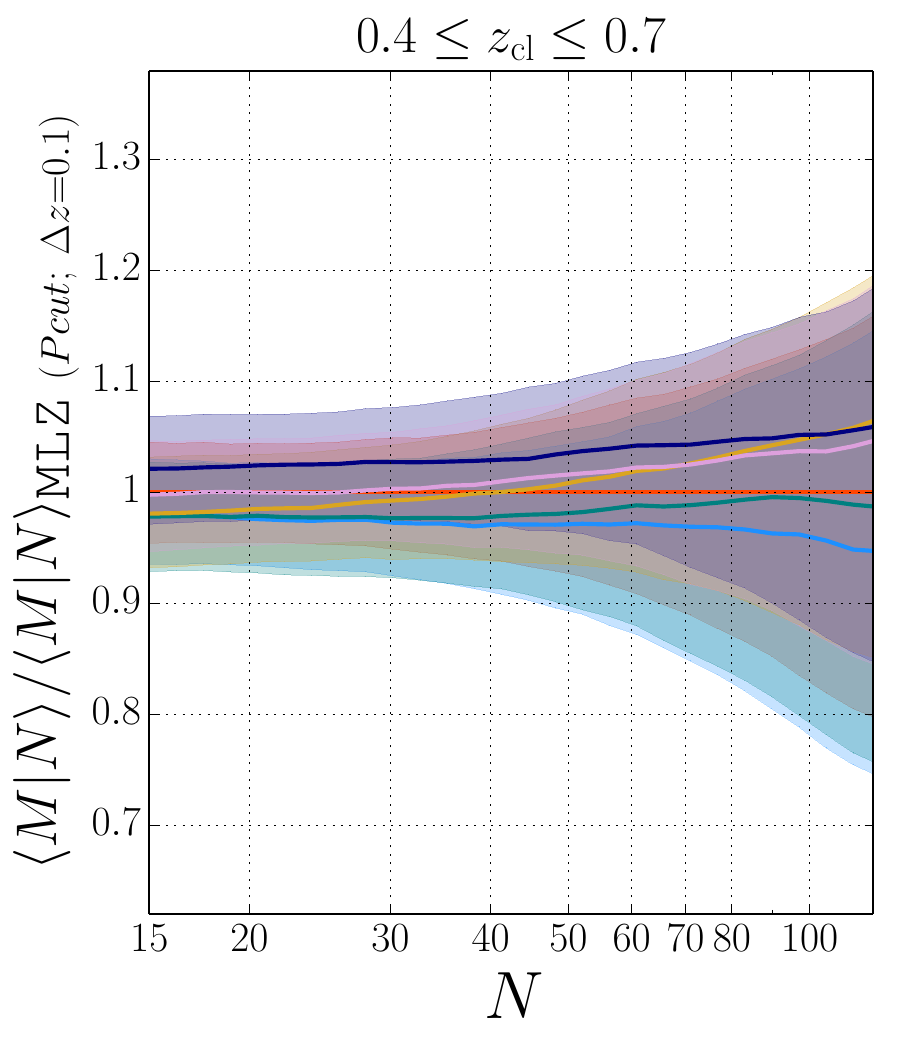}
      \includegraphics[width=5.6cm]{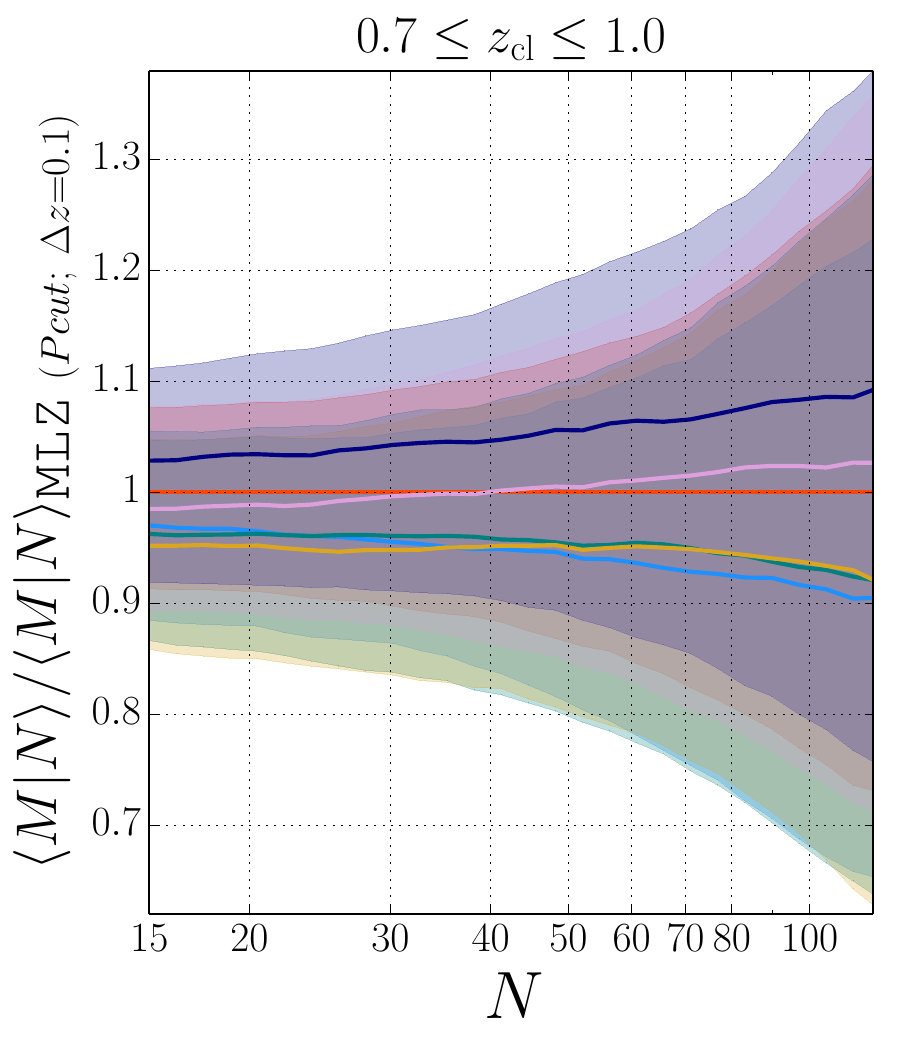}
      \\
      \includegraphics[width=5.6cm]{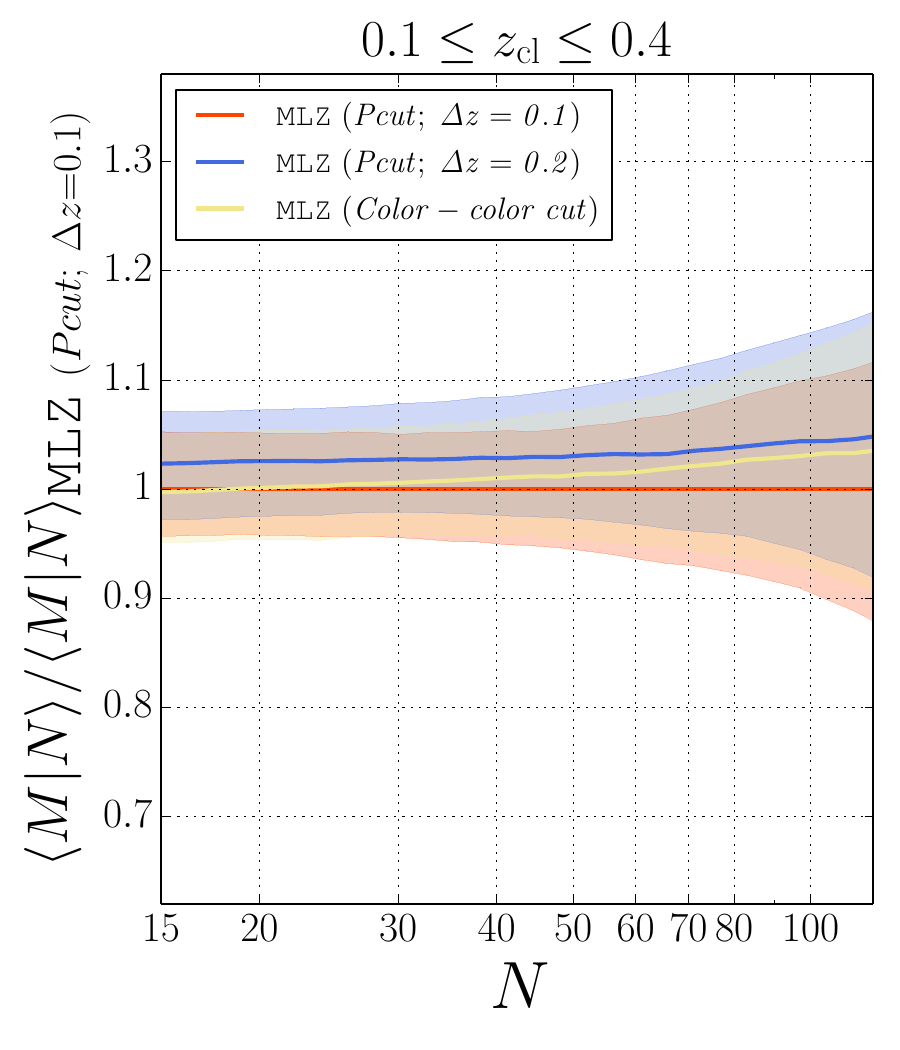}
      \includegraphics[width=5.6cm]{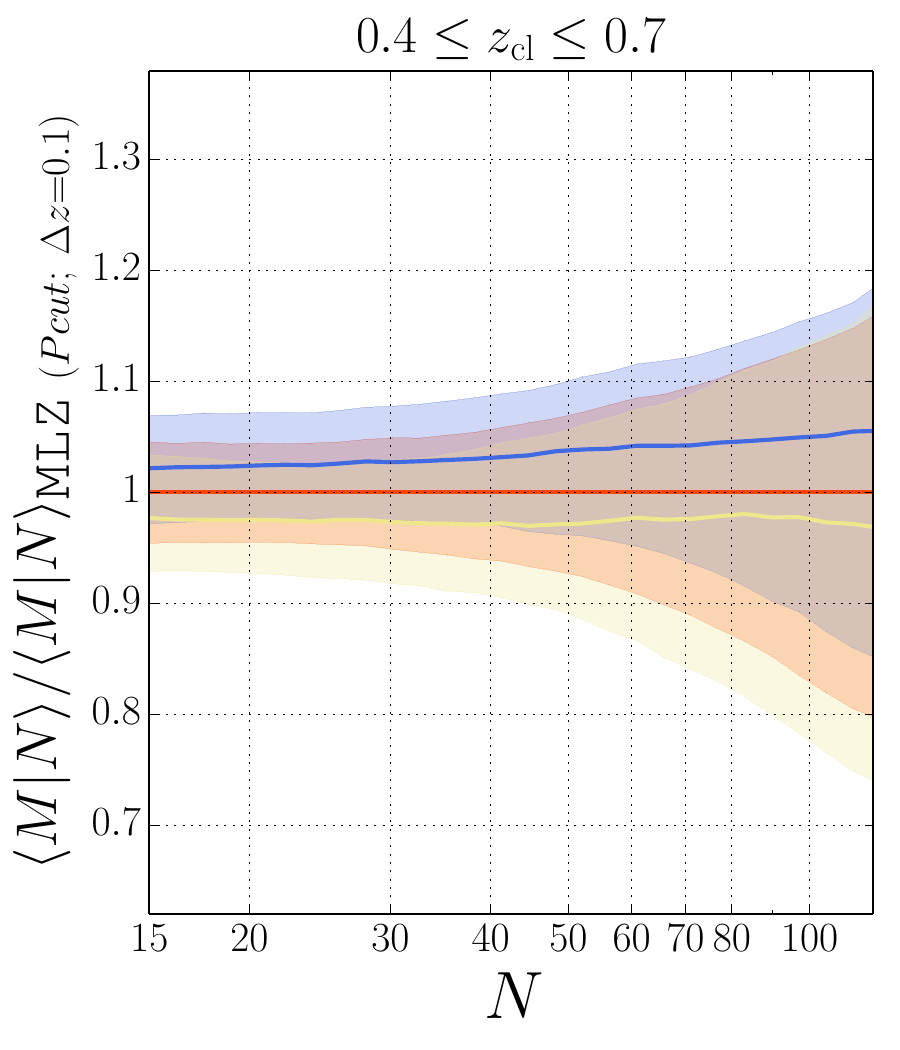}
      \includegraphics[width=5.6cm]{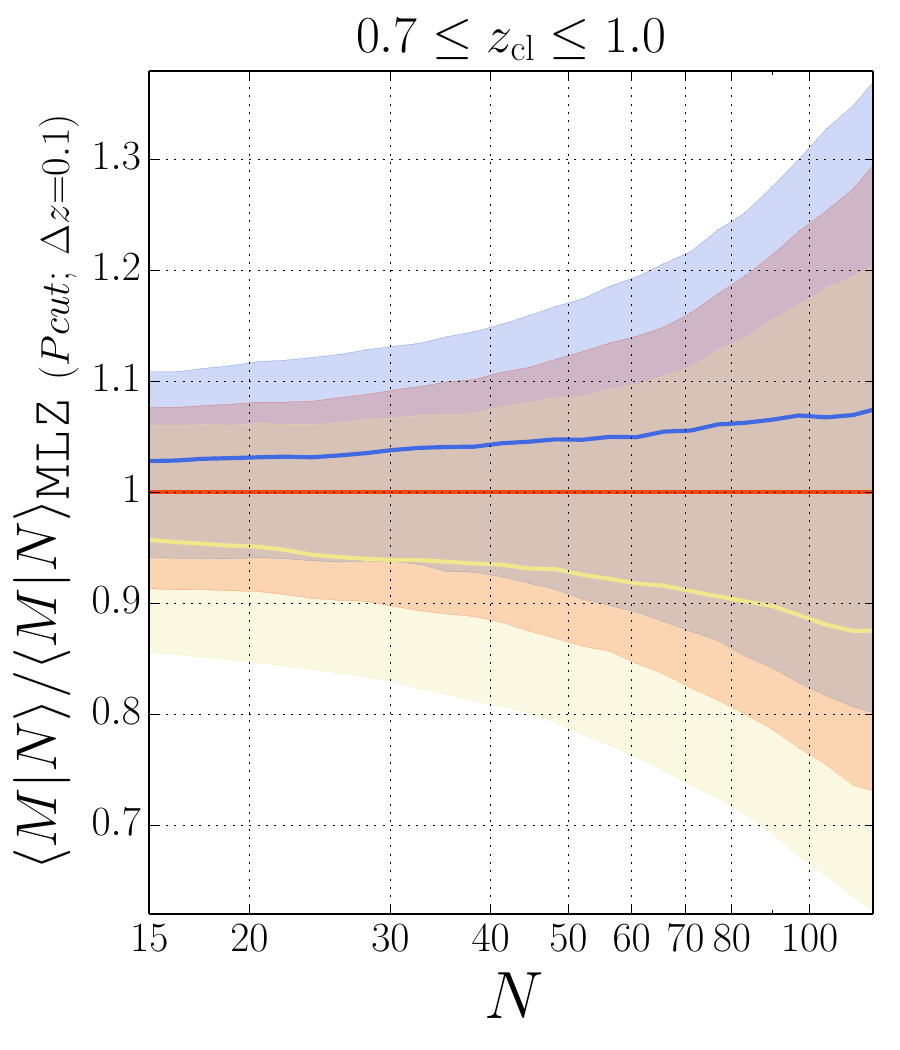}
      }
  \end{center}
  \caption{
    Comparison of the mean mass relation $\langle M|N \rangle$ 
    defined in equation~(\ref{eq:MgivenN}) for each redshift bin
    among the different photometric redshift catalogs (upper panels)
    or the different source selection cuts (lower panels)
    with the {\it Planck} cosmological parameters.
    We show the median and the 16th and 84th percentiles from the MCMC chains,
    with respect to the median values of $\langle M|N \rangle$ for the fiducial catalog $\texttt{MLZ}$
    and the fiducial cut ${\it Pcut}$ with $\Delta z=0.1$.
  }
\label{fig:different_photozcodescuts}
\end{figure*}
\subsection{Richness-mass relation $P(\ln N|M, z)$}\label{sec:result:richnessmass}
As complementary results to Figure~\ref{fig:meanscatterchains_fid_PlanckWMAP} on $P_{\beta}(\ln M|N)$,
we show the mean and scatter relations in equations~(\ref{eq:mean_relation}) and (\ref{eq:scatter_M}) of
the richness-mass relation $P(\ln N|M, z)$ in Figure~\ref{fig:lnNsigmalnN_fid_PlanckWMAP}
for typical masses as a function of redshift.
We find that, for both {\it Planck} and {\it WMAP} cosmologies,
the mean relation has a minimum around $z\sim 0.5$ for all typical masses.
This is also true for the scatter relation especially for {\it WMAP} cosmology.
We also discuss the origin of these constraints in Section~\ref{sec:discussion:redshiftevolution}.
\section{Discussion} \label{sec:discussion}
We discuss the robustness of the fiducial results in Section~\ref{sec:discussion:robustness},
and the complementarity of lensing profile and abundance measurements to constrain the richness-mass relation parameters 
in Section~\ref{sec:discussion:complementarity}.
In Section~\ref{sec:discussion:redshiftevolution},
we discuss redshift evolution in the richness-mass relation and 
the difference between the middle redshift bin versus the lower and higher redshift bins 
shown in Sections~\ref{sec:result:joint} and \ref{sec:result:richnessmass}.
%
\subsection{Robustness of our results}\label{sec:discussion:robustness}
\subsubsection{Shape and photometric redshift measurement uncertainties}
In the fiducial analysis, we marginalize over the shape and photometric redshift uncertainties 
on the lensing measurements by 
including the nuisance parameter $m_{\rm lens}$ in equation~(\ref{eq:lensingbias}).
To check how these uncertainties affect the model parameter constraints, 
we repeat the MCMC analysis ignoring these errors (i.e., setting $m_{\rm lens}=0$) 
for the {\it Planck} cosmology with the same measurements and covariance as in the fiducial analysis.
We find no significant shift in the best-fit parameters with $\chi^{2}_{\rm min}=106.9$ 
compared to $\chi^{2}_{\rm min}=107.0$ for the fiducial analysis with $m_{\rm lens}$. 
We also find the 68$\%$ percentile error widths do not change significantly from the fiducial analysis.
Specifically, the difference between the error widths in this modified versus the fiducial analysis
is smaller than 5$\%$ of the error widths in the fiducial analysis for all parameters.
This result shows that the impact of these shape and photometric redshift measurement errors 
on the joint analysis of current lensing and abundance measurements 
is negligible when constraining the richness-mass relation.
%
\begin{figure*}[t]
  \begin{center}
    \includegraphics[width=0.9 \textwidth]{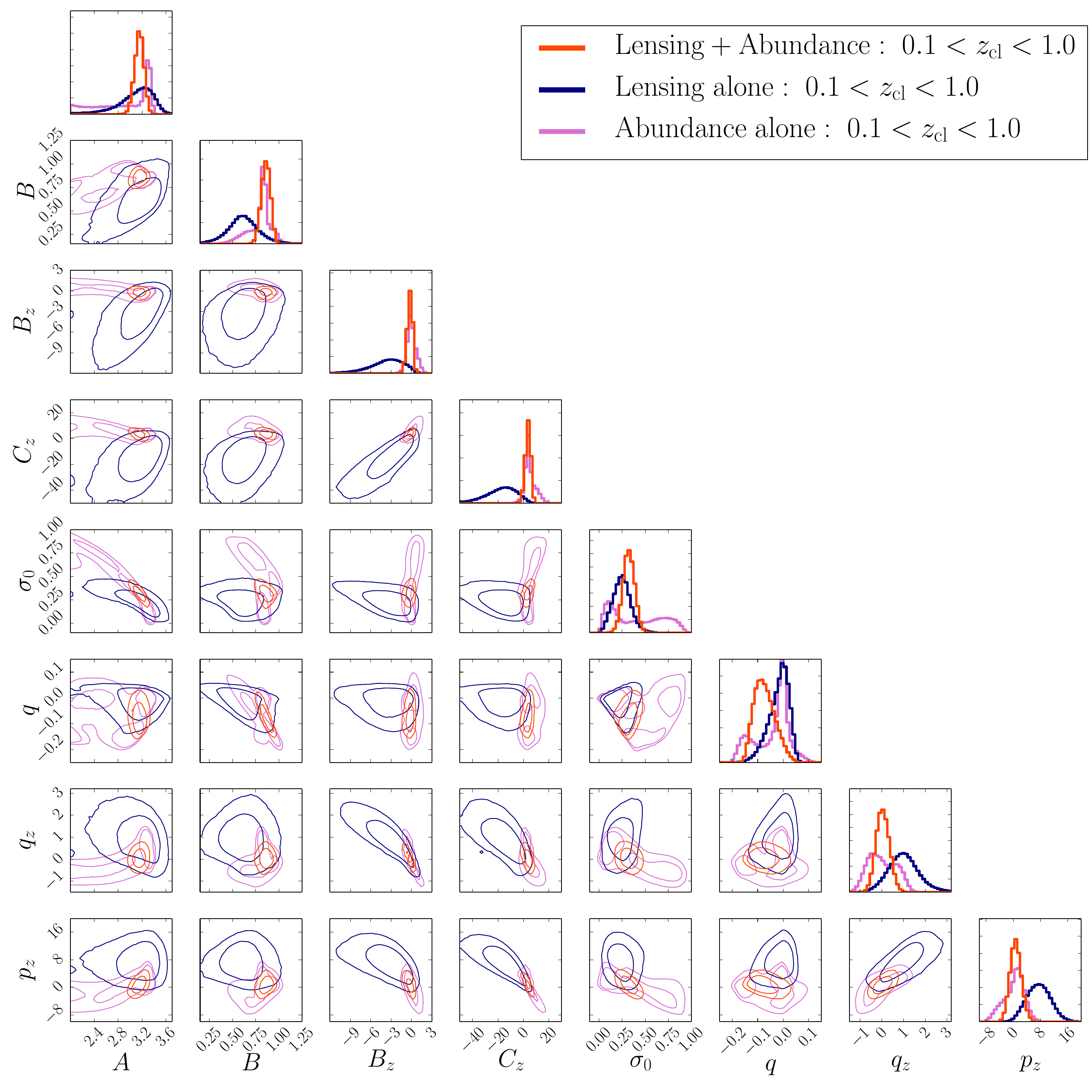}
  \end{center}
  \caption{
    Posterior distributions of the richness-mass relation parameters and the 68$\%$ and 95$\%$ credible level intervals
    in each two-parameter subspace from the MCMC chains of
    the fiducial (red), lensing-alone (navy), and abundance-alone
     (magenta) analyses 
    with the {\it Planck} cosmological parameters.
    The minimal prior range of the parameter $A$ is conservatively set to be $2$
    as shown in Table~\ref{tab:fidparams},
    but the abundance-alone analysis has a tail of the distribution at the minimum value of $A$,
    mainly due to the substantial flexibility in the adopted model for the richness-mass relation
    and the weaker constraining power of the abundance on the mean normalization $A$.
  }
\label{fig:MCMC_lensingalone_abundancealone}
\end{figure*}
\subsubsection{Different photometric redshift catalogs or different photometric redshift cuts}\label{subsubsec:diffphotoz}
We use the photometric catalog \texttt{MLZ} 
and the photometric redshift cuts of {\it Pcut} with $\Delta z=0.1$ for the fiducial analysis 
in Section~\ref{sec:results}.
We check the robustness of our results by using different photometric redshift catalogs 
(see Section~\ref{sec:data:photoz}) or 
different photometric redshift cuts (see Section~\ref{sec:measurements:lens})
for the lensing measurements.
Here we assume the {\it Planck} cosmological parameters for this test, using  
the same model parameters and priors as shown in Table~\ref{tab:fidparams}.
We repeat the same procedure for the lensing measurements and lensing covariance calculation 
for different photometric redshift catalogs or cuts to constrain the model parameters 
by jointly fitting to the lensing and abundance measurements.

In Figure~\ref{fig:different_photozcodescuts}, 
we show the median and the 16th and 84th percentiles of 
the mean relation $\langle M|N \rangle$ for each redshift bin 
with respect to the median from the fiducial result of \texttt{MLZ} and the {\it Pcut} method with $\Delta z=0.1$.
The top panels in Figure~\ref{fig:different_photozcodescuts} show 
the results with different photometric redshift catalogs, 
but with the same photometric redshift cuts of the {\it Pcut} method with $\Delta z=0.1$.
We show that the mean relation $\langle M|N \rangle$ 
is consistent between the difference photometric redshift catalogs for all redshift bins. 
The lower panels in Figure~\ref{fig:different_photozcodescuts} show
the results with different photometric redshift cuts, but with the same photometric redshift catalog \texttt{MLZ},
showing that the results are consistent with each other given the error bars for all of the redshift bins.
This is partly due to our conservative choice of 
the radial range from $0.5 h^{-1} {\rm Mpc}$ in comoving coordinates
for the lensing measurements to avoid possible dilution effects on the lensing measurement
based on \cite{Medezinskietal2018b}.
These results show the robustness of our result to photo-$z$ differences.
%
 \subsection{Complementarity of abundance and stacked lensing profile measurements}\label{sec:discussion:complementarity}
We constrained the model parameters by jointly fitting to the lensing profiles and abundance measurements
in Section~\ref{sec:results}.
Here we study how lensing profile or abundance measurements alone constrain the model
parameters, which helps explain
how the joint analysis lifts the model parameter degeneracies.
We note that we use the same measurements and covariance matrix 
as in the fiducial analysis.

Figure~\ref{fig:MCMC_lensingalone_abundancealone} shows the 68$\%$ and 95$\%$ credible level contours 
in each two-parameter subspace of the richness-mass relation parameters.
The figure shows that the two observables are complementary to each other, which is why their
combination can efficiently lift model parameter degeneracies. 
The lensing measurements constrain the mean normalization parameter $A$ better than the abundance measurements,
mainly because the lensing measurements are more sensitive to the mass scale in each redshift and richness bin.
The lensing measurements also constrain the scatter normalization parameter $\sigma_0$ better than the abundance.
On the other hand, other parameters such as the mass dependence parameter $B$ and 
the redshift evolution parameters $B_z$, $C_z$ in the mean relation are relatively better constrained 
by the abundance measurements.
In summary, the two observables combine effectively to break complex degeneracies in the model parameters.
\subsection{Redshift evolution in the richness-mass relation} \label{sec:discussion:redshiftevolution}
\begin{figure}
  \begin{center}
     \includegraphics[width=8.0cm]{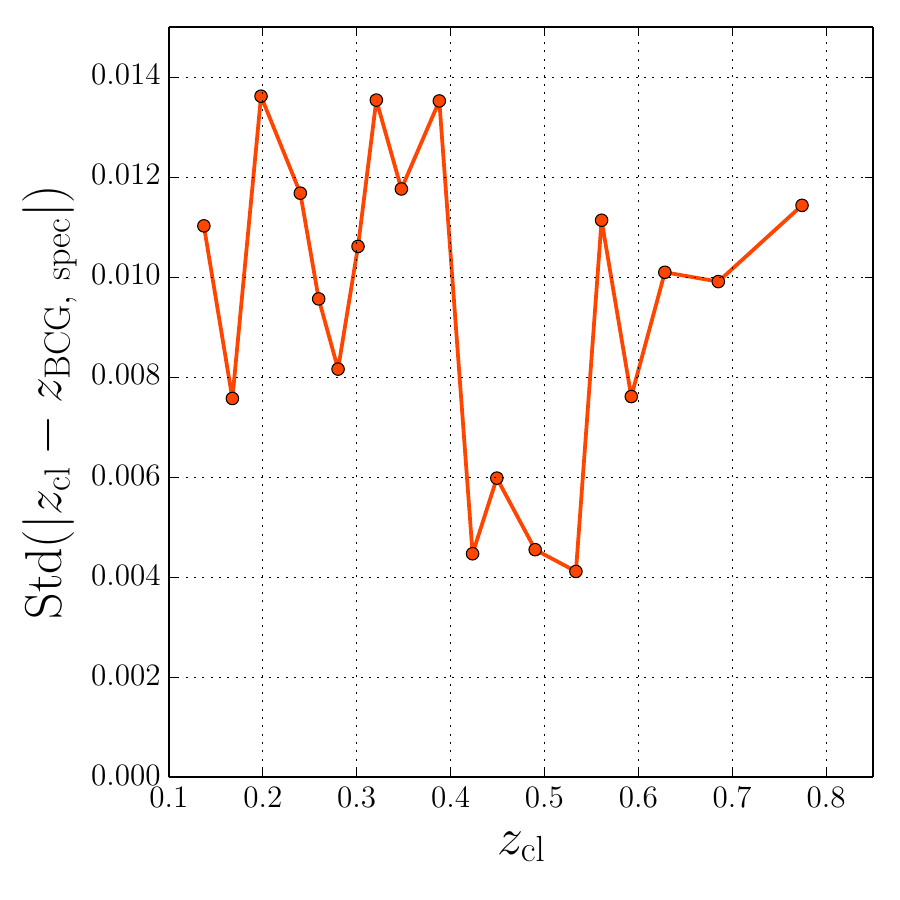}
  \end{center}
  \caption{
    Comparison between the photometric cluster redshifts from the CAMIRA algorithm and
    the available spectroscopic redshifts of BCGs without any clipping.
    We show standard deviations of the difference between the photometric
    and spectroscopic redshifts in each photometric redshift bin.
    Here we use 841 clusters with $N \geq 15$, $0.1 \leq z_{\rm cl} \leq 1.0$, and spectroscopic redshifts.
    We use 19 bins 
    that are defined such that 
    each bin includes almost the same number of clusters.
    The standard deviations around $0.4 \leq z_{\rm cl} \leq 0.55$ are smaller 
    than those in the lower and higher redshift bins (see text for more details).
    }
\label{fig:camira_zclzspec}
\end{figure}
%
\begin{figure*}
  \begin{center}
    {
    \includegraphics[width=8.5cm]{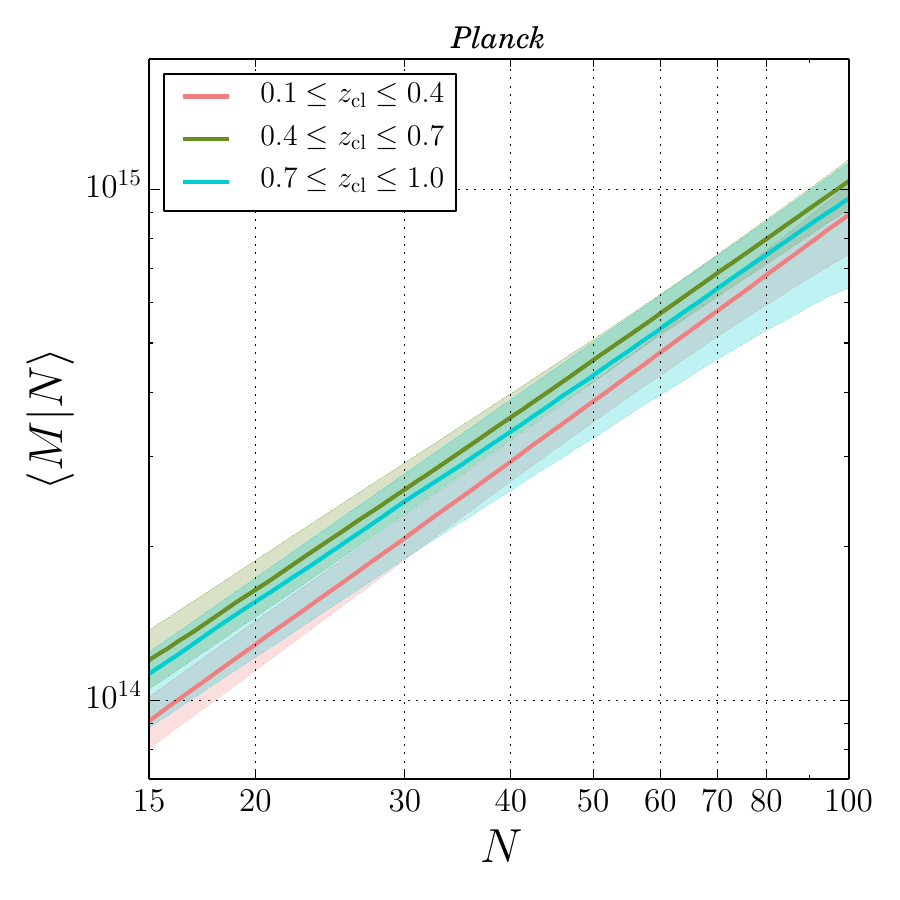}
    \includegraphics[width=8.5cm]{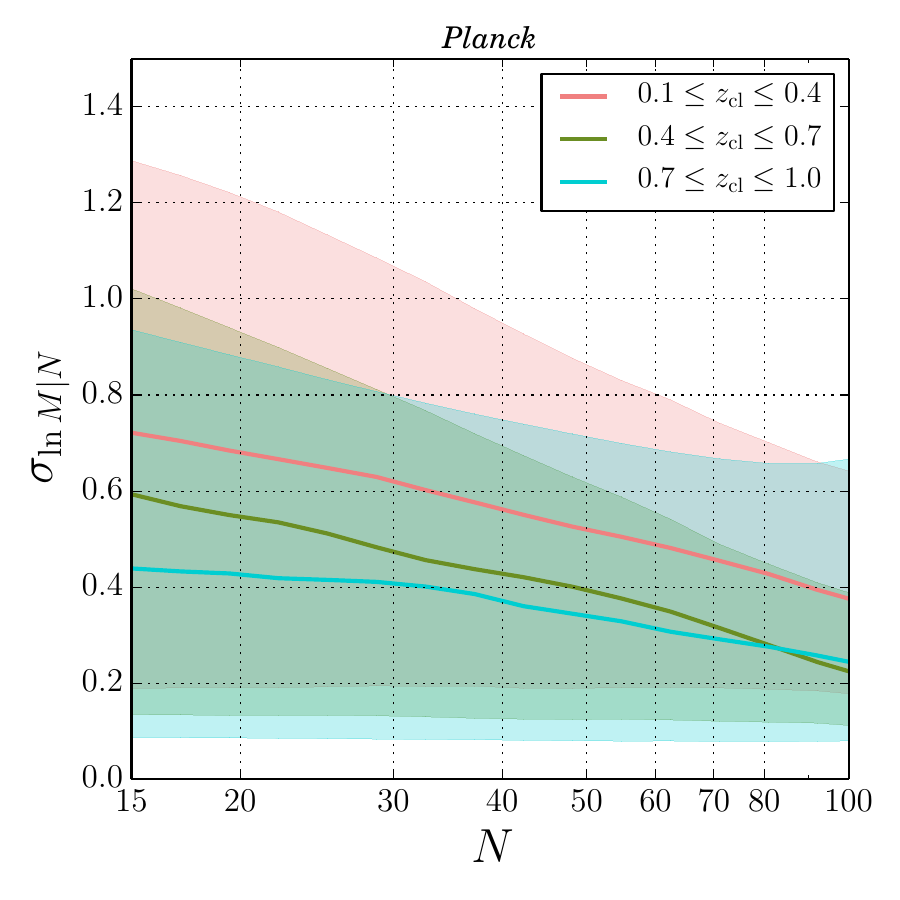}
    }
  \end{center}
  \caption{
    The median and the 16th and 84th percentiles of $\langle M|N \rangle$ and $\sigma_{\ln M|N}$ 
    from the analysis of individual redshift bins 
    without redshift evolution parameters
    in the richness-mass relation,
    assuming {\it Planck} cosmology.
    These results are consistent with 
    the result of our fiducial analysis shown in Figure~\ref{fig:meanscatterchains_fid_PlanckWMAP}.
  }
\label{fig:meanscatter_onez_Planck}
\end{figure*}
In this paper, we use the richness-mass relation model with linear and square redshift evolution parameters 
($B_z$, $C_z$, $q_z$, and $p_z$)
in the fiducial analysis.
When we use the richness-mass relation model without
any redshift evolution parameters (i.e., without 
$B_z$, $C_z$, $q_z$, and $p_z$) we
obtain $\chi^{2}_{\rm min}/{\rm dof}=151.3/101$ ($p$-${\rm value}=8\times 10^{-4}$),
and when we adopt the model without square 
redshift evolution parameters (i.e., without $C_z$
and $p_z$) we obtain
$\chi^{2}_{\rm min}/{\rm dof}=138.6/99$ ($p$-${\rm value}=5\times 10^{-3}$), 
both of which are unacceptable $p$-values.
Here
we use the {\it Planck} cosmological parameters with the same covariance as the fiducial analysis
for these analyses.
Given the acceptable $p$-value of $\chi^{2}_{\rm min}/{\rm dof}=107.0/97$ 
in Table~\ref{tab:fidparams},
we use the richness-mass model with both linear and square redshift evolution parameters 
in the fiducial analysis.

With such linear and square redshift evolution parameters,
the model prediction can include a non-monotonic dependence
on redshift for a fixed mass in $P(\ln N|M, z)$ or a fixed richness in $P_{\beta}(\ln M|N)$.
These non-monotonic behaviors in $P(\ln N|M, z)$ and $P_{\beta}(\ln M|N)$
are preferred
given the significant improvements of $\chi^{2}_{\rm min}$ by adding the redshift evolution parameters
as shown in Figures~\ref{fig:joint_fid_PlanckWMAP}, \ref{fig:PMgivenN_best_fid_PlanckWMAP},
\ref{fig:meanscatterchains_fid_PlanckWMAP}
and \ref{fig:lnNsigmalnN_fid_PlanckWMAP}.
More specifically,
the mean relation $\langle M|N \rangle$ for the middle redshift bin
($0.4 \leq z_{\rm cl} \leq 0.7$)
has the $\sim$$20\%$ higher amplitude than those for the other
redshift bins
for both the {\it Planck} and {\it WMAP} cosmologies
as shown in Figure~\ref{fig:meanscatterchains_fid_PlanckWMAP} with relatively high significance,
whereas the scatter relation $\sigma_{\ln M|N}$ for the middle redshift bin
is slightly smaller than those for the other redshift bins.
In addition, Figure~\ref{fig:lnNsigmalnN_fid_PlanckWMAP}
shows the non-monotonic behaviors in the mean and scatter relations of $P(\ln N|M, z)$
as a function of redshift.

A possible explanation for the non-monotonic behavior
as a function of redshift is different impacts of
 projection effects at different cluster redshifts.
To illustrate this point, in Figure~\ref{fig:camira_zclzspec} we show
a comparison between the photometric cluster redshifts measured from the CAMIRA algorithm
and the available spectroscopic redshifts of BCGs without the $4 \sigma$ clipping
done in \cite{Ogurietal2018a}.
The standard deviations of the difference between the photometric and spectroscopic redshifts
are smaller around $0.4 \leq z_{\rm cl} \leq 0.55$ than for lower and higher cluster redshifts,
which is also a non-monotonic behavior in terms of redshift.
The larger errors at high redshifts can be
understood by larger errors on galaxy magnitudes,
whereas the larger errors at low redshifts are
most likely due to the lack of $u$-band in the HSC
survey as well as too-bright galaxy magnitudes
(leading to the saturation in some cases) for such
low redshifts.

Optical cluster-finding algorithms in imaging surveys
essentially use photometric redshifts of individual
galaxies to identify cluster member galaxies. The larger
cluster photometric redshift errors imply that photometric
redshift errors of individual galaxies are also larger,
leading to larger contaminations of non-member
galaxies along the line-of-sight direction in estimating the cluster
richness.
Since the photometric redshift errors in the middle redshift bin ($0.4 \leq z_{\rm cl} \leq 0.7$)
are smaller than in the other redshift bins,
we expect that CAMIRA separates true member galaxies from non-member galaxies
along the line-of-sight direction more effectively 
in the middle redshift bin.
In this case,
the mean richness values $\langle \ln N(M,z) \rangle$
should be smaller than
those in the other redshift bins for a fixed mass
even when non-monotonic behaviors in terms of redshift
do not exist for intrinsic richness values (i.e., without non-member galaxies along the line-of-sight direction).
This is consistent with the result in Figure~\ref{fig:lnNsigmalnN_fid_PlanckWMAP}
for the mean relation $\langle \ln N(M,z) \rangle$ for the typical masses.
In addition,
the non-monotonic behaviors in the order of the mean relation $P_{\beta}(\ln M|N)$
in Figure~\ref{fig:meanscatterchains_fid_PlanckWMAP}
might also be interpreted by the explanation above.
Specifically, the smaller observed richness for a fixed mass
leads to the higher mass for a fixed richness,
since we expect a smaller contribution
of non-member galaxies to observed richness values
at the redshift
with smaller photometric redshift errors.

The projection effect modifies not only the mean relation
but also the scatter of the richness-mass relation.
In particular, the larger projection effect implies
larger scatter because the projection effect depends
sensitively on the projection direction such that a
large projection effect is expected when it is
projected along the direction of the filamentary structure.
This is also consistent with the results
in Figures~\ref{fig:meanscatterchains_fid_PlanckWMAP} and \ref{fig:lnNsigmalnN_fid_PlanckWMAP},
which show slightly smaller scatter values $\sigma_{\ln M|N}$ in the middle redshift bin
than the lower and higher redshift bins,
and smaller scatter values $\sigma_{\ln N|M,z}$ around $z_{\rm cl}\sim 0.5$
for typical masses, respectively.
\begin{table}
  \caption{
    The median and 68$\%$ percentile uncertainties of the model parameters
    with the {\it Planck} cosmological parameters
    when we use the measurements only from one redshift bin.
    $^{*}$
  }
  \begin{center}
    \begin{tabular*}{0.48 \textwidth}{cccc} \hline\hline
      \centering
      Parameter &  Low-$z$  & Middle-$z$ & High-$z$ \\ \hline
                                                                                                                                                                       
      $A$ & $3.34^{+0.25}_{-0.20}$ & $3.19^{+0.20}_{-0.15}$ & $3.31^{+0.15}_{-0.26}$ \\
      $B$ & $0.85^{+0.08}_{-0.07}$ & $0.94^{+0.09}_{-0.07}$ & $0.88^{+0.08}_{-0.05}$ \\
      $\sigma_0$ & $0.36^{+0.07}_{-0.21}$ & $0.33^{+0.06}_{-0.21}$ & $0.27^{+0.14}_{-0.20}$ \\
      $q$ & $-0.06^{+0.09}_{-0.11}$ & $-0.08^{+0.09}_{-0.09}$ & $-0.03^{+0.04}_{-0.11}$ \\
      $f_{\rm cen}^{1, \beta}$ & $0.52^{+0.18}_{-0.27}$ & $0.43^{+0.12}_{-0.21}$ & $0.64^{+0.21}_{-0.17}$ \\
      $f_{\rm cen}^{2, \beta}$ & $0.75^{+0.13}_{-0.35}$ & $0.62^{+0.13}_{-0.16}$ & $0.39^{+0.20}_{-0.21}$ \\
      $f_{\rm cen}^{3, \beta}$ & $0.73^{+0.13}_{-0.45}$ & $0.86^{+0.09}_{-0.12}$ & $0.62^{+0.19}_{-0.24}$ \\
      $R_{1, \rm off}$ & $0.33^{+0.20}_{-0.24}$ & --- & --- \\
      $R_{2, \rm off}$ & --- & $0.47^{+0.19}_{-0.20}$ & --- \\
      $R_{3, \rm off}$ & --- & --- & $0.50^{+0.22}_{-0.25}$ \\
      $m_{\rm lens}$ & $0.00^{+0.01}_{-0.01}$ & $0.00^{+0.01}_{-0.01}$ & $0.00^{+0.01}_{-0.01}$ \\ \hline
      $\chi^2_{\rm min}/{\rm dof}$ & $30.6/29$ & $25.2/29$ & $34.4/29$ \\ \hline
    \end{tabular*}
  \end{center}
  \tabnote{
    $^{*}$
    The ``Low-$z$'' column shows the results from the lensing and abundance measurements only from $0.1 \leq z_{\rm cl} \leq 0.4\ (\beta=1)$,
    ``Middle-$z$'' only from $0.4 \leq z_{\rm cl} \leq 0.7\ (\beta=2)$, and ``High-$z$'' only from $0.7 \leq z_{\rm cl} \leq 1.0\ (\beta=3)$.
    Here we use the same prior ranges for the model parameters shown in Table~\ref{tab:fidparams} for the model parameters,
    whereas we use different parameters of $f_{\rm cen}^{1, \beta}$, $f_{\rm cen}^{2, \beta}$ 
    and $f_{\rm cen}^{3, \beta}$ in equation~(\ref{eq:fcenparam})
    instead of $f_0$, $f_N$, and $f_z$ in Table~\ref{tab:fidparams}.
    We use a flat prior between 0 and 1 for $f_{\rm cen}^{1, \beta}$, $f_{\rm cen}^{2, \beta}$ and $f_{\rm cen}^{3, \beta}$.
    For these analyses we do not include redshift evolution parameters when fitting the richness-mass relation.
  }
  \label{tab:params_onezbin}
\end{table}

To check the robustness of the fiducial result
to its parametrization for the redshift evolution,
we repeat the MCMC analysis by using only one of the three redshift bins
with a simpler richness-mass relation model without redshift evolution parameters
(i.e., only $A$, $B$, $\sigma_0$, and $q$ for the richness mass relation),
assuming {\it Planck} cosmology with the same covariances as the fiducial analysis.
This model is similar to the one used in \cite{Murataetal2018} for SDSS redMaPPer clusters over $0.10 \leq z_{\rm cl} \leq 0.33$.
Table~\ref{tab:params_onezbin} shows the parameter constraint from this model for each redshift bin.
We find that the $p$-values are acceptable for all redshift bins.
Figure~\ref{fig:meanscatter_onez_Planck}
shows the median and the 16th and 84th percentiles of $\langle M|N \rangle$
and $\sigma_{\ln M|N}$ from 
the parameter constraints shown in 
Table~\ref{tab:params_onezbin}.
The mean and scatter constraints shown in Figure~\ref{fig:meanscatter_onez_Planck}
are consistent with the fiducial result 
shown in Figure~\ref{fig:meanscatterchains_fid_PlanckWMAP}
within the errors.
It is worth noting that the mean relation in the middle redshift bin 
also favors higher values
than 
for the lower and higher redshift bins, which is similar to
the fiducial results.
This result supports the non-monotonic 
redshift evolution of the richness-mass
relation found in our fiducial analysis,
although the significance is not very high given the larger errors.
%
\section{Conclusion} \label{sec:conclusion}
In this paper, 
we present the results of the richness-mass relation analysis of 1747 HSC CAMIRA clusters
in a wide redshift range ($0.1 \leq z_{\rm cl} \leq 1.0$) with 
a richness range of $N \geq 15$
by jointly fitting to the stacked weak lensing profiles and abundance measurements
from the HSC-SSP first-year data ($\sim$232 ${\rm deg^2}$ for the cluster catalog and $\sim$140 ${\rm deg^2}$ for the shear catalog).
The exquisite depth and image quality of the HSC survey enables us to measure 
stacked weak lensing signals 
even for high-redshift clusters at
$0.7 \leq z_{\rm cl} \leq 1.0$ with a total signal-to-noise ratio of $19$.

We constrain the richness-mass relation 
defined in equations~(\ref{eq:mean_relation}) and~(\ref{eq:scatter_M})
assuming a log-normal distribution for the relation $P(\ln N|M,z)$
for both the {\it Planck} and {\it WMAP} cosmological parameters,
based on a forward modeling method.
We constrain the richness-mass relation parameters without informative priors  
when marginalizing over off-centering effects on the stacked lensing profiles.
We employ an $N$-body simulation-based 
halo emulator
for theoretical predictions of the halo mass function and the lensing profiles.
We also use an analytic model for the sample covariance matrix, which is validated 
against the HSC mock shear and halo catalogs.
We find that our model simultaneously fits the stacked lensing profiles and abundance measurements quite well 
for both the {\it Planck} and {\it WMAP} cosmological parameters 
with $\chi^{2}_{\rm min}/{\rm dof}=107.0/97$ and $\chi^{2}_{\rm min}/{\rm dof}=106.6/97$, respectively.
We check the robustness of the results against the choice of 
different photo-$z$ catalogs and source selection cuts.
We also show how 
the stacked lensing and abundance measurements 
individually constrain the model parameters,
and show that the joint analysis efficiently breaks the richness-mass parameter degeneracies.

We then derive
the mass-richness relation $P_{\beta}(\ln M|N)$ in each redshift bin,
using Bayes theorem from the constraint on $P(\ln M|N,z)$.
We show that the mean relations $\langle M|N \rangle$ in each redshift bin
are consistent between the {\it Planck} and {\it WMAP} cosmological parameters within the errors,
but the scatter relation values $\sigma_{\ln M|N}$ 
for the {\it Planck} cosmological model are larger than those for the {\it WMAP} model.
In addition, scatter values for the {\it Planck} model increase
toward lower richness values, whereas those for the {\it WMAP} model 
are consistent with constant values as a function of richness.

We also show that 
we need to include the linear and square redshift-dependent parameters 
in terms of $\ln(1+z)$
for the mean and scatter relations in $P(\ln N|M,z)$ 
to have acceptable $p$-values.
The models without such redshift-dependent parameters 
resulted in much worse $p$-values.
By including the square redshift-dependent parameters,
we show that the mean relation $\langle M|N \rangle$ in the middle redshift bin 
has $\sim$$20\%$ higher amplitude 
than in the lower and higher redshift bins, 
whereas
the scatter relation $\sigma_{\ln M|N}$ in the middle redshift bin is slightly smaller 
than in the other bins.
We ascribe this non-monotonic redshift dependence to
the non-monotonic behavior of the projection effect 
as a function of redshift, which is supported by the 
redshift dependence of cluster photometric redshift 
errors.
Redshift evolution in the mean relation $\langle M|N \rangle$ should be properly accounted for
when one uses the stacked weak lensing signals around 
the HSC CAMIRA clusters to validate shear and photo-$z$ catalogs, 
or to define source selection cuts \citep[e.g.,][]{Medezinskietal2018b},
by matching the cluster weight distributions in terms of cluster redshift and richness values.
We also check the consistency of our fiducial results based on the richness-mass relation
including the redshift-dependent parameters
with those from the analysis of individual redshift bins
without redshift-dependent parameters.

Our results indicate that we cannot distinguish between
{\it Planck} and {\it WMAP} cosmological models from 
the current abundance and lensing profile 
measurements. This is partly because of our choice 
of a flexible richness-mass relation model without
any informative prior constraints on the model parameters. 
However, we find that the predicted scatter 
values are clearly different between {\it Planck} and 
{\it WMAP} cosmologies, suggesting that any additional
constraints on the scatter of the mass-richness relation
may break the degeneracy between these two 
cosmological models.
For instance, one could add other independent probes 
(e.g., X-ray temperature, the Sunyaev--Zel'dovich effect, or lensing magnification effect)
for constraining the scatter values
in order to distinguish cosmological models from 
cluster observables. 
The analysis of lensing magnification effect might provide complementary information, and
will constrain the richness-mass relation of the HSC CAMIRA clusters (Chiu et al. {\it in prep.}).

Another possible observable to break the degeneracy is spatial clustering of clusters. 
Here we briefly investigate the difference of the real-space three-dimensional halo-halo correlation function $\xi_{\rm hh}(r)$ in each redshift and richness bin
between the {\it Planck} and {\it WMAP} models
by using the best-fit richness-mass relation parameters
from the fiducial analyses in this paper,
to roughly assess 
its power to break the degeneracy.
We find that the 
predicted amplitudes of the halo-halo correlation function for 
the {\it WMAP} model are $\sim$$20$--$30\%$ larger than those for the {\it Planck} model
for all richness and redshift bins at $3 \lesssim r \lesssim 50 h^{-1}{\rm Mpc}$.
This implies that clustering of clusters 
adds useful information that is complementary to abundance and lensing, although
careful investigations
of cluster photo-$z$ accuracy, redshift-space distortions, and projection effects 
should be 
conducted in order to combine our results with 
the cluster clustering analysis to obtain tight cosmological constraints. 
We leave this exploration for future work.

In addition, our result should be useful for other cluster-related observational studies
\citep[e.g.,][]{Linetal2017, Jianetal2018, Nishizawaetal2018, Miyaokaetal2018, Otaetal2018, Hashimotoetal2018}, 
including
galaxy formation in cluster regions and the mass scale estimation 
for clusters detected via strong-lensing, X-ray, and Sunyaev--Zel'dovich effect with richness values.
We can also use our constraint on $P(\ln N|M,z)$ 
to construct mock CAMIRA cluster catalogs 
with richness values by using halo mass and redshift in 
$N$-body 
simulations, which may be useful for testing 
the performance of cluster-finding algorithms with 
simulations
\citep[e.g.,][]{Dietrichetal2014, Ogurietal2018a, Costanzietal2018}.

Our analysis involves several assumptions.
Most critically, we have assumed 
that the CAMIRA clusters
are randomly oriented with respect to the 
line-of-sight direction
in the forward modeling method to compute the cluster observables 
from the mass function and 
the spherically averaged halo-matter cross-correlation function 
from $N$-body simulation outputs.
However, 
this assumption is inaccurate if the CAMIRA clusters are affected 
by projections effects such as misidentification of non-member galaxies along the line-of-sight direction
as member galaxies in the richness estimation \citep[e.g.,][]{Cohnetal2007, Zuetal2017, BuschandWhite2017, Costanzietal2018, Sunayama&More2019}.
Since our cluster selection is based on richness values, 
CAMIRA could preferentially detect clusters with filamentary structure along the line-of-sight direction.
Given the correlation between the halo 
orientation and surrounding large-scale structure, 
this effect can change
the lensing profile from the spherically-symmetric case \citep{Osatoetal2018}.
This investigation is beyond the scope of this paper.
Further careful investigations of projection effects for the mass-richness relation are warranted.
In order to properly address the projection effects, 
we need to construct realistic mock catalogs of the CAMIRA clusters with intrinsic richness values
and to derive
cluster observables accounting for the projection effects.
Our results 
might be useful to check the consistency of such simulation setups with the observations
by comparing our measurements and constraints on the mean and scatter relations 
with ones from such mock catalogs with projection effects.

We have presented a richness-mass relation analysis from HSC first-year data.
We will have $\sim$1000 ${\rm deg^2}$ area for cluster and shear catalogs when the HSC-SSP survey is complete in 2020.
The final HSC cluster analysis has the potential to provide stronger constraints 
on the richness-mass relation 
and its better applications for cosmological, galaxy formation, and cluster-related studies 
particularly when combined with
other probes.
There is also room for improving 
the measurement methods 
e.g., improving the shear measurement technique to include more galaxy shapes, 
and increasing the sample of galaxies with spectroscopic redshifts independent from the COSMOS 30-band catalog
to improve and understand the accuracy of cluster photometric
redshifts, and also improving 
the model framework 
by further accounting for possible systematic effects 
such as projection effects.
\bigskip
\begin{ack} \label{sec:ack}
We thank the anonymous referee for helpful comments that improved the quality of this work.

RyM acknowledges financial support from
the University of Tokyo-Princeton strategic partnership grant,
Advanced Leading Graduate Course for Photon Science (ALPS),
Research Fellowships of the Japan Society for the Promotion of Science
for Young Scientists (JSPS), and JSPS Overseas Challenge Program for Young Researchers.
This work was supported by JSPS KAKENHI Grant Numbers JP15H05892, JP17J00658, JP17K14273, JP18K03693, and JP18H04358,
and by Japan Science and Technology Agency CREST JPMHCR1414.
This work was supported by World Premier International Research Center Initiative (WPI Initiative), MEXT, Japan.
Numerical computations were in part carried out on Cray XC30 and XC50 at Center for Computational Astrophysics, National Astronomical Observatory of Japan.

The Hyper Suprime-Cam (HSC) collaboration includes the astronomical communities of Japan and Taiwan, and Princeton University. The HSC instrumentation and software were developed by the National Astronomical Observatory of Japan (NAOJ), the Kavli Institute for the Physics and Mathematics of the Universe (Kavli IPMU), the University of Tokyo, the High Energy Accelerator Research Organization (KEK), the Academia Sinica Institute for Astronomy and Astrophysics in Taiwan (ASIAA), and Princeton University. Funding was contributed by the FIRST program from Japanese Cabinet Office, the Ministry of Education, Culture, Sports, Science and Technology (MEXT), the Japan Society for the Promotion of Science (JSPS), Japan Science and Technology Agency (JST), the Toray Science Foundation, NAOJ, Kavli IPMU, KEK, ASIAA, and Princeton University.

This paper makes use of software developed for the Large Synoptic Survey Telescope. We thank the LSST Project for making their code available as free software at http://dm.lsst.org.

The Pan-STARRS1 Surveys (PS1) have been made possible through contributions of the Institute for Astronomy, the University of Hawaii, the Pan-STARRS Project Office, the Max-Planck Society and its participating institutes, the Max Planck Institute for Astronomy, Heidelberg and the Max Planck Institute for Extraterrestrial Physics, Garching, The Johns Hopkins University, Durham University, the University of Edinburgh, Queen’s University Belfast, the Harvard-Smithsonian Center for Astrophysics, the Las Cumbres Observatory Global Telescope Network Incorporated, the National Central University of Taiwan, the Space Telescope Science Institute, the National Aeronautics and Space Administration under Grant No. NNX08AR22G issued through the Planetary Science Division of the NASA Science Mission Directorate, the National Science Foundation under Grant No. AST-1238877, the University of Maryland, and Eotvos Lorand University (ELTE) and the Los Alamos National Laboratory.

Based on data collected at the Subaru Telescope and retrieved from the HSC data archive system, which is operated by Subaru Telescope and Astronomy Data Center, National Astronomical Observatory of Japan.
\end{ack}
\appendix
%
\section{Covariance}
We use analytic calculations of the sampling variance contribution to the covariances,  
assuming that the distribution of clusters and lensing fields obeys the Gaussian statistics.
We describe the analytic model and the detailed estimation procedure in Appendix~\ref{appendix:analyticcov},
and validate it against realistic shear and cluster mock catalogs in Appendix~\ref{sec:appendix:compmock}.
We also use an analytic model for Poisson shot noise in the abundance covariance.
On the other hand,  
we do not use an analytic model for the shape noise covariance in the lensing profiles,
but rather estimate it 
directly from the data catalogs as described below.
\subsection{Analytic model of the covariance matrix} \label{appendix:analyticcov}
We employ an analytic covariance model for cluster abundances
\citep{HuandKravtsov2003, Takada&Bridle2007, OguriandTakada2011} as
\begin{equation}
{\bf C}[ N_{\alpha, \beta}, N_{\alpha',\beta' } ]
= N_{\alpha, \beta}\delta_{\alpha \alpha' }^{K} \delta_{\beta \beta' }^{K}
+ S_{\beta, \alpha \alpha' } \delta_{\beta \beta' }^{K},
\end{equation}
where $\delta_{\alpha \alpha'}^{K}$ denotes the Kronecker delta function.
The first term denotes the Poisson shot noise from the finite number of available clusters 
and the second term gives the sample covariance as
\begin{eqnarray}
S_{\beta, \alpha \alpha' }&=&  N_{\alpha, \beta} N_{ \alpha', \beta } 
\int_{z_{\beta, {\rm min}}}^{z_{\beta, {\rm max}}} \frac{{\rm d}z}{H(z)} W^{\rm h}_{\alpha, \beta}(z) 
W^{\rm h}_{ \alpha', \beta}(z) \chi^{-2}(z) \nonumber \\
&\times& \int \frac{\ell {\rm d} \ell}{2 \pi} \left| \tilde{W}_s(\ell \Theta_s) \right|^2 P_{\rm mm}^{\rm L}\left(k=\frac{\ell}{\chi}; z\right),
\end{eqnarray}
where $\tilde{W}_s(\ell \Theta_s)$ is the Fourier transform of the survey window function,
for which we assume a circular survey geometry with survey area $\Omega_{\rm tot}=\pi \Theta_s^{2}$ for simplicity:
$\tilde{W}_s(\ell \Theta_s)=2 J_1(\ell \Theta_s)/(\ell \Theta_s)$.
We use \textsc{CAMB} \citep{Lewisetal2000} for computing the linear matter power spectrum $P_{\rm mm}^{\rm L}(k; z)$.
The halo weight function is defined as
\begin{eqnarray}
W^{\rm h}_{\alpha, \beta}&(&z)= \frac{ \Omega_{\rm tot} }{ N_{\alpha, \beta} } \chi^{2}(z)  \nonumber \\
&& \times \int {\rm d}M \frac{ {\rm d}n }{ {\rm d}M } S(M, z|N_{\alpha, {\rm min}}, N_{\alpha, {\rm max}}) b_{\rm h}(M; z).
\end{eqnarray}
Here $b_{\rm h}(M; z)$ is the bias parameter for halos with mass $M$ at redshift $z$, 
for which
we employ a halo bias function 
presented in \cite{Tinkeretal2010} calculated using the {\it colossus} package 
\citep{Diemer2018}.

We calculate the covariance model 
for the stacked lensing profiles among different redshift, richness, and radial bins 
by decomposing it into the shape noise covariance ${\bf C}^{\rm SN}$ and the sample covariance ${\bf C}^{\rm SV}$
as
\begin{eqnarray}
{\bf C}= {\bf C}^{\rm SN}+{\bf C}^{\rm SV},
\label{eq:full_cov}
\end{eqnarray}
where we compute the shape noise covariance by randomly rotating background galaxies \citep[e.g.,][]{Murataetal2018}.
More specifically, we measure the lensing estimators around the clusters in the data catalog
with all the multiplicative biases
after randomly rotating background shapes in the HSC data catalog, repeating the process 15,000 times.
We then calculate the covariance among different richness, redshift, and radial bins based on these measurements.

For the sample covariance of the lensing profiles, 
we use a Gaussian covariance \citep{OguriandTakada2011, Shirasaki&Takada2018} as
\begin{eqnarray}
&&{\bf C}^{\rm SV}[ \Delta\Sigma_{\alpha, \beta}(R_m), \Delta\Sigma_{ \alpha', \beta' }(R_n)] \nonumber \\ 
&=& \frac{1}{\Omega_{\rm lens} \langle \chi_{l,\beta} \rangle \langle \chi_{l,\beta'} \rangle} 
\int \frac{ k {\rm d}k }{ 2 \pi } \nonumber \\
&\times& \left[  C_{\kappa\kappa,\alpha\alpha'\beta}^{mn}(k) C_{\rm hh,\alpha\alpha'\beta}^{\rm obs}(k) \delta_{\beta \beta'}^{K} 
+ C_{\Delta\Sigma, \alpha\beta}(k) C_{\Delta\Sigma, \alpha' \beta'}(k)\right] \nonumber\\
&\times& \widehat{J}_2( k R_m) \widehat{J}_2(k R_n),
\label{eq:analyticLensCV}
\end{eqnarray}
where the power spectrum of convergence fields with $\langle \Sigma_{\rm cr} \rangle_{ls}$ terms are defined as
\begin{eqnarray}
& &C_{\kappa\kappa, \alpha\alpha'\beta}^{mn}(k)= \left< \Sigma_{\rm cr} \right>_{ls, \alpha \beta}(R_m) \left< \Sigma_{\rm cr} \right>_{ls, \alpha' \beta }(R_n) \times \nonumber \\
&&\int \frac{ {\rm d}z}{ H(z) } \left( \bar{\rho}_{\rm m 0} \Sigma_{\rm cr}^{-1}(z, z_{s, \beta} )
\frac{ \langle \chi_{l,\beta} \rangle }{ \chi(z) }\right)^{2} 
P_{\rm mm}\left(k'=\frac{ \langle \chi_{l,\beta} \rangle}{\chi(z)} k, z\right), \nonumber \\
\end{eqnarray}
and $\langle \chi_{l,\beta} \rangle$ is the average of $\chi(z)$ in $\beta$-th cluster redshift bin from the data with $N \geq 15$.
Here $z_{s, \beta}$
is the weighted mean of $z_{s,{\rm best}}$ ($z_{\rm best}$ for a source galaxy, $s$)
in $\beta$-th cluster redshift bin over all the radial bins as
\begin{equation}
z_{s, \beta} =\frac{ \displaystyle \sum_{l,s; z_l\in z_\beta} z_{s, \rm{best}} w_{ls} }{ \displaystyle \sum_{l,s; z_l\in z_\beta} w_{ls} }.
\end{equation}
We find $z_{s,1}=1.09$, $z_{s,2}=1.30$, and $z_{s,3}=1.57$ 
for the fiducial photo-$z$ catalog and source selection cut with the {\it Planck} cosmological parameters.
The weighted mean critical surface mass density is calculated as
\begin{equation}
\left< \Sigma_{\rm cr} \right>_{ls, \alpha\beta}(R)=\frac{ \displaystyle \left. \sum_{l,s; N_l\in N_\alpha, z_l\in z_\beta} \left< \Sigma_{\rm cr}^{-1} \right>_{ls}^{-1} w_{ls}\right|_{R=\chi_l|\btheta_l-\btheta_s|} }{ \displaystyle \left. \sum_{l,s; N_l\in N_\alpha, z_l\in z_\beta} w_{ls} \right|_{R=\chi_l|\btheta_l-\btheta_s|} }.
\end{equation}
We use \textsc{Halofit} \citep{Smithetal2003} for the nonlinear matter power spectrum based on \cite{Takahashietal2012}. 
The terms $C_{\rm hh,\alpha\alpha'\beta}^{\rm obs}$ and $C_{\Delta\Sigma, \alpha\beta}$ are defined as
\begin{equation}\label{eq:Chhobs}
C_{\rm hh,\alpha\alpha'\beta}^{\rm obs}(k)=C_{\rm hh,\alpha\alpha'\beta}(k) + \frac{ \langle \chi_{l,\beta} \rangle^2 \Omega_{\rm tot} }{ N_{\alpha, \beta} }\delta_{\alpha \alpha'}^{K},
\end{equation}
with
\begin{equation}
C_{\rm hh,\alpha\alpha'\beta}(k)=\int_{ z_{\beta, \rm min} }^{ z_{\beta, \rm max} } \frac{{\rm d}z}{H(z)} W_{\alpha,\beta}^{\rm h}(z) W_{\alpha',\beta}^{\rm h}(z) P^{\rm L}_{\rm mm}(k; z),
\end{equation}
and
\begin{eqnarray}
&C&_{\Delta\Sigma,\alpha\beta}(k)
=\frac{\Omega_{\rm tot}}{N_{\alpha,\beta}} 
\bar{\rho}_{\rm m0} 
\int_{ z_{\beta, \rm min} }^{ z_{\beta, \rm max} } \frac{ {\rm d}z}{H(z)}~\chi^2(z) \nonumber \\
&&\times \int {\rm d}M \frac{ {\rm d}n }{ {\rm d}M } 
S(M,z|N_{\alpha, \rm min},N_{\alpha, \rm max}) P_{\rm hm}(k; M,z). 
\end{eqnarray}
The second-order Bessel function after averaging within radial bins is given as
\begin{equation}
\widehat{J}_2(k R_n)=\frac{2}{R^2_{n, {\rm max}}-R^2_{n, {\rm min}} } 
\int_{ R_{n, {\rm min}} }^{  R_{n, {\rm max}} } {\rm d}R~R~J_2(k R).
\end{equation}

We do not account for the window function effect 
of the cluster and shear catalogs in this analytic model. 
In Appendix~\ref{sec:appendix:compmock}, 
we validate this analytical covariance 
by using realistic mock shear and cluster catalogs.
We also ignore 
the cross-covariance between the stacked lensing profiles and abundance measurements 
since 
this cross-covariance does not have a large impact in the parameter estimation,
which we confirm by using 
the cross-covariance estimated from the mock catalogs.
Specifically, we repeat the MCMC analysis 
based on the fiducial covariance 
with the cross-covariance from the mock catalogs derived in Appendix~\ref{sec:appendix:compmock}
to find that the $68\%$ percentile widths are consistent with the fiducial ones 
and the shift of the $\chi^{2}_{\rm min}$ value from the fiducial value is $\sim 0.1$.

In the parameter estimation, we fix the richness-mass relation parameters
for the analytic covariance model
to reduce the model calculation time (especially of the lensing covariance).
For each setup of the photo-$z$ catalog, source selection cut, and the cosmological parameters,
we estimate the analytic covariance model as follows.
First, we perform the MCMC analysis with a simpler covariance model 
which does not include the richness-mass relation dependent terms of 
$C_{\Delta\Sigma, \alpha\beta}(k) C_{\Delta\Sigma, \alpha' \beta'}(k)$ in equation~(\ref{eq:analyticLensCV})
and $C_{\rm hh,\alpha\alpha'\beta}(k)$ in equation~(\ref{eq:Chhobs}).
We do not fix the richness-mass relation parameters
for other terms in the abundance and lensing profiles.
We obtain the best-fit parameters as
$\{A, B, B_z, C_z, \sigma_0, q, q_z, p_z\}=\{ 3.16, 0.92, -0.13, 4.17, 0.29, -0.12, -0.02, 0.52 \}$
for the {\it Planck} model, and
$\{A, B, B_z, C_z, \sigma_0, q, q_z, p_z\}=\{ 3.37, 0.84, -0.14, 4.47, 0.17, -0.02, 0.19, 0.50  \}$
for the {\it WMAP} model
with the fiducial photo-$z$ catalog (\texttt{MLZ}) and source selection cut ({\it Pcut} with $\Delta z=0.1$).
Second, we calculate the covariance with the analytic model based on these parameters from the simpler covariance
to derive the parameter constraints.
For the case of the fiducial photo-$z$ catalog and source selection cut,
the richness-mass relation parameters for the covariance calculation above
are consistent with our final results shown in Table~\ref{tab:fidparams},
and the $\chi^{2}_{\rm min}$ values are not very different 
from the 
final values shown in 
Table~\ref{tab:fidparams}.
We show our covariance matrix for the fiducial setup 
with the {\it Planck} cosmological parameters 
in Figures~\ref{fig:Cij_fid_Planck} and \ref{fig:rij_fid_Planck}.
\subsection{Validation against realistic mock shear and cluster catalogs}\label{sec:appendix:compmock}
We validate 
our model of the covariance matrix presented in
Appendix~\ref{appendix:analyticcov}
against the realistic HSC shear and halo catalogs \citep[][see Section~\ref{sec:simulation:mockcat} for more details]{Shirasakietal2019}.
We use 2268 realizations of the mock catalogs 
that share
the same footprints of the shear and halo catalogs as the {\it real} data catalogs. 
The cosmological parameters for the mock catalogs are the same as those for {\it WMAP} used in this paper.
We assign richness values for halos with $M\geq 10^{12} h^{-1}M_{\odot}$ 
to create the mock catalogs of the CAMIRA clusters with the richness values.
The richness values are assigned according to the
richness-mass relation parameters consistent with our results shown in Table~\ref{tab:fidparams} for the {\it WMAP} model
as $\{A, B, B_z, C_z, \sigma_0, q, q_z, p_z\}=\{3.37, 0.85, -0.14, 4.47, 0.18, -0.05, 0.19, 0.50\}$.
We repeat the measurements of the abundance and lensing profiles for each realization
to calculate the covariance matrix from the 2268 realizations.
Here we measure the lensing profiles from the shear values without shape noise as
we use the shape noise covariance estimated in Appendix~\ref{appendix:analyticcov}.
We also calculate the covariance contribution ${\bf C}^{\rm R}$ from the random subtraction \citep{Singhetal2017}.
As shown in Section~2.3 of \cite{Singhetal2017}, we subtract ${\bf C}^{\rm R}$ from the covariance above
to account for the random subtraction.
\cite{Murataetal2018} found that this term is negligible (${\bf C}^{\rm R}/{\bf C} \sim 0.01$ for the diagonal terms)
for the SDSS redMaPPer clusters.
Similarly, we find that these values for the HSC CAMIRA clusters are similar to those for the SDSS redMaPPer clusters
and thus are negligible for the HSC CAMIRA clusters.

\begin{figure*}
  \begin{center}
    {
     \includegraphics[width=8.0cm]{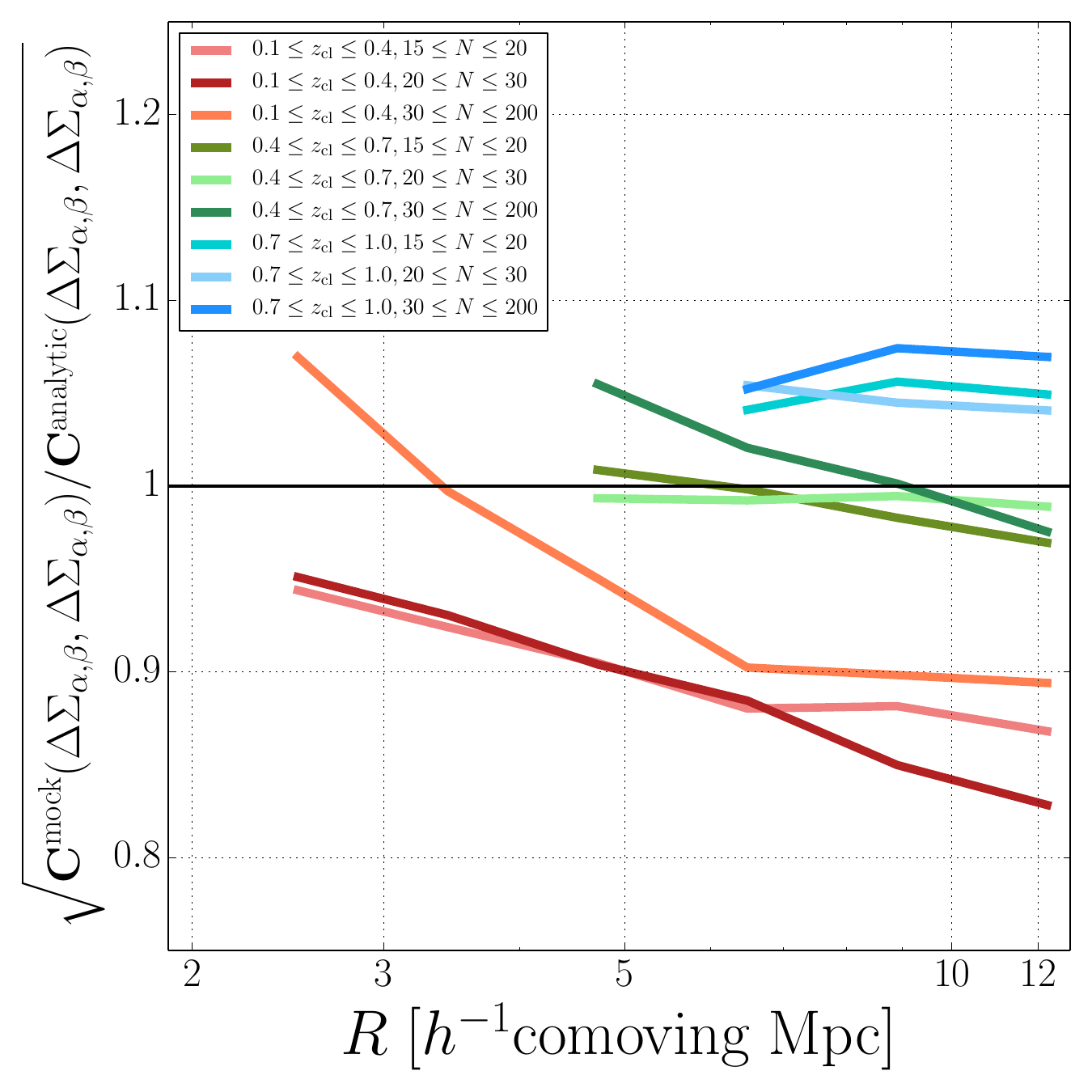}
     \includegraphics[width=9.0cm]{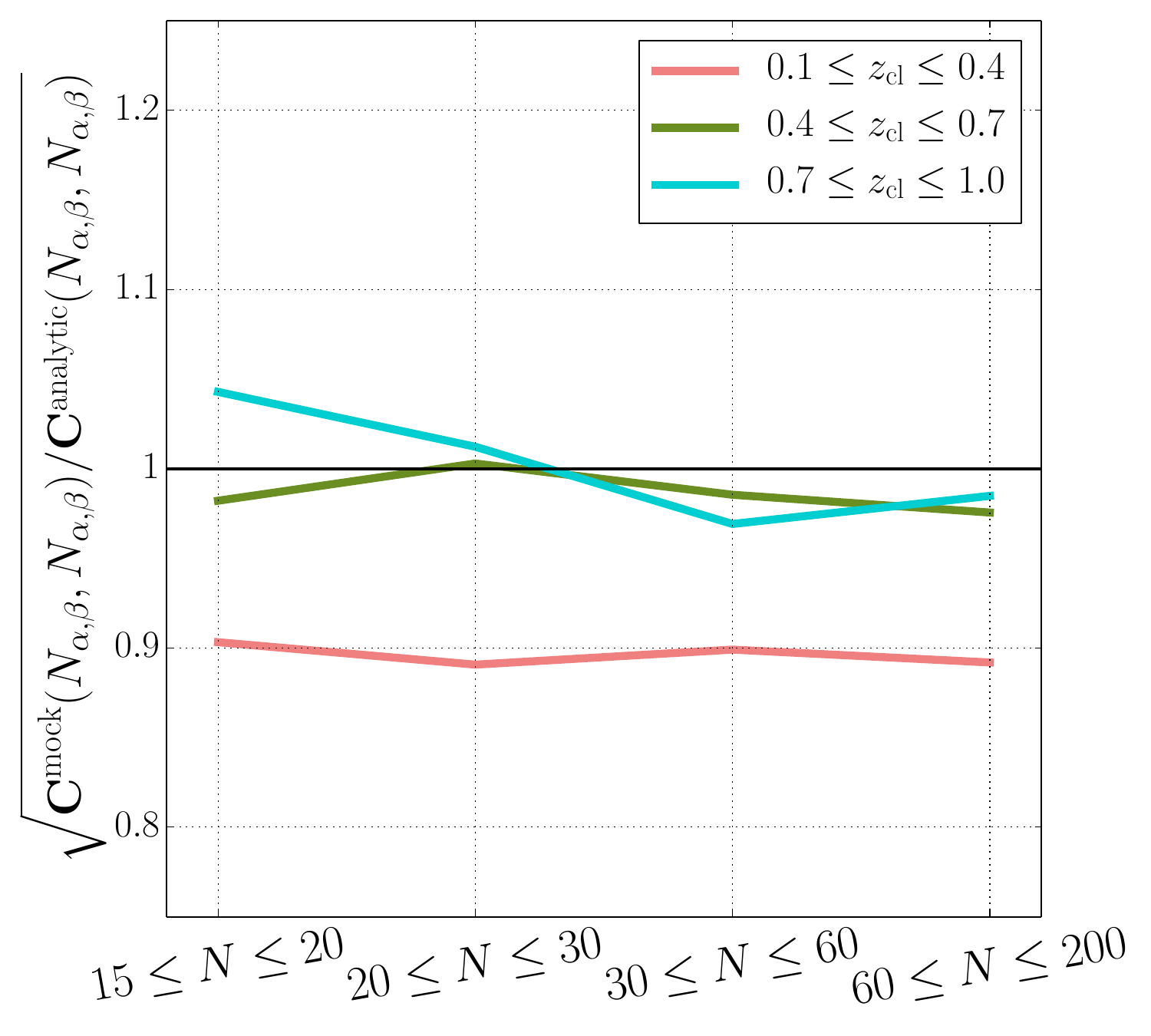}
    }
  \end{center}
  \caption{ 
    The comparison of diagonal components between the covariance estimated from the mock catalogs 
    and from the analytic model presented in Appendix~\ref{appendix:analyticcov}
    for the {\it WMAP} cosmological parameters. 
    Here we use the richness-mass relation parameters of 
    $\{A, B, B_z, C_z, \sigma_0, q, q_z, p_z\}=\{3.37, 0.85, -0.14, 4.47, 0.18, -0.05, 0.19, 0.50\}$,
    which are consistent with our fiducial result for the {\it WMAP} cosmological parameters shown in Table~\ref{tab:fidparams}.
    The left panel shows the comparison of the lensing covariance for each redshift and richness bin. 
    We only show the result on 
    the radial scales that are larger than the resolution limits in the mock catalog for each redshift bin. 
    We include the shape noise covariance estimated from randomly rotating galaxy shapes in the data catalog for both covariances.
    The right panel shows the comparison of the abundance covariances in each redshift bin.
    The diagonal parts of the analytic covariances match the mock covariance to better than $\sim$$10\%$.
    }
\label{fig:comparison_mockcov_analyticcov}
\end{figure*}

In Figure~\ref{fig:comparison_mockcov_analyticcov},
we show the comparison 
between the covariance estimated from mock catalogs
and the covariance with the analytic model for the {\it WMAP} model using the same richness-mass relation parameters.
Since the resolution of the lensing shear in the mock catalogs is limited to $0.43$ arcmin,
we compare the lensing covariance only above an effective resolution limit for each cluster redshift bin.
Here we set the resolution limit by comparing the mean of the lensing profiles from the mock catalogs with the model prediction 
(see Section~\ref{sec:modeling:lensing}).
The figure shows that the diagonal components of the covariance with the analytic model 
agree well with those from the mock catalogs 
at better than the $\sim$$10\%$ level for both the abundance and lensing profiles measurements.
We use the covariance with the analytic model for our parameter
estimation because the covariance matrix from the 
mock catalogs is affected by the resolution effect
as mentioned above.
%
\section{Model parameter constraint contours} \label{appendix:mcmccontour}
\begin{figure*}
  \begin{center}
    \includegraphics[width=17.0cm]{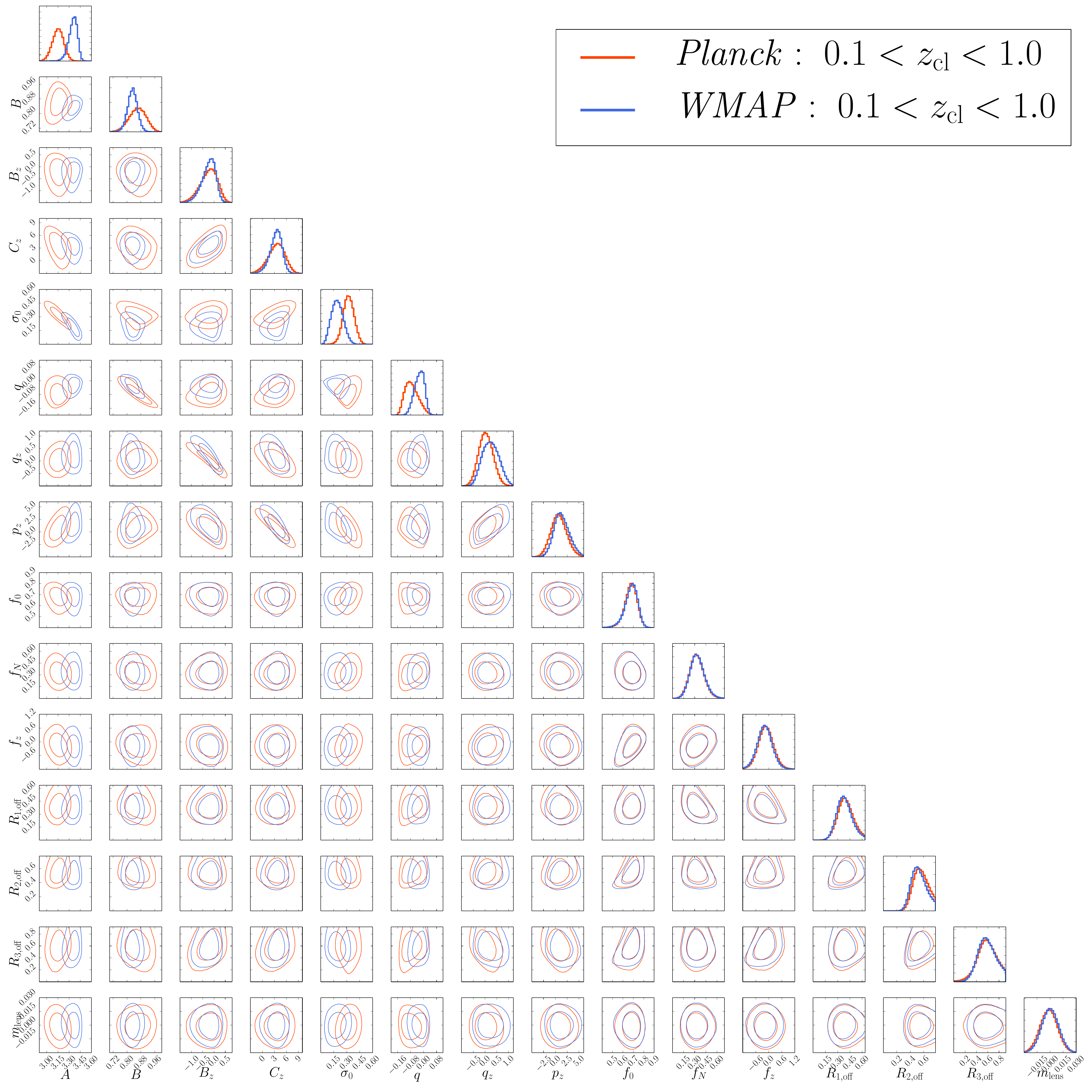}
  \end{center}
  \caption{ The model parameter constraints in the fiducial analysis for both the {\it Planck} and {\it WMAP} cosmological parameters.
            Diagonal panels show the posterior distributions of the model parameters,
            and the other panels show the 68$\%$ and 95$\%$ credible levels contours in each two-parameter subspace from the MCMC chains.
          }
\label{fig:MCMC_fid_PlanckWMAP}
\end{figure*}

We show the model parameter constraint contours in Figure~\ref{fig:MCMC_fid_PlanckWMAP} 
from the fiducial analysis
to show the marginalized one-dimensional posterior distributions for each parameter
and the 68$\%$ and 95$\%$ credible levels contours 
for each two-parameter subspace from the MCMC chains.
%
\bigskip
\noindent
%

\label{lastpage}
\end{document}